\documentclass{jfm}

\usepackage{natbib}

\usepackage{epstopdf, epsfig}
\usepackage{amsmath,amssymb,amsfonts,bm}
\usepackage{xcolor}
\usepackage{graphicx}
\usepackage{tikz}

\newcommand{\axialVelocity}{U_3}
\newcommand{\rayleigh}{Ra}
\newcommand{\prandtl}{\sigma}
\newcommand{\convectiveReynolds}{\widetilde{Re}_\ell}
\newcommand{\ekmanf}{E_f}

\newcommand{\pd}[1]{\partial_{#1}}
\newcommand{\wRa}{\widetilde{Ra}}

\newcommand{\horizResLap}{\nabla_\perp^{\prime 2}}

\newcommand{\cov}[1]{\boldsymbol{\widehat{e}}_{#1}}
\newcommand{\contra}[1]{\boldsymbol{\widehat{e}}^{#1}}
\newcommand{\baseCov}[1]{\boldsymbol{\widehat{e}}_{#1}}
\newcommand{\baseContra}[1]{\boldsymbol{\widehat{e}}^{#1}}
\newcommand{\nablaPerpPrime}{\boldsymbol{\nabla}_\perp^\prime}
\newcommand{\nablaPerpTwo}{\nabla_\perp^{\prime 2}}
\newcommand{\boldDot}{\boldsymbol{\cdot}}
\newcommand{\coords}{\xi}
\DeclareMathOperator{\divergence}{\mathrm{div}}

\newcommand{\tangle}{\vartheta_f}
\newcommand{\temperature}{T}
\newcommand{\bdot}{\boldsymbol{\cdot}}

\newcommand{\Sstress}{\mathcal{S}_{\mathrm{vorticity\, stress}}}
\newcommand{\Sbuoy}{\mathcal{S}_{\mathrm{buoyancy\, torque}}}

\def\haz{\boldsymbol{\widehat z}}
\def\hy{\boldsymbol{\widehat y}}
\def\hx{\boldsymbol{\widehat x}}

\DeclareMathOperator{\curl}{curl}

\def\lb{\left ( }
\def\rb{\right ) }
\def\lsq{\left [ }
\def\rsq{\right ] }
\def\lbr{\left \langle }
\def\rbr{\right \rangle }

\def\ub{\boldsymbol{u}}

\def\hz{\boldsymbol{\widehat \eta}}

\shorttitle{Quasi-geostrophic convection on the tilted $f$-plane}
\shortauthor{B. Miquel et al.}

\title{Quasi-geostrophic Rayleigh-B\'enard convection on the tilted $f$-plane}

\author{Benjamin Miquel\aff{1}\corresp{\email{benjamin.miquel@cnrs.fr}}, Abram Ellison\aff{2}, Michael A. Calkins\aff{3}, Keith Julien\aff{2}, Edgar Knobloch\aff{4}
}

\affiliation{\aff{1} CNRS, Ecole Centrale de Lyon, INSA Lyon, Universit\'e Claude Bernard Lyon 1, LMFA, UMR5509, 69130 Ecully, France,
\aff{2}Department of Applied Mathematics, University of Colorado,
Boulder, CO 80309, USA, 
\aff{3}Department of Physics, University of Colorado, 
Boulder, CO 80309, USA,
\aff{4}Department of Physics, University of California at Berkeley,
Berkeley, CA 94720, USA
}

\begin{document}

\maketitle

\begin{abstract}
Rapidly rotating Rayleigh-Bénard convection on a $f$-plane at colatitude $\vartheta_f$ is investigated 
numerically using an asymptotically reduced equation set valid in the limit of very 
rapid rotation. The equations provide a non-hydrostatic but quasi-geostrophic 
description in a non-orthogonal coordinate system. The tilt changes the structure of 
the large-scale barotropic condensate from large-scale vortices to zonal flows as the 
colatitude of the $f$-plane increases, with bistable states present for certain parameter 
ranges, extending prior work to a geophysically significant parameter regime. This 
behaviour is understood through the impact of broken rotation symmetry on the barotropic 
source terms resulting from baroclinic vortical stresses and baroclinic torque. 
As the tilt angle $\vartheta_f$ increases, global 
heat and momentum transport is reduced relative to upright-polar convection, a result 
that is explained through linear theory and nonlinear power maps both of which demonstrate 
increased attenuation of the domain of dynamically active spatial scales as the convective 
modes depart from a North-South alignment in the horizontal plane. A key finding is that 
the predominance of lateral thermal mixing allows for the maintenance of a persistent 
unstable mean temperature gradient that saturates at increasing forcing levels and remains 
insensitive to the colatitude.

\end{abstract}

\begin{keywords}
Authors should not enter keywords on the manuscript, as these must be chosen by the author during the online submission process and will then be added during the typesetting process (see http://journals.cambridge.org/data/\linebreak[3]relatedlink/jfm-\linebreak[3]keywords.pdf for the full list)
\end{keywords}

\newpage
\section{Introduction}

Buoyantly driven convection that is constrained by the Coriolis force is a ubiquitous
phenomenon occurring within planetary and stellar interiors where it acts as the power source for sustaining large scale magnetic fields \citep{cJ11b,pR13,jmA15,kS25,dormyGAFD25},
and large scale zonal winds \citep{uC02,vasavada2005,mH05,yK20,mH22,jN24}
and vortices \citep{aA2018,lS22}. Rotating convection is 
also thought to be an important source of turbulent mixing within the subsurface oceans
of icy moons \citep{kS2019,Bire2020,pagnoscinICARUS2026}, where it controls the
ice thickness~\citep{zengGRL2026}. Estimates of the non-dimensional parameters that 
characterise rotating turbulence are extreme \citep{gS11,kS2019}. Specifically, the bulk
global scale Reynolds number measuring turbulent intensity is large, 
\begin{equation}
Re_H\equiv\frac{\tau_\nu}{\tau_u}=\frac{U H}{\nu}\gg1 \nonumber,  \label{eqn:Re_H}
\end{equation}
while the Ekman and bulk Rossby numbers measuring the importance of rotation are small, 
\begin{equation}
\label{eqn:geo}
E\equiv\frac{\nu}{2\Omega H^2}=\frac{\tau_\Omega}{\tau_\nu}\ll1, \qquad
Ro_H\equiv\frac{U}{2\Omega H} = Re_H E =\frac{\tau_\Omega}{\tau_u}\ll 1 .
\end{equation}
Here, $\Omega$ denotes the rotation rate, $H$ the domain height, $\nu$ the kinematic
viscosity, and $U$ the characteristic flow speed. Evidently $E \ll Ro_H \ll 1$ implying
the relative time ordering $\tau_\Omega \ll \tau_u \ll \tau_\nu$ for the rotation
time $\tau_\Omega = (2\Omega)^{-1}$, the eddy turnover time $\tau_u = H/U$, and the 
viscous diffusion time $\tau_\nu= H^2/\nu$. In this regime of parameter space the fluid 
flow is dominated by geostrophy -- a pointwise balance between the Coriolis and pressure 
gradient forces.  As an example, for the Earth’s outer core estimates suggest 
$Re_H = \textit{O}(10^8)$, $E = \textit{O}(10^{-15})$ and $Ro_H = \textit{O}(10^{-8})$
\citep{pR13}. Importantly, this parameter regime is far beyond the current investigative
capabilities of laboratory experiments and direct numerical simulations (DNS) in both 
global spherical or local planar domains which remain limited to 
$E\gtrsim \textit{O}(10^{-8})$ and $Re_H\lesssim \textit{O}(10^4)$ (see the regime 
diagram in \citet{jmA15} or \citet{vanKan24}).
 
\begin{figure}
	\centering
        \includegraphics[height=0.37\textwidth]{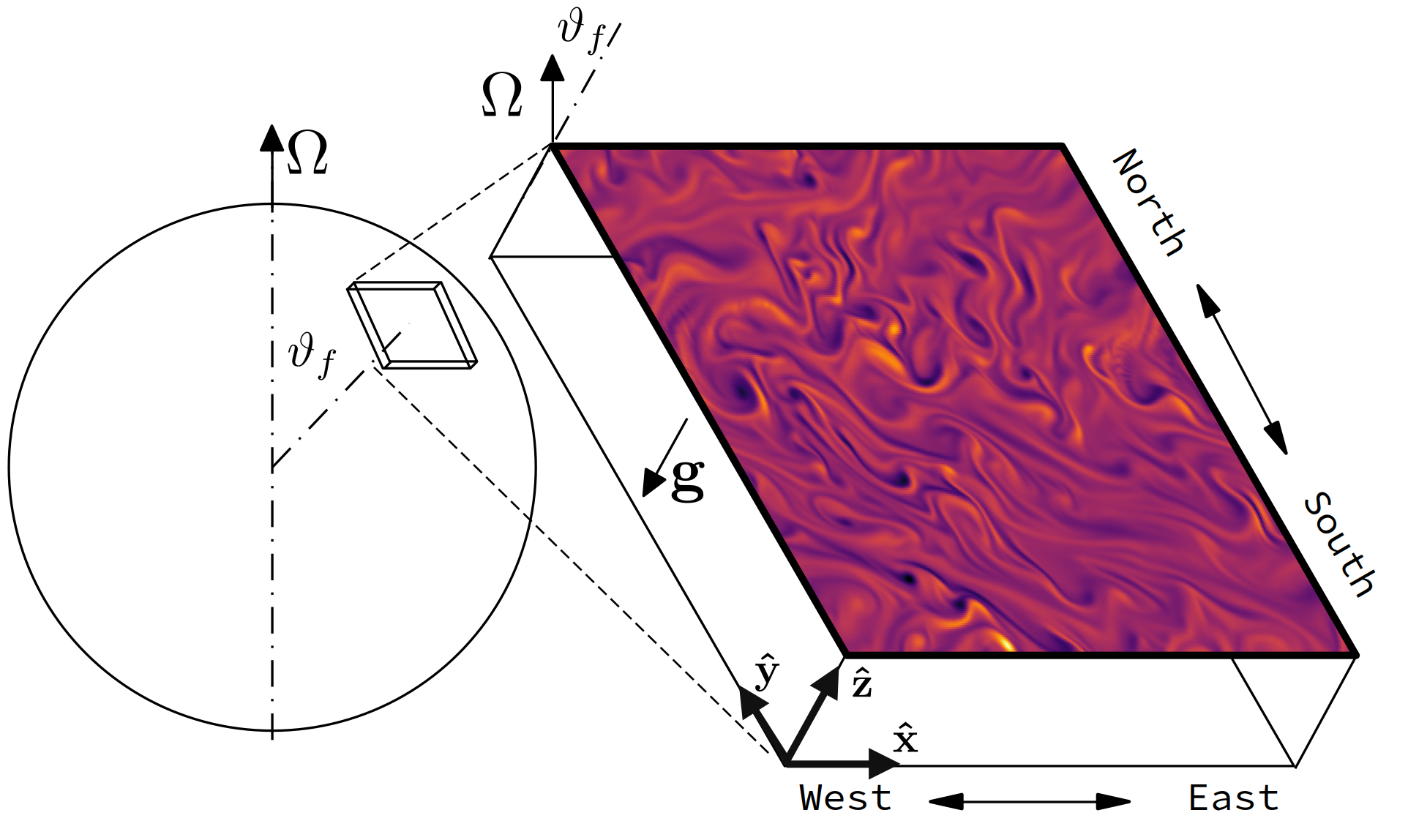}
        \includegraphics[height=0.37\textwidth]{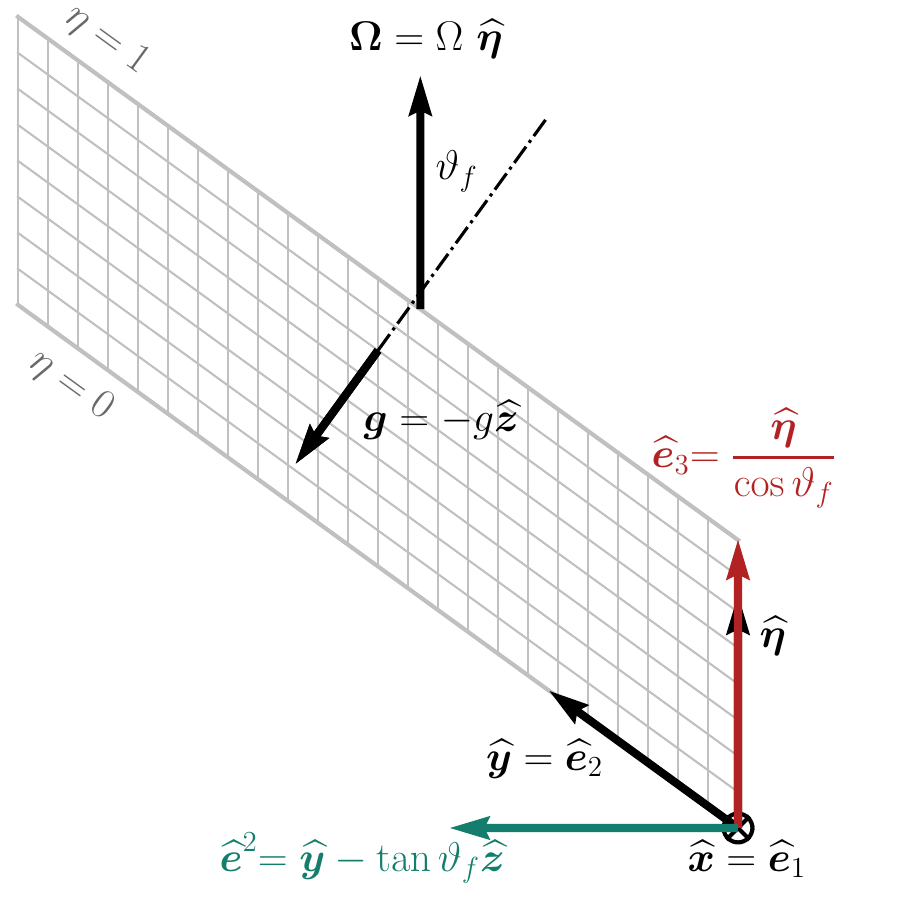}
	\caption{
	Local area $f$-plane approximation on the sphere at colatitude $\vartheta_f$
	(left plot) and the explicit computational domain (right plot)
	highlighting the non-orthogonality of the coordinate directions. The 
	rotation vector is given by $\boldsymbol{\Omega}$ and $\eta$ is the 
	coordinate in this direction. Coordinates $(x,y,\eta)$ increase in the
	eastward, northward and axial directions, respectively.
	}
        \label{fig:Schematic}
\end{figure}

The investigative barrier that constrains laboratory experiments and DNS can be overcome 
theoretically by employing the quasi-geostrophic approximation which restricts the dynamics 
to the vicinity of the geostrophic manifold. This modeling paradigm is an asymptotically
rigorous reduction of the governing fluid equations that filters out fast inertial waves 
and viscous Ekman boundary layers \citep{mS06, kJ06}. The  quasi-geostrophic approach has
been particularly valuable for the case of upright rotating Rayleigh-B\'enard convection,
i.e., the canonical paradigm of a plane layer of fluid with an imposed destabilizing 
temperature gradient and aligned  gravity and rotation. In the context of applications 
to planetary and stellar interiors this case is pertinent to thermal convection in the
polar regions of spherical geometries in which the colatitude $\vartheta_f =0$  (see 
figure~\ref{fig:Schematic}). Numerical simulations of this model have elucidated the entire 
range of possible flow morphologies, ranging from cellular motions near the onset of convection
through to geostrophic turbulence \citep{kJ12}. It has also enabled the exploration of 
the fidelity of dissipation-free power scaling laws for turbulent heat and momentum 
transport~\citep{kJ12b,cG19,sM21,tO23} given, respectively, by
\begin{equation}
Nu\sim \sigma^{-1/2} Ra^{3/2} E^2,\quad
Re_H \sim \sigma^{-1} Ra \, E, \quad
Ra \equiv \frac{g\alpha\Delta_T H^3}{\nu\kappa}.
\end{equation}
Here, the Nusselt number $Nu$ denotes the non-dimensional heat transport and
$\sigma\equiv\nu/\kappa$ denotes the Prandtl number measuring the importance of viscous 
dissipation relative to thermal dissipation; $Ra$ is the Rayleigh number measuring the
strength of the buoyancy forcing with $g$ denoting the gravitational acceleration, 
$\Delta_T$ the temperature jump across the layer, and $\alpha$ the thermal expansion 
coefficient. Importantly, investigation of the quasi-geostrophic regime in the limit 
of very rapid rotation reveals the existence of a nonlocal inverse kinetic energy transfer 
resulting in the appearance of a domain scale condensate in the form of a large scale dipolar
vortex (LSV) \citep{kJ12,aR14,sM21}, a conclusion confirmed in
DNS studies at finite values of $E$ \citep{bF14,cG14,sS14}.

Extensions of the quasi-geostrophic approximation to the case where rotation and gravity 
are not aligned, i.e., the tilted $f$-plane at finite colatitude $\vartheta_f>0$, have 
received less attention. This is particularly so for strongly nonlinear turbulent 
flows in a similarly extreme parameter regime, although DNS studies with $E\gtrsim 10^{-5}$ 
have recently been performed \citep{novi2019, barker2020, kannan26, zengGRL2026}, following earlier
work by \citet{hathawayGAFD80} and \citet{hathawayJFM83}. 
Laboratory investigations are challenging, however, although non-vertical gravity resulting from a centrifugal force has been used to study single plume dynamics~\citep{sheremet2004} and jet formation on a laboratory $\beta$-plane~\citep{cabanes2017}.
Recently, preliminary results on a laboratory realization of the $f$-plane problem were reported by~\citet{keQingRBC10th_ref1} and \citet{keQingRBC10th_ref2} at the 10th International Conference on Rayleigh-B\'enard Turbulence (Lyon 2026).

It has been demonstrated through laboratory experiments \citep{jmA15,rK2021,bouillautPNAS2021,jA23,hadjerciJFM2024}, numerical simulations
\citep{tG16,novi2019,barker2020} and theory \citep{kJ98} that rotating convection is characterised
by strongly anisotropic, axially aligned dynamics. Specifically, when compared with the
$\textit{O}(H)$ axial  convective turnover scales, the cross-axial scales $\ell$  satisfy
the condition $\ell\ll H$. This is a necessary prerequisite for unstably stratified geostrophic 
motions to overcome the Taylor-Proudman constraint that suppresses axial variations.
Of specific interest is the turbulent, rotationally constrained (quasi-geostrophic) regime
on the $f$-plane as characterised by the Coriolis parameter 
	$2\Omega \cos\vartheta_f$ 
which captures
the projection of the angular velocity onto the local vertical direction
at colatitude $\vartheta_f$. Here, for fluid parcels characterised by velocity scale $U$,
the Coriolis acceleration ($\sim 2U\Omega \cos\vartheta_f$) is in geostrophic balance with
the pressure gradient ($\sim P/\ell$) and dominates over fluid inertia ($\sim U^2/\ell$).
A necessary further requirement is the dynamical balance known as the Coriolis-Inertia-Archimedean
(CIA) balance between axial-scale vortex stretching $(\sim 2U\Omega\cos\vartheta_f /H)$, 
vortical advection $(\sim (U/\ell)^2)$, and buoyancy (Archimedean) torque $(\sim g\alpha 
\theta^*/\ell)$ \citep{AHJPRR20}. %<- notation issue with section 2.1
A buoyancy force subdominant to the geostrophic 
force balance is a necessary prerequisite for the CIA balance, i.e.,
\begin{equation}
\label{eq:ghold}
g\alpha \theta^* \ll 2 U \Omega \cos \vartheta_f \sim \frac{P}{\ell}
\end{equation}
with $\theta^*/\Delta_T \sim \ell/H$ \citep{AHJPRR20}. %<- heretoo
Non-dimensionally, these requirements 
are captured by the  columnar limit with small Rossby number and large (global-scale)
Reynolds number:
\begin{equation}
\frac{\ell}{H} 
\sim Ro_\ell,\quad
Ro_\ell = \frac{U^2/\ell}{2U \Omega\cos\vartheta_f} = \frac{U}{2\ell \Omega\cos\vartheta_f} \ll 1,\quad 
Re_H  \gg 1.
\label{eqn:ndp0}
\end{equation}
By definition, $Ro_\ell \ll 1$ implies that the $f$-plane is bounded away from the equator
where the Coriolis parameter vanishes. \textit{A priori} estimates for these internal
control parameters may be obtained upon explicit specification of the characteristic 
velocity scale $U$. Lower bound estimates are provided by the diffusive velocity scaling
$U_\nu=\nu/\ell$ associated with the local Reynolds number
$\convectiveReynolds = U\ell/\nu \sim \textit{O}(1)$ appropriate near the onset of convection. Upper
bound estimates are based upon the dissipation-free thermal wind scaling with
$U_\Omega=g\alpha \Delta_T/2\Omega\cos\vartheta_f$ resulting from the CIA force balance
appropriate for the turbulent regime \citep{AHJPRR20}. 

For the diffusive velocity scaling $U_\nu$,
\begin{subequations}
\begin{equation}
Ro_\ell\sim\frac{\ell}{H} =  E_f^{1/3}, \quad
Re_H = E_f^{-1/3}, \quad 
E_f\equiv\frac{\nu}{2H^2\Omega\cos\vartheta_f},
\label{eqn:ndp1}
\end{equation}
where $E_f$ is the Ekman number appropriate to colatitude $\vartheta_f$.\footnote{We 
emphasize that the definition of the Ekman number used throughout is based on the locally
projected vertical component of rotation, the Coriolis parameter $f=2\Omega\cos\vartheta_f$, 
such that $E_f=\nu/2H^2\Omega\cos\vartheta_f$.} These relations indicate an explicit
dependence on the \textit{latitudinal} Ekman number $E_f$, an external control parameter
defining the relative importance of momentum dissipation and the Coriolis force at 
colatitude $\vartheta_f$. Moreover, the constraint \eqref{eq:ghold} may now be reformulated as 
\begin{equation}
Ro_{\rm conv} = \sqrt{\frac{Ra}{\sigma}} E_f
= \sqrt{\frac{\widetilde{Ra}}{\sigma}} E_f^{1/3}
= o(E_f^{1/6}),
\end{equation}
where $Ro_{\rm conv}$ is the convective Rossby number, an external measure utilized in
many prior investigations of rotating convection \citep[see][]{AHJPRR20}, and 
$\widetilde{Ra}\equiv Ra \, E_f^{4/3}$ is the reduced \textit{latitudinal} Rayleigh number,
assumed to be of order one. Likewise, for the dissipation-free scaling $U_\Omega$,
\begin{equation}
\frac{\ell}{H}\sim Ro_\ell 
\sim \sqrt{\frac{\widetilde{Ra}}{\sigma}}\,E_f^{1/3} = Ro_{\rm conv},
\quad  Re_H = \frac{\widetilde{Ra}}{\sigma}  E_f^{-1/3},
\label{eqn:ndp2}
\end{equation}
\end{subequations}
indicating that both the diffusive and the dissipation-free scalings lead to the same asymptotic dependence of $Re_H$ on $E_f$. Moreover, noting that the small scale Reynolds number and large scale Reynolds number are related via $\convectiveReynolds = Re_H E_f^{1/3}$, we see that both Reynolds numbers exhibit the same (linear) dependence on $\widetilde{Ra}$ (see \cite{sM21,tO23} 
for a detailed assessment of this scaling when $\vartheta_f=0$). The requirement 
\eqref{eq:ghold} now implies the self-consistent result $\theta^*/\Delta_T\ll1$. 

Equations (\ref{eqn:ndp1}) and (\ref{eqn:ndp2}) suggest that
\begin{equation}
\varepsilon= E_f^{1/3}
\end{equation}
is a natural external parameter for investigating quasi-geostrophic dynamics and
specifically the dependence on $\vartheta_f$, which is the focus of this paper.
In the regime of interest convective motions occur for $Ra=\textit{O}(E_f^{-4/3})$ 
or equivalently, $\wRa=\textit{O}(1)$ \citep{sC61,kJ98}. Thus the constraint 
\eqref{eq:ghold} now implies $Ra=o(E_f^{-5/3})$, or $\wRa=o(E_f^{-1/3})$, indicating
a large range of thermal forcing over which convection remains rotationally constrained. 

The organization of this paper is as follows. In Section~\ref{sec:prelim}, the
rotating Rayleigh-B\'enard problem on the tilted $f$-plane is formulated within 
the incompressible Navier-Stokes equations. Particular attention is paid to identifying
the relevant physics that poses challenges to exploring the extreme parameter regimes
of geophysical and astrophysical relevance. In Section~\ref{sec:QG-RB}, the asymptotically 
reduced equations describing Rayleigh-B\'enard convection on an $f$-plane in the rapid 
rotation limit (the non-hydrostatic quasi-geostrophic equations, hereafter $f$NHQGE) are
presented, with the specific focus on familiarizing the reader with the non-orthogonal 
coordinate system that arises as a natural representation for the $f$-plane problem, 
while retaining a connection to the Cartesian velocity variable representation; 
a complete derivation of the $f$NHQGE is detailed in Appendix~\ref{Appdx:A}. The linear asymptotic
solutions of the $f$NHQGE are presented in Section~\ref{sec:QG-RB} for steady convection,
highlighting the broken symmetries that hold for upright rotating convection and the 
existence of the different ranges of dynamically unstable modes as a function of the
horizontal orientation of local eddies. The numerical algorithm used in our simulations 
is discussed in Section~\ref{sec:num}. The simulation results are presented in 
Section~\ref{sec:sim} with a focus on the flow morphology and the mechanics underlying
the observed inverse kinetic energy transfer. Discussion and concluding remarks are found in 
Section~\ref{sec:conc}.

%========================================================================
 
\section{Preliminaries}\label{sec:prelim}

% ... FIGURE 1
\subsection{Setup and problem formulation}
Thermal convection driven by a destabilizing temperature drop  $\Delta_T$ across a 
spherical shell rotating at angular velocity $\Omega$ and confined between inner and 
outer impenetrable surfaces at radii $r^*_i$ and $r^*_o$ is considered. A thin shell
approximation is adopted, where the depth $H \equiv r^*_o - r^*_i $ satisfies the 
narrow gap criterion $H\ll (r^*_o + r^*_i)/2$. This leads to the utilization of the 
$f$-plane formulation wherein the dynamics are considered in a local area plane layer 
located at fixed colatitude $\vartheta_f$ and azimuth $\phi_f$ (see fig.~\ref{fig:Schematic}). 
Locations within the plane layer can be identified by a right-handed Cartesian system 
with unit vectors $(\hx,\hy,\haz)$ and vector position coordinates 
$x^* = (\phi -\phi_f) r^*_i \sin \theta_f$, $y^* = -(\vartheta -\vartheta_f) r^*_i$, $z^*= r^* - r^*_i$, 
respectively denoting the zonal, meridional and radial directions. The gravitational 
acceleration $\boldsymbol{g}=-g\haz$ is assumed to be uniform and along the local vertical 
direction $\haz$.

We assume an incompressible flow and a linear equation of 
state for the temperature-dependent density of the fluid,
$\rho^*=\rho^*_i(1 -\alpha (T^*- T^*_i))$,  where the temperature 
$T^*=T_b^* + \Theta^*$. Here, $T^*_i$ denotes the fixed temperature at the lower 
boundary, and $\Theta^*$ 
denotes the temperature fluctuation about the motionless,
thermally conducting, base state $\ub_b^*=0$, $T_b^*(z^*)=T^*_i - z^*\Delta_T/H$.

Planetary rotation is captured locally on the $f$-plane by a constant 
vector $2\Omega\hz$ with unit vector
\begin{equation}
\hz =  \sin\vartheta_f \hy + \cos\vartheta_f \haz \triangleq\eta_2 \hy + \eta_3 \haz 
\end{equation}
in the axial direction. In the following we also make use of the (non-unit) vector
\begin{equation}\label{eq:def_covariant3}
	\baseCov{3} = \hz / \cos \vartheta_f = \tan \vartheta_f\, \hy + \haz\, ,
\end{equation}
a covariant base vector naturally associated with our non-orthogonal coordinates, and show that this vector facilitates the problem formulation.

The governing equations considered are the incompressible Navier-Stokes equations
(iNSE), non-dimensionalized with the diffusive length $\ell= \varepsilon H$,
time scale $\ell^2/\nu$ and velocity $\nu/\ell$, together with the pressure scale
$\rho_i\nu^2/\varepsilon \ell^2$ and the temperature scale $\Delta_T$:
\begin{subequations}
\label{eqn:gov}
\begin{align}
	\label{eqn:gov_momentum}
	D_{t} \ub  +   \frac{1}{\varepsilon}\baseCov{3}\times\ub &
	= -   \frac{1}{\varepsilon}  \nabla p 
	+  \frac{1}{\varepsilon} \frac{\widetilde{Ra}}{\sigma}  \Theta \boldsymbol{\widehat z}
	+ \nabla^2 \ub,
	\\
	\label{eqn:gov_temperature}
D_{t}  \Theta  -  \varepsilon w  &
	=  \frac{1}{\sigma} \nabla^2 \Theta,
	\\
	\label{eqn:gov_incompressibility}
\boldsymbol{\nabla\cdot u} &= 0,
\end{align}
\end{subequations}
where $D_t \equiv \partial_t + \boldsymbol{u\cdot\nabla}$ is the material derivative.
In dimensionless units, the vertical extent of the fluid domain corresponds to
$0\le z \le \varepsilon^{-1}$. On both the lower and the upper bounding surfaces, 
eqs.~(\ref{eqn:gov}) are accompanied by impenetrable kinematic and fixed temperature
boundary conditions for perturbations of the conducting state $1-\varepsilon z$,
\begin{equation}
\haz\cdot\ub = 0, \quad \Theta=0 \qquad \mathrm{for}\quad z=0,~\varepsilon^{-1}
\end{equation}
together with non-slip and/or stress-free mechanical boundary conditions. It is 
noted that the specification of the mechanical boundary conditions is not required 
in the asymptotic approach described below.

The primary challenges that exist for direct numerical simulations of (\ref{eqn:gov}) in
the relevant geophysical and astrophysical quasi-geostrophic limit $\varepsilon\rightarrow 0$ 
are fourfold: 
(i) the presence of fast inertial waves propagating with characteristic dimensional length 
scale $\textit{O}(\ell)$ and time scale $\textit{O}(\Omega^{-1})$ that respectively 
translate to $\textit{O}(1)$ and $\textit{O}(\varepsilon)$ scales in our non-dimensional units;
(ii) the presence of Ekman boundary layers of thickness $\textit{O}(E^{1/2}H)$ 
(non-dimensionally, $\textit{O}(\varepsilon^{3/2})$) adjacent to all mechanical
boundaries \citep{kJ98}; 
(iii) the presence of an $\textit{O}(E^{1/3} H)$ (non-dimensionally, 
$\textit{O}(\varepsilon)$) vertical crossing scale at non-polar latitudes due to
rotational alignment of columnar convective eddies; and
(iv) the thermal relaxation time scale to achieve a statistically stationary state
occurring on the $\textit{O}(H^2/\nu)$ vertical diffusive scale compared to the
$\textit{O}(\ell/U)$ convective scale (non-dimensionally,
$\textit{O}(\varepsilon^{-2})$ and  $\textit{O}(1)$, respectively). 
Each challenge imposes a prohibitive $\varepsilon$-dependent numerical constraint
on the DNS. In turn, partial abatement of these constraints may occur through: 
1) the utilization of implicit time-stepping algorithms for the coupled linear 
components of the iNSE that 
	filters the fastest and dynamically irrelevant inertial waves without compromising numerical stability,
2) the imposition of boundary conditions that parameterize linear Ekman boundary
layers and the associated pumping or suction \citep{sS14,jCKMSV16,pJSM16}, and 
3) the utilization of axially aligned spatial grids and associated $\eta$-dependent
spectral basis functions. However, the issue of thermal relaxation, issue (iv), remains.  
Indeed, a multi-decade  literature survey of DNS of rotating convection to the
present date indicates that an average lower bound of order
$\varepsilon_{lb} \sim 10^{-8/3}$ (i.e. $E \gtrsim 10^{-8}$) on accessible Ekman numbers for upright convection
still remains \citep{kJ96,schmitzTilgnerPRE2009,rK2021,song24}. Examples of recent DNS of RBC on a tilted
$f$-plane at arbitrary latitudes have reported results for $E\gtrsim 10^{-5}$ \citep{novi2019, barker2020}. 

Owing to the aforementioned prohibitive constraints, we pursue here an asymptotic 
approach that investigates a system of reduced PDEs valid on the $f$-plane in the 
quasi-geostrophic limit $\varepsilon\rightarrow 0$. This approach is known to overcome 
the prohibitive spatio-temporal constraints on DNS in extreme regimes using the iNSE.

\subsection{Non-orthogonal coordinates suggested by the Taylor-Proudman constraint}
Foundational to the derivation of the reduced system is
the assumption of geostrophy on small spatial scales, i.e. the pointwise balance 
between the Coriolis and pressure gradient forces at leading order.
This force balance implies the Taylor-Proudman constraint \citep{jP16, gT23} in which 
axial invariance is satisfied on small axial scales.
It is this preference for rotationally aligned dynamics that suggests the necessity for a 
representation of all fluid variables in the  non-orthogonal coordinate system $(x,y,\eta)$ where the 
meridional coordinate is obtained by shearing the cartesian coordinates $(X,Y,Z)$:
\begin{equation}
	\label{def:sheared_y}
	y = Y - \gamma Z\, \quad \text{where}\quad \gamma = \tan \vartheta_f\,.
\end{equation}
and $x=X$, $\eta=Z$. The associated covariant and contravariant bases
are respectively
$\left( \baseCov{1}, \baseCov{2}, \baseCov{3}\right) = \left( \hx, \hy, \baseCov{3}\right)$
and $\left( \baseContra{1}, \baseContra{2}, \baseContra{3} \right) = \left(\hx,\baseContra{2}, \haz \right)$,
where $\baseCov{3}$ has been defined above in (\ref{eq:def_covariant3}) and
\begin{equation}
	\label{eq:def_contravariant2}
	\baseContra{2} =  \hy - \gamma \haz\, .
\end{equation}
Note that these dual bases possess orthogonality properties $\baseCov{i}\boldsymbol{\cdot} \baseContra{j} = \delta_{ij}$ that will facilitate projections.

In the following we express all fluid variables as functions of the fast $\textit{O}(\varepsilon)$ horizontal coordinates
$x,y$, the fast axial coordinate $\varpi\sim \eta/\varepsilon$ and the slow $\textit{O}(1)$ axial scale $\eta$. With this notation, the gradient and the divergence operators are expressed using the nabla operator
\begin{equation}
	\label{eq:nonOrthogonal_nabla}
	\boldsymbol{\nabla} = \hx \partial_x + \baseContra{2} \partial_y + \haz \partial_\varpi + \varepsilon \haz \partial_\eta
	\equiv \boldsymbol{\nabla}^\prime + \varepsilon \haz \partial_\eta\, ,
\end{equation}
where the prime denotes derivatives on fast spatial scales. It follows that the geostrophic balance at leading order $\textit{O}(\varepsilon^{-1})$ in equation~(\ref{eqn:gov_momentum}) becomes:
\begin{equation}
	\label{eq:quasi_geostrophy}
	\baseCov{3}\times\ub = -  \boldsymbol{\nabla}^\prime p, \quad \boldsymbol{\nabla^\prime\cdot u}  =   0.
\end{equation}
Upon projection onto $\baseCov{3}$, it now follows that small scale gradients cannot arise in the axial direction: 
\begin{equation}
	\baseCov{3} \boldsymbol{\cdot \nabla}^\prime (\boldsymbol{u}, p) = \partial_\varpi  (\boldsymbol{u}, p) = 0 \, ,
\end{equation}
thereby guaranteeing strong anisotropy of the flow: fluid variables only vary on the slow $\textit{O}(1)$ axial scale $\eta$ in this direction. Thus, the geostrophic balance
implies that gradients and the divergence are dominated by the horizontal component $\nablaPerpPrime$ such that 
expression (\ref{eq:nonOrthogonal_nabla}) becomes:
\begin{equation}
\label{eq:nablaPerpPrime}
\boldsymbol{\nabla} = \nablaPerpPrime + \varepsilon \haz \partial_\eta\, 
\quad \text{where} \quad
\nablaPerpPrime =  \hx \partial_x + \baseContra{2} \partial_y \, .
\end{equation}

\subsection{Quasi-geostrophic solutions}
Before expressing general solutions to the geostrophic balance equation (\ref{eq:quasi_geostrophy}),
we recall that a velocity field with Cartesian expression $\ub = u\, \hx + v\, \hy + w\, \haz$
is expressed in the non-orthogonal basis as:
\begin{equation}
	\ub 
	=  u\,\hx + \widetilde{v} \hy
	 + w \baseCov{3}\,\quad \text{with} \quad \widetilde{v} = v-\gamma w,
\end{equation}
where $u,v,\widetilde{v}, w$ are all recast as functions of the non-orthogonal coordinates $(x,y,\eta)$. 
To emphasize that employing $\widetilde{v}$ is advantageous with these coordinates, 
we note that the Coriolis force now becomes
\begin{equation}
	\label{eq:Coriolis}
	\baseCov{3}\times\ub=-\widetilde{v} \hx +  u \baseContra{2}\, 
\end{equation}
while the leading order incompressibility condition takes the compact form
\begin{equation}
\nablaPerpPrime \boldsymbol{ \cdot u} = \partial_x u + \partial_y \widetilde{v} = 0\, .
\end{equation}
Thus, unless explicitly stated otherwise,
references to zonal and meridional components of the velocity field will be taken 
to mean the velocity components $(u, \widetilde{v})$ in the covariant basis 
$(\hx, \hy, \baseCov{3})$. Solutions to the geostrophic balance equation (\ref{eq:quasi_geostrophy}) are thus  
\begin{equation}
\label{eq:geostrophy_solution_components}
p= \Psi(x,y,\eta,t), \quad
u = -\pd{y} \Psi, \quad 
\widetilde{v} = \partial_x \Psi
\end{equation}
with $w=w(x,y,\eta, t)$.
Equivalently, geostrophically balanced flows on the tilted $f$-plane in the non-orthogonal covariant basis take the form
\begin{equation}
\label{eq:geostrophy_solution_covariant}
	\boldsymbol{u}=- \curl{\left(\Psi\haz\right)}+w\baseCov{3}=- \partial_y \Psi \, \hx + \partial_x \Psi\, \hy + w\, \baseCov{3}.
\end{equation}

\subsection{Axial velocity-vorticity representation}
By analogy with the convective NHQG equations~\citep{kJ12} that
govern the dynamics of the geostrophic streamfunction and the vertical velocity
in a plane layer at the North pole (i.e., with rotation and gravity aligned), we define the axial vorticity as
\begin{equation}
\label{def:axial_vorticity}
\zeta\equiv\baseCov{3} \boldsymbol{\cdot}\nablaPerpPrime \times \boldsymbol{u} = \nablaPerpTwo \Psi\, ,
\end{equation}
where the anisotropic horizontal diffusion operator $\nablaPerpTwo$ is given by
\begin{equation}
	\nablaPerpTwo = \partial_{xx} +\frac{1}{\cos^2 \vartheta_f} \partial_{yy}\, .
\end{equation}
To conclude our description of the velocity field, we introduce the axial velocity
\begin{equation}
\label{def:axial_velocity}
\axialVelocity \equiv \baseCov{3}\bdot\ub =
	\frac{1}{\cos^2 \vartheta_f}  w + \tan \vartheta_f  \, \partial_{x}\Psi\,,
\end{equation}
and note that, reciprocally,
\begin{equation}
	\label{eq:cartesian_axial_velocity}
w = \haz \bdot \ub = \cos^2 \vartheta_f \,\axialVelocity - \sin \vartheta_f \cos \vartheta_f \,  \partial_{x}\Psi\,.
\end{equation}
In the following section, we show that the use of the variables $\axialVelocity$ and $\Psi$ leads to a natural formulation of the 
equations governing the dynamics of geostrophically balanced flows (equation~(\ref{eq:geostrophy_solution_covariant})).

\section{Reduced quasi-qeostrophic equations on the $f$-plane}
\label{sec:QG-RB}
In the limit of strong rotational constraint 
$\varepsilon\equiv Ro_\ell = E_f^{1/3}\rightarrow 0$, 
the non-orthogonal decomposition introduced in Section~\ref{sec:prelim} leads to the following reduction of
the governing iNSE (\ref{eqn:gov})
on a $f$-plane located at colatitude $\vartheta_f$ within the spherical shell
(fig.~\ref{fig:Schematic}), hereafter
the $f$NHQGE:
\begin{subequations}
\label{eqn:qgf}
\begin{gather}
	\label{eqn:qgfPSI}
\partial_{t}  \nablaPerpTwo \Psi
	+ J\left[ \Psi, \nablaPerpTwo \Psi \right]
        -  \partial_{\eta} \axialVelocity 
	= 
        - \gamma  \displaystyle{\frac{\widetilde{Ra}}{\sigma}} \partial_{x} \theta	
        +  \nabla^{\prime 4}_\perp  \Psi ,
\\
	\label{eqn:qgfW}
\partial_{t} \axialVelocity
	+  J\left[ \Psi, \axialVelocity \right] 
	+  \partial_{\eta}\Psi
	=  \displaystyle{\frac{\widetilde{Ra}}{\sigma}} \theta 
	+  \nablaPerpTwo  \axialVelocity , 
\\
	\label{eqn:qgftheta}
\partial_{t} \theta
	+ J\left[ \Psi, \theta \right] 
	+ w \left( \partial_{\eta} \overline{\Theta} - 1 \right) 
	= 
         \frac{1}{\sigma}   \nablaPerpTwo   \theta ,
\\
	\label{eqn:qgfTHETA}
\varepsilon^{-2} \partial_{t} \overline{\Theta}  
	+ \partial_{\eta} \left(\overline{ w \theta} \right) 
	=
	\frac{1}{\sigma} \partial_{\eta\eta} \overline{\Theta}.
\end{gather}
\end{subequations}
Here $\gamma\equiv\tan\vartheta_f$.
A summary of the derivation of the $f$NHQGE is 
provided in Appendix~\ref{Appdx:A}, following \citet{kJ06}, eqs.~(3.4). 
The system 
(\ref{eqn:qgf}) evolves the geostrophic streamfunction $\Psi$ 
through the axial vorticity $\zeta \equiv 
\nablaPerpTwo \Psi$, the axial velocity $\axialVelocity \equiv \baseCov{3}\cdot\ub $,
and temperature $\Theta \equiv\overline{\Theta} +\varepsilon \theta$ 
decomposed into mean and fluctuating components, $\overline{\Theta}$ 
and $\theta$. Thus, temperature fluctuations about the mean are asymptotically small. Here the mean is defined as a horizontal	average over horizontal
spatial scales $x,y$. The Jacobian operator
\begin{equation}
J[\Psi,h] \triangleq
	\partial_{x}\Psi\partial_{y} h - \partial_{y}\Psi\partial_{x} h = \ub_\perp\boldDot \nablaPerpPrime h
	= \nablaPerpPrime \boldDot \left( \boldsymbol{u}_\perp h \right)
	\end{equation}
describes small-scale non-axial advection of a scalar field $h$ by the horizontal velocity:
\begin{equation}
	\label{def:horizontal_velocity}
	\boldsymbol{u}_\perp = \boldsymbol{u} - w \baseCov{3} = -\partial_y \Psi\, \hx + \partial_x \Psi \, \hy\,.
\end{equation}

The reduced system is accompanied by impenetrable, fixed-temperature boundary conditions 
\begin{equation}
\label{eqn:abcs}
w = 0, \ \  \overline{\Theta}=0, \quad \eta=0,1.
\end{equation}
On the boundaries, equation (\ref{eqn:qgftheta}) implies that temperature fluctuations 
satisfy an advection-diffusion equation with the thermal variance constraint
$\lim_{t\rightarrow\infty}\overline{\theta^2} =0$, indicating that 
$\lim_{t\rightarrow\infty}\theta =0$. The boundaries are thus isothermal or evolving 
towards isothermality. It is noted that the specific form of the thermal boundary 
conditions is unimportant in the limit of rapid rotation due to an equivalence mapping
between fixed temperature and fixed heat flux models \citep{mc15c}. Moreover, the 
reduction in the axial spatial order of the momentum equations (\ref{eqn:qgf}a,b) 
compared with iNSE (\ref{eqn:gov_momentum}) precludes the imposition of mechanical boundary 
conditions and therefore impenetrability suffices. Inspection of the $f$NHQGE on 
the lateral boundaries indicates that they are incompatible with both no-slip (where
$\Psi =0$) and stress-free
conditions (see equation~\eqref{AppEq:stressFree_BC} in Appendix~\ref{Appdx:boundaryConditions}). 
This property implies the existence of Ekman boundary layers for which 
finite $\varepsilon$-effects may be captured through parameterized Ekman pumping 
boundary conditions \citep{jCKMSV16}. This correction has been considered for the 
upright case \citep{pJSM16}, and its effect on both linear stability and single mode nonlinear
solutions has recently been modelled using matched asymptotics~\citep{troJFM24};
however, this aspect of the problem is not pursued in this initial investigation of quasi-geostrophic turbulence on a tilted $f$-plane.

The mean temperature $\overline{\Theta}$ evolves on the slow time scale 
$\tau = \varepsilon^2 t$ associated with the axial diffusion time $H^2/\nu$ and is found 
to reach a statistically stationary state. Moreover, the slow time evolution term in 
(\ref{eqn:qgfTHETA}) is observed to be subdominant to the extent that without impacting the
time-averaged $\overline{\Theta}$ it may be neglected or even replaced with an effective
$\varepsilon^*>\varepsilon$ that relaxes the temporal stiffness of the $f$NHQGE
(\citet{kJ98a}; see also~\citet{julienJCP2025}, Appendix D).

Finally, before analyzing the details of the energy budget of system (\ref{eqn:qgf}), 
we observe that global-scale incompressibility is upheld through the solenoidal condition for the sub-dominant ageostrophic horizontal velocity:
\begin{equation}
\nabla'_\perp \cdot\ub^{\prime ag} + \varepsilon \pd{\eta} w = 0.
\end{equation}
Thus $\ub^{\prime ag} =\textit{O}(\varepsilon)$ and while it is not part of the 
leading order system, it can be determined from the above equation. The reduced
$f$NHQG system represents an asymptotic reduction of the iNSE (\ref{eqn:gov}) provided
$Ro_\ell= o(1)$, or equivalently from \eqref{eqn:ndp1}, $\wRa = o(\varepsilon^{-1})$. 
Fast  inertial waves with $\textit{O}(\varepsilon)$ vertical scale and dimensionless
frequency $\textit{O}(\varepsilon^{-1})$ are filtered out from the above equations,
allowing for substantial computational savings compared with DNS in the small 
$\varepsilon$ limit. However, slow inertial waves with $\textit{O}(1)$ spatial scales
and $\textit{O}(1)$ frequencies on par with the convective eddy turnover time are 
retained \citep{kJ12}. It is emphasized that this characteristic is distinct from 
stably stratified quasi-geostrophy where all inertial waves are filtered out \citep{jC71}.   

\subsection{Validity}

The $f$NHQGE are formally valid at all finite Prandtl numbers $\sigma$, provided these are are not as large as $E^{-1/3}$ and not as small as $\sigma\sim E$. In the former case additional terms have to be retained on the right side of equations (\ref{eqn:qgf}) while in the latter case the assumption that the convective scale is $O(E^{1/3})$ breaks down. Indeed \citet{zhang97} and \citet{bassom98} demonstrate that when $\sigma\sim E$ the onset wavenumber is $O(1)$ and hence no longer small. \citet{dawes01} examines the case $\sigma=sE^{\alpha}$ at $\vartheta_f=0$, where $\alpha>0$ and $s$ is an $O(1)$ constant, for which $k_c\sim E^{\gamma}$, where
  $$
  -\frac{1}{3}<\gamma\equiv\frac{1}{3}(\alpha-1)<0.
  $$
Our theory thus remains formally valid for $\alpha=0$, i.e. $E\to 0$ at fixed $\sigma$, but would have to be modified if $\sigma$ is so small as to be comparable to a power of $E$, and in particular in the case $\sigma\sim E$ where $k_c\sim 1$.
Despite this remark, the theory remains widely applicable to natural flows: icy moons' oceans typically feature $\prandtl \gtrsim 1$, while their gaseous atmospheres are characterized by $\prandtl \lesssim 1$. 
Even for planetary cores, composed of liquid metal characterized by relatively small 
Prandtl numbers, $\prandtl=O(10^{-2})$, this value remains comfortably large 
compared to typical Ekman numbers, preserving the validity of the asymptotic hierarchy, and this is so under solar conditions as well, where the thermal diffusivity is dominated by photon diffusivity, leading to $\prandtl=O(10^{-6})$.

\subsection{Energetics} 

Like the iNSE, in a statistically steady state, the $f$NHQGE conserve the
volume-averaged power integrals for kinetic energy dissipation and thermal dissipation:
\begin{equation}
\label{eqn:dissrate}
\epsilon_{u} = \frac{\wRa}{\sigma^2}\left( Nu -1 \right), \quad 
\epsilon_{\vartheta} = Nu -1, \quad
Nu -1 = \sigma \left \langle   \overline{w  \theta }  \right\rangle_{\eta,t} , 
\end{equation}
where $Nu$ denotes the Nusselt number measuring the non-dimensional heat transport 
evaluated via depth and time-averaging $\left \langle \ \ \right \rangle_{\eta,t}$ and 
\begin{equation}
\epsilon_{u} = \left \langle
	\eta^2_3 \left[ 
	\overline{ \left( \nabla^{\prime 2}_\perp   \Psi \right)^2} 
	+ \overline{\left\| \nabla_\perp^{\prime} \axialVelocity\right\|^2}
	\right]
	\right \rangle_{\eta,t},
	\qquad
\epsilon_{\vartheta} = \left \langle
	\left( 	\pd{\eta} \overline{\Theta} \right)^2
      + \overline{\left\| \nabla^{\prime}_\perp 
	                                        \theta \right\|^2}
	\right \rangle_{\eta,t},
\end{equation}
see Appendix \ref{appC} for detailed derivation. The local scale Reynolds number based on the rms vertical velocity, 
\begin{equation}
	\label{def:convectiveReynolds}
    \convectiveReynolds \equiv \sqrt{\left\langle\overline{w^2}\right \rangle_{\eta,t}},
\end{equation} 
is also utilized.
The $f$NHQGE conserve pointwise the potential vorticity in the inviscid limit, namely, 
\begin{equation}
D^\perp_t q = 0 \quad\mathrm{where} \quad q = \haz\bdot\boldsymbol{\omega} +
\lb \boldsymbol{\omega} \bdot\nabla'  +\pd{\eta}\rb \lb \frac{\theta}{\pd{\eta}\overline{\Theta}} \rb 
\end{equation}
and $D^\perp_t = \pd{t} + \ub\cdot\nabla'$. More explicitly,
\begin{equation}
q = \eta^2_3 \lb \nabla^{\prime 2} \Psi + \gamma \pd{x} \axialVelocity \rb - J\lsq \axialVelocity,   \lb \frac{\theta}{\pd{\eta}\overline{\Theta}} \rb  \rsq +  \pd{\eta}   \lb \frac{\theta}{\pd{\eta}\overline{\Theta}} \rb,
\end{equation}
demonstrating the nonlinear complexity and diminished utility of potential vorticity (PV) compared to the linear PV that arises in stably stratified quasi-geostrophic layers. Here $q$ is the quasi-geostrophic reduction of Ertel's potential vorticity $q_E \equiv \boldsymbol{\omega}_a\bdot\nabla \Theta$ in the Boussinesq limit, where $\boldsymbol{\omega}_a \equiv\boldsymbol{\omega} + \hz/Ro$ is the absolute vorticity. Conservation of potential vorticity following fluid elements is equivalent to conservation of volume of a fluid element because  $\boldsymbol{\omega}_a$ evolves as a line element and $\nabla \Theta$ evolves as a surface element \citep{kJ06}.

\subsection{Symmetry breaking}
It is important to understand the symmetries, or indeed the broken symmetries, associated with the tilted $f$-plane about the mean temperature state $1-\eta + \overline{\Theta}$.
For upright convection, where $\gamma =0$, the $f$NHQGE  (\ref{eqn:qgf}) possess the following horizontal rotational and vertical midplane reflection symmetries:  
\begin{subequations}
\label{eqn:sym}
\begin{align}
\label{eqn:sm1}
\mathcal{R}_\phi:&  \quad (\boldsymbol{x}_\perp,\eta) \mapsto (\boldsymbol{R}[\phi]\boldsymbol{x}_\perp,\eta), \quad 
(\Psi, w, \theta )\mapsto (\Psi, w, \theta) ,
\\
\label{eqn:sm2}
\mathcal{R}_\eta:&  \quad (\boldsymbol{x}_\perp,\eta) \mapsto (\boldsymbol{x}_\perp, 1- \eta), \quad 
(\Psi, w, \theta )\mapsto (\Psi, -w, -\theta)
\end{align}
\end{subequations}
with the rotation matrix 
$\boldsymbol{R}[\phi] =$ {\small {$  \lb
\begin{array}{rr}
 \cos\phi  &  -\sin\phi   \\
 \sin\phi   &  \cos\phi     
\end{array}
 \rb$}}.
In particular, the symmetry $\mathcal{R}_\phi$ 
implies that there is no preferred horizontal orientation. The following reflection symmetries in the horizontal plane also hold 
\begin{subequations}
\label{eqn:rf}
\begin{align}
\label{eqn:rf1}
\mathcal{R}_x:&  \quad (x,y,\eta) \mapsto (-x, y,\eta), \quad 
(\Psi, w, \theta )\mapsto (-\Psi, -w, -\theta),
\\
\label{eqn:rfs}
\mathcal{R}_y:&  \quad (x,y,\eta) \mapsto (x, -y,\eta), \quad 
(\Psi, w, \theta )\mapsto (-\Psi, -w, -\theta).
\end{align}
\end{subequations}
together with the following central symmetry:
\begin{equation}
\mathcal{R}_{\boldsymbol{x}_\perp,\eta}: \quad ( \boldsymbol{x_\perp}, \eta)
                                       \mapsto (-\boldsymbol{x_\perp},-\eta), 
				       \quad
(\Psi, w, \theta )\mapsto (\Psi, -w, -\theta).
\end{equation}
Each of these reflection symmetries, absent in the full governing iNSE (\ref{eqn:gov}), 
enforces an equivalence between cyclonic and anticyclonic vortical motions. This property
is a well-known feature of quasi-geostrophic dynamics \citep{kJ98,vanKan24}. 

Inspection of the tilted $f$-plane equations, where $\gamma\ne 0$, reveals that symmetries
held by the upright case are singular to that specific case: the rotational and midplane
reflection symmetries (\ref{eqn:sym}) and the reflection symmetry \eqref{eqn:rf1} no longer
hold. The broken rotational symmetry $\mathcal{R}_\phi$ indicates that different horizontal
orientations are no longer equivalent. The tendency for fluid structures to align with 
rotation also breaks the midplane reflection symmetry $\mathcal{R}_\eta$.  The sole remaining
reflection symmetry $\mathcal{R}_y$ maps the zonal velocity $u:=-\pd{y}\Psi\mapsto u$; thus 
no preference for eastward over westward zonal flows or vice versa is present. Moreover, the 
cyclonic-anticyclonic equivalence survives through the unbroken symmetry $\mathcal{R}_y$.

\subsection{Linear stability}\label{sec:linear_stability}

For Prandtl numbers $\sigma > 0.68$, appropriate for this study with $\sigma=1$, the onset of convection occurs via a steady state bifurcation of the conduction state. 
On substituting the normal mode ansatz for convective rolls of the form $F(\eta)\exp {i (k_x x + k_y y)}$ with 
$(k_x,k_y) = k_\perp (\cos\chi,\sin\chi)$ into the linearized version of (\ref{eqn:qgf}), an analytical expression for the marginal stability curve follows:
 \begin{equation}
 \label{eqn:marg}
\widetilde{Ra}_s =\frac{ \pi^2 +  k_\perp^6 \lb 1 + \gamma^2 \sin^2 \chi\rb^3}{k_\perp^2}, 
 \end{equation}
where $\chi=0$ and $\chi=\pi/2$ respectively denote North-South (N-S) and East-West (E-W) convective rolls.
The critical Rayleigh number and the corresponding critical wavenumber are thus
\begin{equation}
 \label{eqn:crit}
\widetilde{Ra}_{c} =\frac{3}{2}\lb 2\pi^4  \rb^{1/3} \lb 1 + \gamma^2 \sin^2 \chi\rb,
\quad
k_{\perp c} =  \frac{\pi^{1/3} }{2^{1/6}\lb 1 + \gamma^2 \sin^2 \chi\rb^{1/2} }.
\end{equation}
Linear convective roll solutions with real amplitude $A$ are given by
\begin{subequations}
 \label{eqn:efunc}
	\begin{align} 
w &= -A   \cos\lb\boldsymbol{k_\perp\cdot x_\perp}\rb   \sin\lb \pi \eta \rb, \quad 
\theta =\lb \frac{2} {\pi^2}\rb^{1/3}  {\sigma} w,
\\ 
      \Psi &= A \lsq   
		\frac{(2\pi)^{1/3}}{k^2_{\perp c} }
    \cos\lb\boldsymbol{k_\perp\cdot x_\perp}\rb   \cos\lb  \pi \eta \rb   +    \frac{\gamma\cos\chi}{ {k}_{\perp c}}
     \sin\lb\boldsymbol{k_\perp\cdot x_\perp}\rb   \sin\lb \pi \eta \rb
     \rsq 
	\end{align}
\end{subequations}
and are all $\mathcal{R}_y$ symmetric, a consequence of invariance under the mapping $\chi\mapsto -\chi$. The analytical simplicity of the linear solutions illustrates the utility of the non-orthogonal coordinate representation in circumventing the appearance of the rapid upright scale in $z$.

% ... FIGURE 2
\begin{figure}
	\centering
        \includegraphics[width=\textwidth]{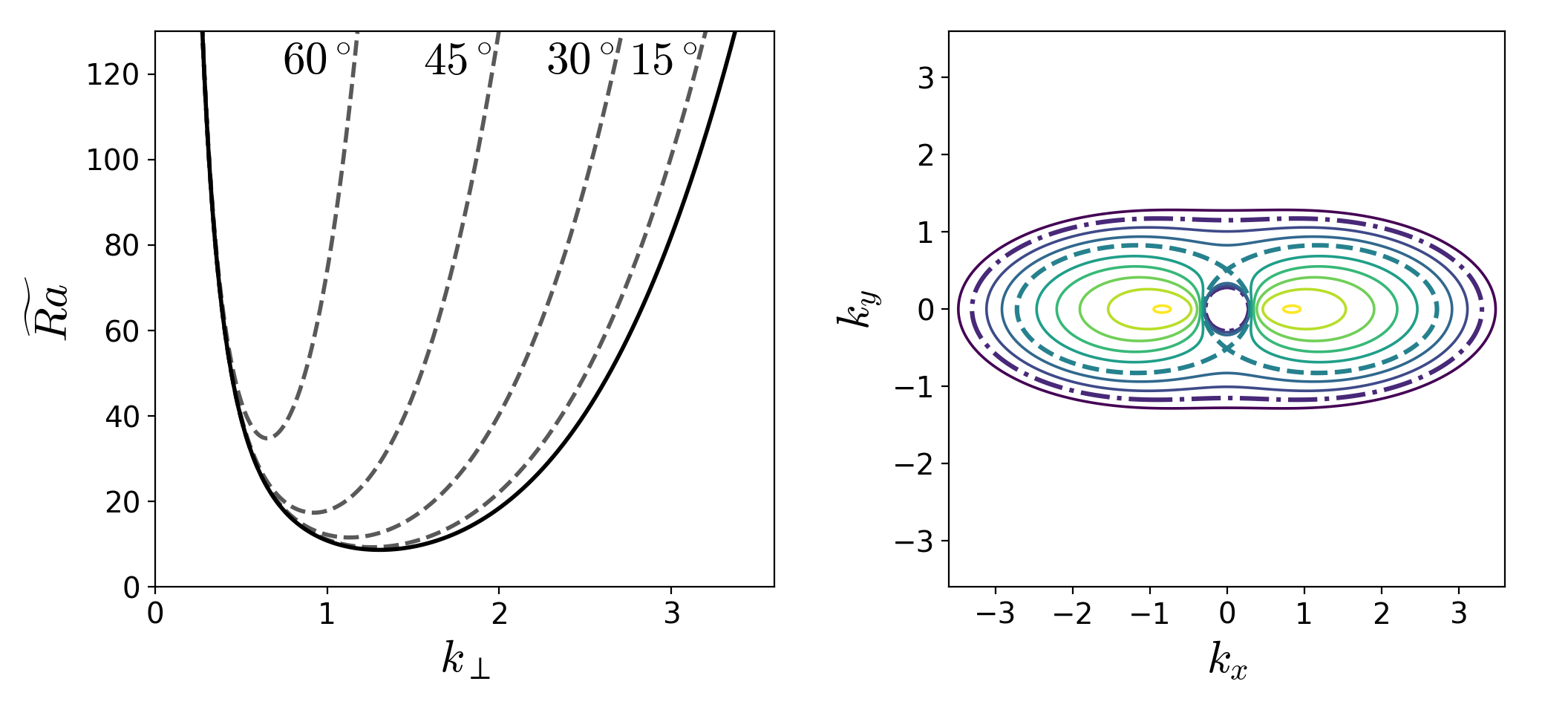}
	\caption{(a) Marginal stability curves showing the reduced latitudinal Rayleigh number $\wRa$ vs the horizontal wavenumber $k_\perp$:  North-South rolls (black curve, $\chi = 0$),  East-West rolls (gray dashed curves, $\chi = \pi/2$) at colatitudes $\vartheta_{f} =\lb 15^\circ, 30^\circ, 45^\circ, 60^\circ\rb$. (b) Contour plot of the linear growth rate at $\wRa=120, \vartheta_f=60^\circ$ with zero contour: dashed-dotted line, separatrix contour: dashed line.}
\label{F:Marg}
\end{figure}

Figure~\ref{F:Marg}(a) illustrates the marginal curve boundaries for N-S ($\chi=0$) and E-W ($\chi=\pi/2$) roll orientations
that respectively bookend the most and the least unstable convective modes. 
Based on the definition of the latitudinal Ekman number that uses the Coriolis parameter $f=2\Omega\eta_3$, N-S roll orientations ($\chi =0$) share the stability boundary for all colatitudes $\vartheta_f$, thereby illustrating the utility of the latitudinal Rayleigh number $\widetilde{Ra}$ for relative comparisons across colatitudes. 
All other orientations are increasingly stabilized at large wavenumbers (small convective scales) as the $f$-plane approaches the equator. For fixed, but sufficiently large $\widetilde{Ra}$, it is observed that all roll orientations share approximately the same low wavenumber (large-scale) stability boundary. Hence, at fixed $\wRa$, the range of convectively unstable length scales and associated supercriticality for a given roll orientation increasingly diminishes from N-S rolls to E-W rolls. Further characteristics are captured by the growth rate dispersion relation. Figure~\ref{F:Marg}(b) shows sample level sets of the linear growth rate in the $(k_x, k_y)$ plane when  $\wRa=120, \vartheta_f=60^\circ$. 
The impact of spatial anisotropy generated by increasing colatitude is evident, as curves of constant growth rate are distorted from circles at $\vartheta_f=0^\circ$ to ellipses of increasing eccentricity. This indicates that length scales on a given contour increase when the convective motions reorient from N-S to E-W.
A measure of this distortion is given approximately by the mapping $k_y\mapsto k_y/\eta_3$ which remaps elliptical to circular contours and the anisotropic diffusion operator to an isotropic operator, i.e., $\nabla^{\prime2}_\perp\mapsto 
\partial_{xx} + \partial_{yy}$. It is demonstrated in Section~\ref{sec:sim} that the dynamics of fully turbulent flow retain some of these characteristic linear features.

\subsection{The inverse energy cascade}
\label{sec:baroman}

In two dimensions, turbulent flow dynamics within the inertial range are known to conserve area averages of the positive-definite energy and all functionals of the potential enstrophy (the mean-squared vorticity). This guarantees that energy and enstrophy are  transported up- and down- scale, respectively \citep{EckeBoffetta2012,ALEXAKIS20181}. In constrast, in three dimensions, enstrophy is not an inviscidly conserved quantity and the guarantee of an upscale energy transport is lost. Nevertheless, the spontaneous development of an inverse cascade in three-dimensional rotating flows does occur and is a topic of much recent interest for upright rotating RBC \citep{kJ12,bF14,cG14,aR14,sS14,sM21,vanKan24}. A key feature of this phenomenon is the existence of a manifold capturing the evolution of the axially-averaged axial vorticity through the barotropic vorticity equation (BVE). For the $f$-plane considered here, with $\vartheta_f > 0^\circ$, the BVE extends 
 to 
 \begin{equation}
 \label{eqn:BVE}
 \pd{t} \nabla^{\prime 2}_\perp \lbr \Psi \rbr + J\lsq \lbr \Psi \rbr, \nabla^{\prime 2}_\perp \lbr \Psi \rbr \rsq
= \nabla^{\prime 4}_\perp  \lbr \Psi \rbr - \lbr J\lsq \Psi', \nabla^{\prime 2}_\perp \Psi' \rsq\rbr 
+\gamma \pd{x} \lbr \pd{\eta}\Psi'  - \frac{\wRa}{\sigma} \theta \rbr.
\end{equation}
Here, $\nabla^{\prime2}_\perp = \pd{xx} + \eta^{-2}_3\pd{yy}$ as before.
The angled brackets refer to axial averaging along the rotation axis $\eta$.
The appropriate quantity remains the axially-averaged axial vorticity (see e.g.~\eqref{def:axial_vorticity}), i.e., the 
barotropic vorticity $\left \langle \zeta \right \rangle 
\equiv \left \langle \baseCov{3}\cdot\boldsymbol{\omega} \right \rangle
=  \nabla^{\prime 2}_\perp \left \langle \Psi \right \rangle$.
The left-hand side (LHS) denotes the material advection of vertical barotropic 
vorticity with the second term referring to energy-conserving barotropic
self-advection. Terms on the right-hand side (RHS) are responsible for the production
or attenuation of barotropic vorticity. The first of these terms represents horizontal diffusion of barotropic
vorticity. The remaining two expressions are potential barotropic production terms.
The first represents the axially integrated divergence of the vortical stresses derived
from convectively driven, equivalently baroclinically driven, modes $\Psi'\equiv \Psi -\lbr \Psi \rbr$, i.e.,
\begin{equation}
\label{eqn:vflux}
	\lbr J\lsq \Psi',  \nabla^{\prime 2}_\perp  \Psi' \rsq \rbr =  \nablaPerpPrime \bdot \lbr \ub'_\perp \zeta' \rbr.
\end{equation}
The third production term on the RHS is a baroclinic torque term arising from the misalignment between gravity and rotation at non-polar colatitudes.

Recalling the expressions for the horizontal divergence
$\nablaPerpPrime \bdot=\partial_x \hx \bdot + \partial_y \baseContra{2} \bdot$
and the horizontal velocity
$\ub_\perp = -\partial_y \Psi \hx + \partial_x \Psi \hy$, cf. eqs.~(\ref{eq:nablaPerpPrime}) and~(\ref{def:horizontal_velocity}), 
we introduce the generalized barotropic-baroclinic decomposition, indicated respectively by angled brackets and primes, i.e.,
\begin{subequations}
	\label{subeqs:baroclinic-barotropic}
\begin{align}
	\left\langle \ub_\perp \right \rangle &= -\partial_y \left\langle \Psi \right \rangle \hx + \partial_x \left \langle \Psi \right \rangle \hy,\\
	             \ub_\perp ^\prime        &= -\partial_y              \Psi ^\prime        \hx + \partial_x               \Psi ^\prime        \hy,
\end{align}
\end{subequations}
and re-express the barotropic vorticity equation \eqref{eqn:BVE} in horizontal divergence form:  
 \begin{equation}
 \label{eqn:BVE_a}
\nablaPerpPrime \bdot\lsq  \pd{t} \nablaPerpPrime \lbr \Psi \rbr + \lbr \ub_\perp \rbr  \nabla^{\prime 2}_\perp \lbr \Psi \rbr 
=  \nabla^{\prime 2}_\perp  \nablaPerpPrime \lbr \Psi \rbr  - \lbr \ub'_\perp  \nabla^{\prime 2}_\perp \Psi'\rbr 
+\gamma \lbr \pd{\eta}\Psi'  - \frac{\wRa}{\sigma} \theta \rbr \hx \rsq .
 \end{equation}
Observing that
\begin{subequations}
\label{eqn:BVE_b}
	\begin{align}
\nablaPerpPrime\lbr \Psi \rbr 
		&	\equiv \pd{x} \lbr \Psi \rbr \hx + \pd{y} \lbr \Psi \rbr 
		\baseContra{2}
		=  \lbr \tilde v \rbr \hx - \lbr u \rbr 
		\baseContra{2},\\
\lbr \ub_\perp\rbr \lbr \zeta \rbr 
	&	=  \lbr u\rbr\lbr \zeta \rbr \hx + \lbr \tilde v\rbr \lbr \zeta \rbr \hy, \\
\lbr \ub^\prime_\perp \zeta' \rbr 
        &	= \lbr u' \zeta' \rbr \hx + \lbr \tilde v' \zeta' \rbr \hy\,,
\end{align}
\end{subequations}
one can now undo the divergence operation in \eqref{eqn:BVE_a} 
by projecting the terms in square brackets on $(\hx,\baseContra{2})$,
yielding 
\begin{subequations}
\label{eqn:BVE_c}
\begin{align}
	\label{eqn:BVE_u}
	\partial_t \left \langle \widetilde{v} \right \rangle 
	+ \left \langle u \right \rangle \left \langle \zeta \right \rangle 
	&= \nabla_\perp^{\prime 2} \left \langle \widetilde{v} \right \rangle 
	- \left \langle u^\prime \zeta^\prime \right \rangle 
	+ \gamma \left \langle \partial_\eta \Psi^\prime - \frac{\wRa}{\sigma} \theta \right \rangle
	- \partial_x G\,, \\
	\label{eqn:BVE_v}
	-\frac{1}{\cos^2\vartheta_f} \partial_t \left \langle u \right \rangle 
	+ \left \langle \widetilde{v} \right \rangle \left \langle \zeta \right \rangle 
	&=-\frac{1}{\cos^2\vartheta_f} \nabla_\perp^{\prime 2} \left \langle u \right \rangle 
	- \left \langle \widetilde{v}^\prime \zeta^\prime \right \rangle 
	+ \partial_y G \,,
\end{align}
\end{subequations}
where $G(x,y)$ is an additive pressure-like gauge function.
It follows from \eqref{eqn:BVE_c} that, in the case of upright rotating convection ($\gamma= 0$), the vorticity stress $\lbr \ub'_\perp \zeta' \rbr$ is the sole baroclinic production term that may overcome viscous dissipation and drive barotropic motions. Specifically, zonal barotropic flows $\lbr u \rbr$ are driven by the meridional vorticity stress $\lbr \tilde v' \zeta' \rbr$ and meridional barotropic flows $\lbr \tilde v \rbr$ are driven by the zonal vorticity stress $\lbr  u' \zeta' \rbr$.
However, on $f$-planes with $\vartheta_f \ne 0$, $\gamma\ne 0$, buoyancy torques also contribute to the meridional force balance. This highlights the importance of spatial anisotropy for the production of the barotropic velocity components and hence axial barotropic vorticity. This issue is explored further in Section~\ref{sec:InvCasBM} where the simulation results are discussed.

In the absence of barotropic dissipation and baroclinic forcing (i.e., omitting all terms on the RHS of (\ref{eqn:BVE})),
the BVE conserves the horizontally-averaged 
barotropic energy\footnote{
	Remarkably, the conserved quantity $\mathcal{E}_{bt}$ corresponds to the kinetic energy of the flow orthogonal to the rotation: 
	$\ub - \left( \ub \bdot \baseCov{3}\right) \haz =  \ub-U_3\haz$, not to be confused 
	with the flow parallel to the surfaces $\ub_\perp= \ub- \left( \ub \bdot \haz\right) \baseCov{3} = \ub-w \baseCov{3}$, as revealed by 
	comparing expression~\eqref{AppEq:kineticEnergy_covariant} with $U_3=0$ and~\eqref{AppEq:kineticEnergy_contravariant} with $w=0$.
            }
$\mathcal{E}_{bt}=\overline{\left\| \nabla'_\perp \lbr \Psi\rbr\right\|^2}$
and the enstrophy $\mathcal{Z}_{bt}=\overline{\lb \nabla^{\prime2}_\perp \lbr \Psi\rbr\rb^2}$, i.e., $\pd{t}\mathcal{E}_{bt}=\pd{t}\mathcal{Z}_{bt} =0$, thereby highlighting the realization of a bi-directional energy-enstrophy cascade within a sub-manifold in the entire state space of the system \citep{aR14, ALEXAKIS20181}. Saturation of the energy cascade, which continues to expand in space unabated to the domain scales $L_x\times L_y$, occurs solely through a force-dissipation balance, as seen from the kinetic energy equation associated with axial vorticity: 
\begin{subequations}
\label{eq:KEB}
\begin{align}
	\pd{t} \mathcal{E}_{bt} =& \hspace{0.5em} \mathcal{D}_{\mathrm{dissipation}} \hspace{0.5em}
	               +   \hspace{0.85em} \Sstress   \hspace{0.6em}
		       +  \hspace{2em}  \Sbuoy 
\\
=& - \overline{ \lb \nabla^{\prime2}_\perp  \lbr \Psi \rbr \rb^2 } 
+ \overline{ \lbr \Psi \rbr \lbr J\lsq \Psi', \nabla^{\prime2}_\perp  \Psi' \rsq \rbr }
 +\gamma \overline{\pd{x}\lbr \Psi \rbr \lbr  \lb \pd{\eta} \Psi' - \frac{\wRa}{\sigma} \theta \rb \rbr}.
\end{align}
\end{subequations}
In line with the BVE from which it is derived, this equation has two baroclinic energy sources $\mathcal{S}_j$. 

Regarding the inverse cascade, the gravest permissible spectral modes that fit into the domain, and thus contribute to the observed condensate profile, have the approximate forms $\cos(2\pi x/L_x)$, representative of a meridional jet, $\cos(2\pi y/L_y)$, a zonal jet, and the superposition $\cos(2\pi x/L_x) + \cos(2\pi y/L_y)$ consistent with a dipolar large scale vortex (LSV). For the upright case with $\vartheta_f=0^\circ$, the BVE possesses the rotational symmetry $\mathcal{R}_\phi$ indicating no preferred spatial or velocity direction within the inverse-cascade.
It is then expected that $\lbr u' \zeta' \rbr \sim \lbr \tilde v' \zeta' \rbr $, so that saturation occurs in the form of a LSV. The $\mathcal{R}_\phi$-symmetry is broken on the tilted $f$-plane when $\vartheta_f\ne 0^\circ$, as evident in the anisotropic Laplacian operator $\nabla^{\prime2}_\perp$. Importantly, vorticity production is now exposed to the additional symmetry-breaking buoyancy torque.
The extent to which the $\mathcal{R}_\phi$-symmetry is broken and its associated role in the observed large scale flow can be estimated through the relative magnitudes of the baroclinic production terms in the kinetic energy equation \eqref{eq:KEB}. The degree to which the vortical stresses contribute to the barotropic momenta can also be assessed through the relative magnitudes of $\lbr u' \zeta' \rbr_{\rm rms}$ and  $\lbr \tilde v' \zeta' \rbr_{\rm rms}$.

\begin{table}
\centering
\begin{tabular}{cccccc}
	$\vartheta_{f}$ & $\widetilde{Ra}$ & $N_x \times N_y \times N_\eta$ & $Nu \pm \sigma_{Nu}$ & $\widetilde{Re} \pm \sigma_{\widetilde{Re}}$ \\
\hline
\hline
%                                                                     Nu                     Re 
$ 0^{\circ} $ & $  10 $ & $ 128 \times 128 \times 256 $ & $   1.27 \pm  0.01 $ & $   0.75 \pm  0.11 $& \\
$ 0^{\circ} $ & $  20 $ & $ 128 \times 128 \times 256 $ & $   4.02 \pm  0.13 $ & $   3.55 \pm  0.79 $& \\
$ 0^{\circ} $ & $  40 $ & $ 128 \times 128 \times 256 $ & $  12.28 \pm  0.60 $ & $  10.67 \pm  2.43 $& \\
$ 0^{\circ} $ & $  60 $ & $ 128 \times 128 \times 256 $ & $  19.88 \pm  1.03 $ & $  17.19 \pm  4.73 $& \\
$ 0^{\circ} $ & $  80 $ & $ 128 \times 128 \times 256 $ & $  30.96 \pm  1.81 $ & $  24.28 \pm  7.39 $& \\
$ 0^{\circ} $ & $ 100 $ & $ 256 \times 256 \times 384 $ & $  43.37 \pm  2.54 $ & $  32.05 \pm  8.24 $& \\
$ 0^{\circ} $ & $ 120 $ & $ 256 \times 256 \times 384 $ & $  58.84 \pm  2.76 $ & $  41.16 \pm 11.50 $& \\
\hline
$ 15^{\circ} $ & $  10 $ & $ 128 \times 128 \times 256 $ & $   1.19 \pm  0.01 $ & $   0.62 \pm  0.02 $& \\
$ 15^{\circ} $ & $  20 $ & $ 128 \times 128 \times 256 $ & $   3.83 \pm  0.10 $ & $   3.44 \pm  0.12 $& \\
$ 15^{\circ} $ & $  40 $ & $ 128 \times 128 \times 256 $ & $  11.56 \pm  0.53 $ & $  10.42 \pm  0.30 $& \\
$ 15^{\circ} $ & $  60 $ & $ 128 \times 128 \times 256 $ & $  20.16 \pm  0.94 $ & $  17.23 \pm  0.58 $& \\
$ 15^{\circ} $ & $  80 $ & $ 128 \times 128 \times 256 $ & $  31.44 \pm  1.79 $ & $  25.27 \pm  1.37 $& \\
$ 15^{\circ} $ & $ 100 $ & $ 256 \times 256 \times 384 $ & $  43.50 \pm  2.71 $ & $  32.43 \pm  1.48 $& \\
$ 15^{\circ} $ & $ 120 $ & $ 256 \times 256 \times 384 $ & $  56.81 \pm  3.14 $ & $  40.57 \pm  2.17 $& \\
\hline
	$ 30^{\circ} $ & $  10 $ & $ 128 \times 128 \times 256 $ & $   1.12 \pm  0.01 $ & $   0.50 \pm  0.03 $& \\
	$ 30^{\circ} $ & $  20 $ & $ 128 \times 128 \times 256 $ & $   3.64 \pm  0.32 $ & $   3.41 \pm  0.19 $& \\
	$ 30^{\circ} $ & $  40 $ & $ 128 \times 128 \times 256 $ & $  10.69 \pm  0.61 $ & $   9.71 \pm  0.28 $& \\
	$ 30^{\circ} $ & $  60 $ & $ 128 \times 128 \times 256 $ & $  18.79 \pm  1.31 $ & $  16.02 \pm  0.68 $& \\
	$ 30^{\circ} $ & $  80 $ & $ 128 \times 128 \times 256 $ & $  28.01 \pm  1.60 $ & $  22.62 \pm  0.90 $& \\
	$ 30^{\circ} $ & $ 100 $ & $ 256 \times 256 \times 384 $ & $  39.40 \pm  2.18 $ & $  30.19 \pm  1.51 $& \\
	$ 30^{\circ} $ & $ 120 $ & $ 256 \times 256 \times 384 $ & $  51.81 \pm  2.67 $ & $  37.58 \pm  2.21 $& \\
\hline
	$ 45^{\circ} $ & $  10 $ & $ 128 \times 128 \times 256 $ & $   1.12 \pm  0.00 $ & $   0.49 \pm  0.02 $& \\
	$ 45^{\circ} $ & $  20 $ & $ 128 \times 128 \times 256 $ & $   3.76 \pm  0.55 $ & $   3.68 \pm  0.61 $& \\
	$ 45^{\circ} $ & $  40 $ & $ 128 \times 128 \times 256 $ & $   9.14 \pm  0.54 $ & $   8.43 \pm  0.40 $& \\
	$ 45^{\circ} $ & $  60 $ & $ 128 \times 128 \times 256 $ & $  16.31 \pm  0.96 $ & $  14.04\pm  0.60 $& (LSV) \\
	       &         &                                        & $  12.49 \pm  0.96 $ & $ 11.42\pm  0.60 $                & (ZJ) \\
	$ 45^{\circ} $ & $  80 $ & $ 128 \times 128 \times 256 $ & $  18.68 \pm  1.39 $ & $  16.35 \pm  0.78 $& \\
	$ 45^{\circ} $ & $ 100 $ & $ 256 \times 256 \times 384 $ & $  25.78 \pm  1.33 $ & $  21.45 \pm  1.26 $& \\
	$ 45^{\circ} $ & $ 120 $ & $ 256 \times 256 \times 384 $ & $  30.77 \pm  2.64 $ & $  25.36 \pm  2.99 $& \\
\hline
	$ 60^{\circ} $ & $  10 $ & $ 128 \times 128 \times 256 $ & $   1.16 \pm  0.01 $ & $   0.55 \pm  0.02 $& \\
	$ 60^{\circ} $ & $  20 $ & $ 128 \times 128 \times 256 $ & $   2.60 \pm  0.17 $ & $   2.43 \pm  0.27 $& \\
	$ 60^{\circ} $ & $  40 $ & $ 128 \times 128 \times 256 $ & $   4.68 \pm  0.31 $ & $   4.73 \pm  0.34 $& \\
	$ 60^{\circ} $ & $  60 $ & $ 128 \times 128 \times 256 $ & $   7.41 \pm  0.88 $ & $   7.49 \pm  0.54 $& \\
	$ 60^{\circ} $ & $  80 $ & $ 128 \times 128 \times 256 $ & $  10.66 \pm  1.56 $ & $  10.51 \pm  0.83 $& \\
	$ 60^{\circ} $ & $ 100 $ & $ 256 \times 256 \times 384 $ & $  14.35 \pm  1.02 $ & $  13.96 \pm  0.91 $& \\
	$ 60^{\circ} $ & $ 120 $ & $ 256 \times 256 \times 384 $ & $  17.64 \pm  1.87 $ & $  16.85 \pm  1.27 $& \\
\hline
\hline
\end{tabular}
\caption{Simulation grid at Prandtl number $\sigma = 1$. Reported are time-averaged values of the Nusselt number ($Nu$) and the convective scale Reynolds number ($\convectiveReynolds$), along with their standard deviations.}
\label{tab:simulation_grid}
\end{table}

% ... FIGURE 3   
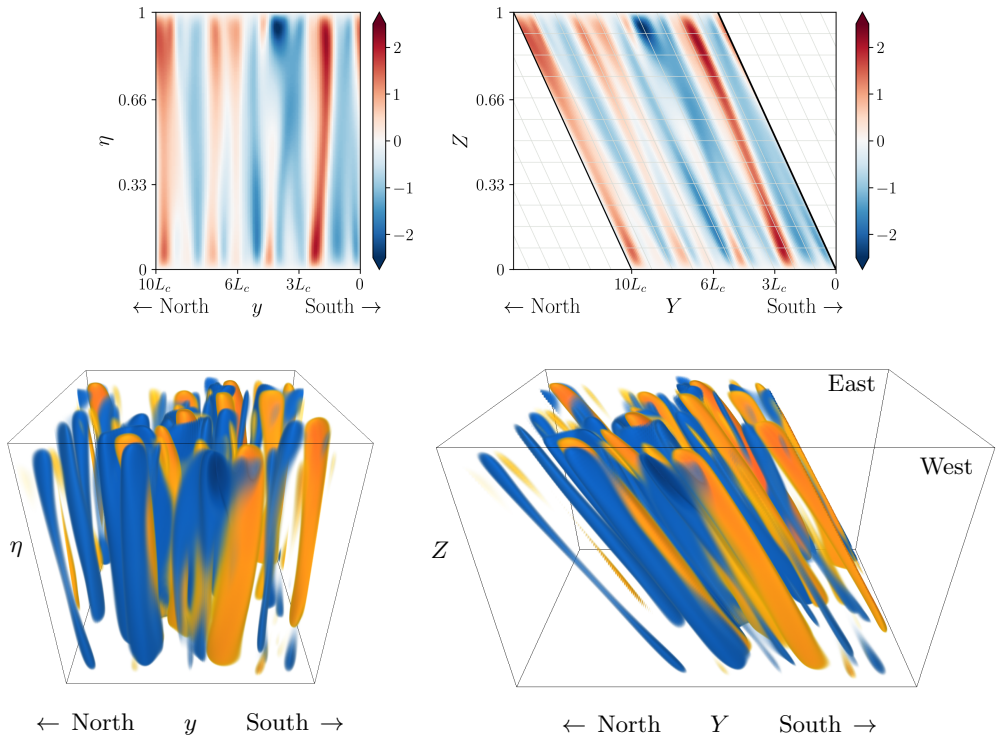
\begin{figure}
\centering
	\begin{tikzpicture}
		\node at (0,0) {\includegraphics[height=4.5cm, trim={0  0 5cm 0}, clip ]{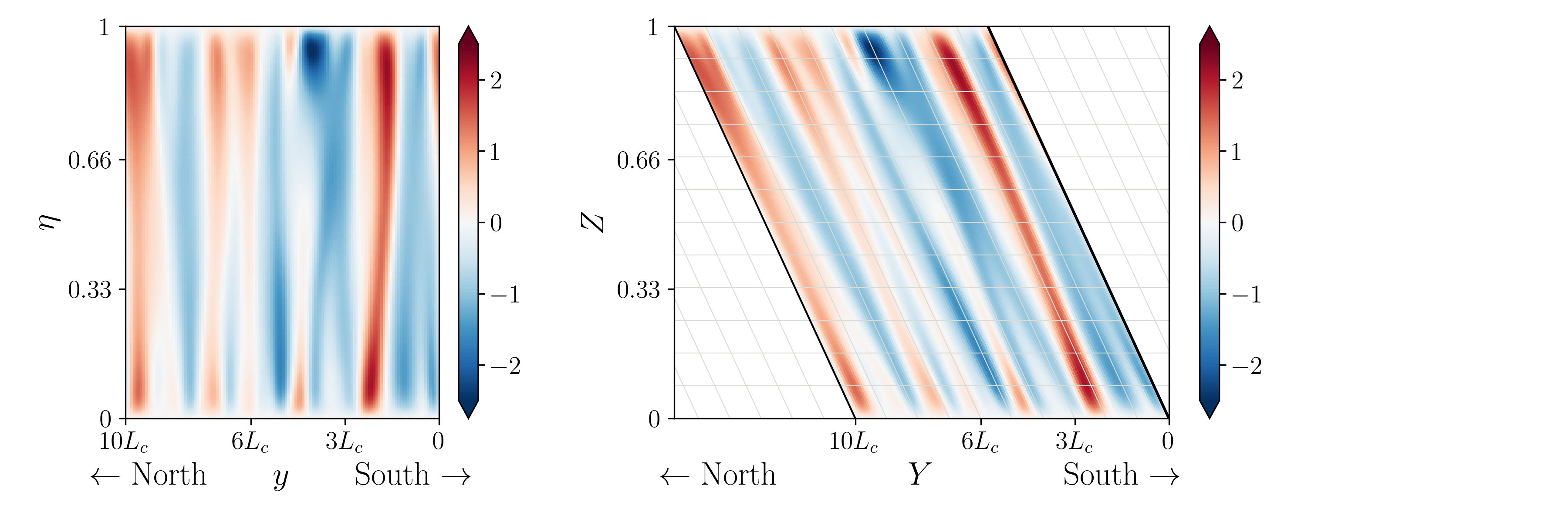}};
		\node at (-4.9,1.7) [anchor=south east] {(a)};
		\node at (-0.1,1.7) [anchor=south east] {(b)};
	\end{tikzpicture}
\\
\vspace{2mm}
	\begin{tikzpicture}
		\node at (0,0) {\includegraphics[height=4.5cm, trim={5cm 0 5cm 0}, clip]{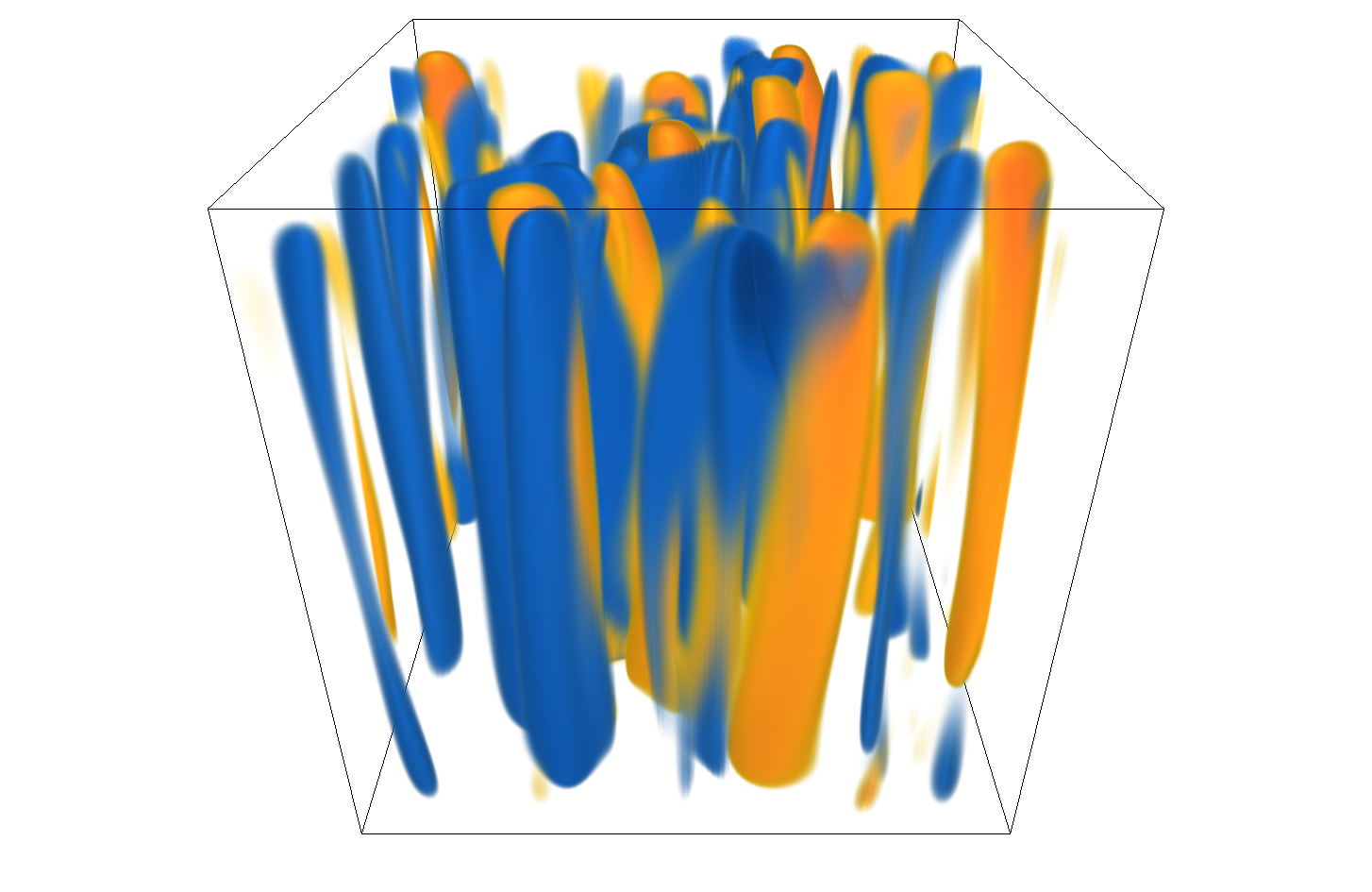}};
		\node at (-2.3,-0.2) { $\eta$};
	        \node at (0, -2.5) { $\leftarrow \, \mathrm{North} \qquad y\qquad \mathrm{ South} \, \rightarrow$};
		\node at (7,0) 
		{\includegraphics[height=4.5cm, trim={0cm 2cm 0cm 1cm}, clip]{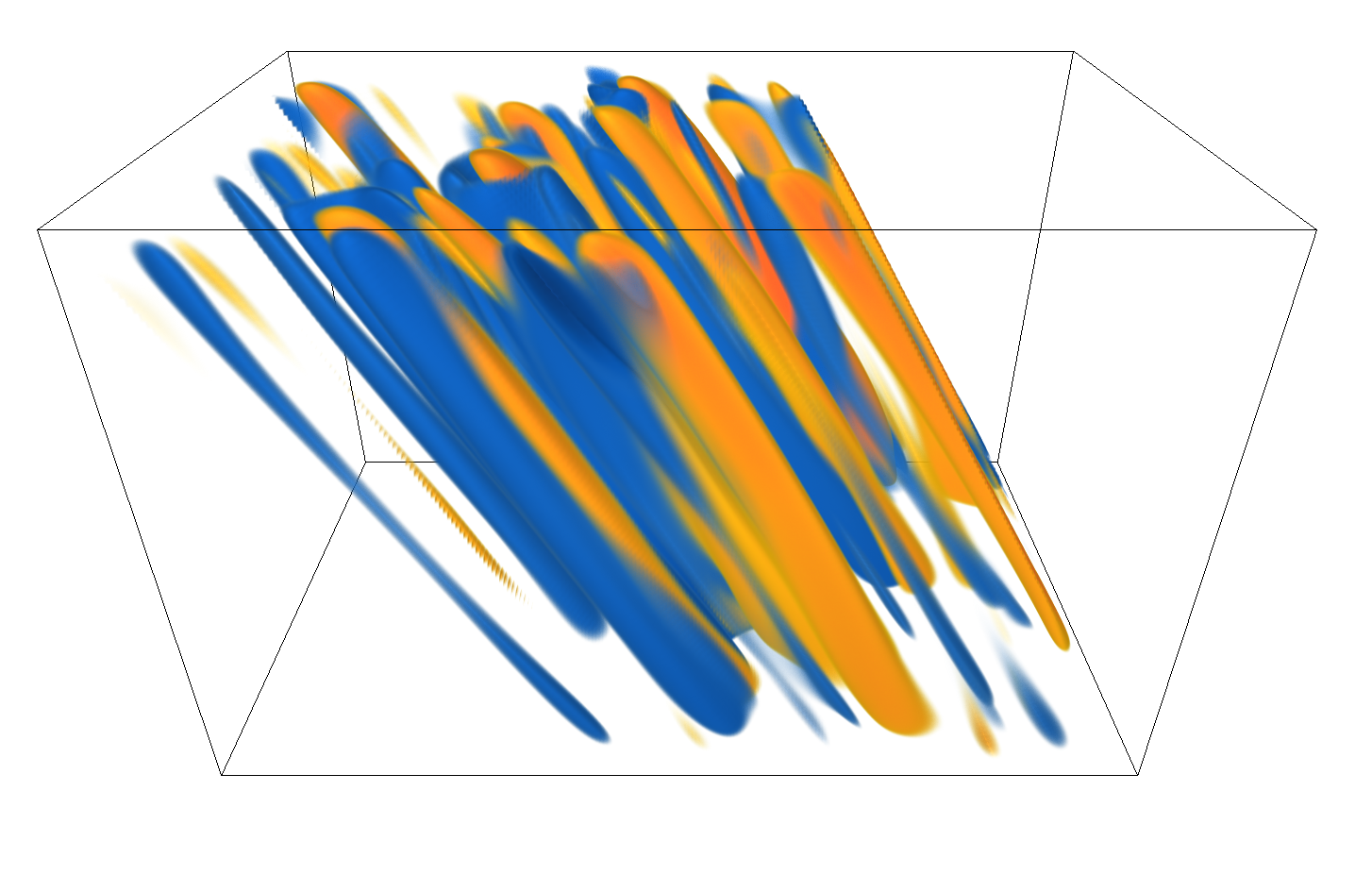}};
		\node at (3.3,-0.2) { $Z$};
	        \node at (7, -2.5) { $\leftarrow \, \mathrm{North} \qquad Y\qquad \mathrm{ South} \, \rightarrow$};
		\node at (8.75, 2) {East};
		\node at (10, 0.9) {West};
		\node at (-2,2) {(c)};
		\node at ( 4,2) {(d)};
	\end{tikzpicture}
\caption{
Snapshot of the fluctuating temperature field $\theta$ at $\wRa = 20$, 
$\vartheta_f = 30^\circ$. Cross-section in (a) the virtual $y$-$\eta$ meridional plane as defined in equation (\protect\ref{def:sheared_y}), and (b) the physical $Y$-$Z$ meridional plane.
 3D volume rendering: (c) virtual, and (d) physical. All visualizations are for 
domain dimensions $10L_c\times 10L_c\times 1$ and are not to scale: horizontal
	spatial scales are in units of $E_f^{1/3}$.}
\label{fig:theta_slice_30deg}
\end{figure}

\section{Numerical algorithm}
\label{sec:num} 

The $f$NHQGE are solved in the non-orthogonal coordinates by a pseudo-spectral method
with implicit-explicit time-stepping and periodic boundary conditions in the horizontal.
This is done with the Coral code \citep{bM21},
a flexible platform for solving partial differential equations with spectral accuracy, i.e.,
with exponential error convergence. All fluid 
variables are discretized with a Fourier-mode expansion in the horizontal and a 
Chebyshev polynomial expansion in the axial direction. The code temporally evolves 
the spectral coefficients of these modes in spectral space via the 3rd-order four-stage
implicit-explicit Runge-Kutta time-stepping scheme RK443 \citep{Ascher97}, where all
linear terms are treated implicitly as unknowns and nonlinear terms are  treated 
explicitly as known quantities from prior sub-timesteps (i.e., pseudo-spectrally). 
All simulations are performed in domains with a square horizontal cross-section to 
avoid the introduction of additional  anisotropy into the problem.

% ... FIGURE 4
\begin{figure}
\centering
	\begin{tikzpicture}
		\node at (0,0) {
			\includegraphics[height=4.5cm]{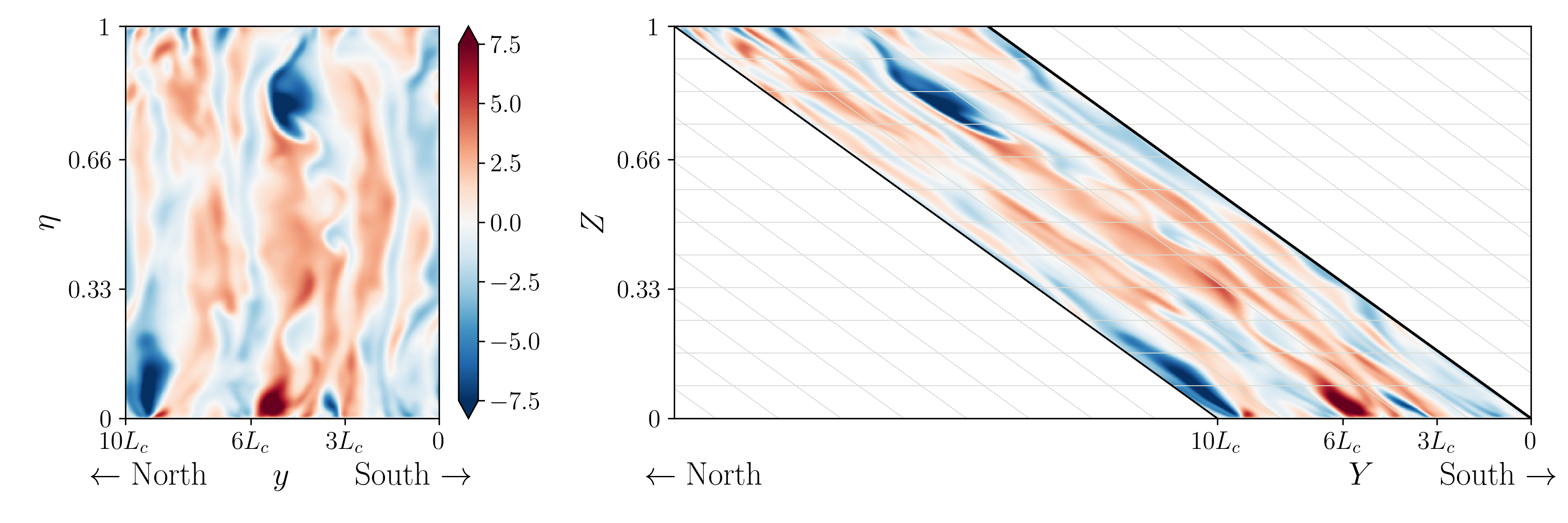}};
		\node at ( -6.,1.7) [anchor=south east] {(a)};
		\node at (-1.2,1.7) [anchor=south east] {(b)};
	\end{tikzpicture}
	\begin{tikzpicture}
		\node at (-0.5,0)
		{\includegraphics[height=3.8cm]{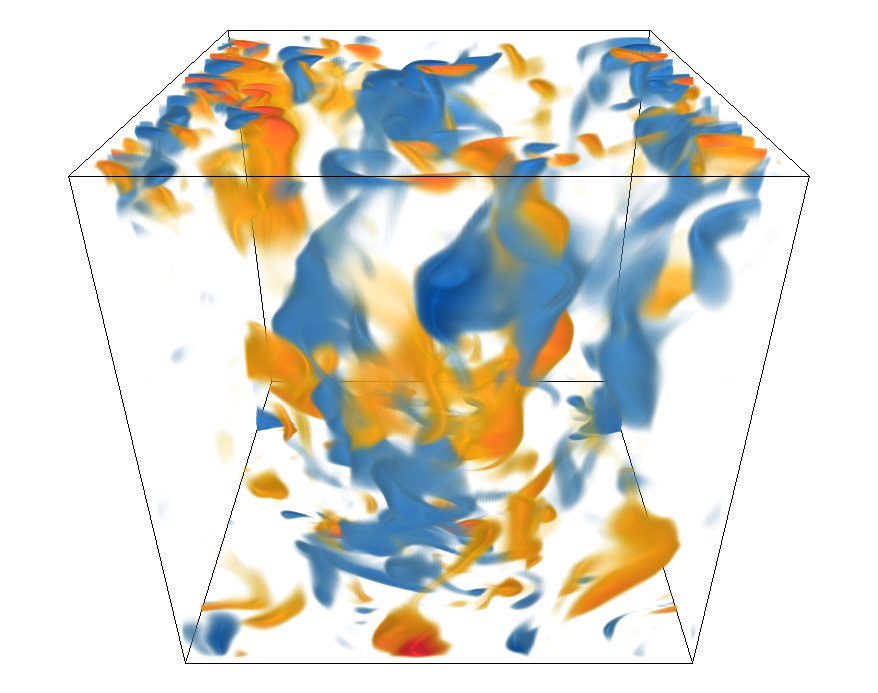}};
		\node at (-2.8,-0.2) { $\eta$};
	        \node at (-0.5, -2.10) { $\leftarrow \, \mathrm{North} \qquad y\qquad \mathrm{ South} \, \rightarrow$};
		\node at (6.5,0) 
		{\includegraphics[height=4  cm, trim={0cm 4cm 0cm 4cm}, clip]{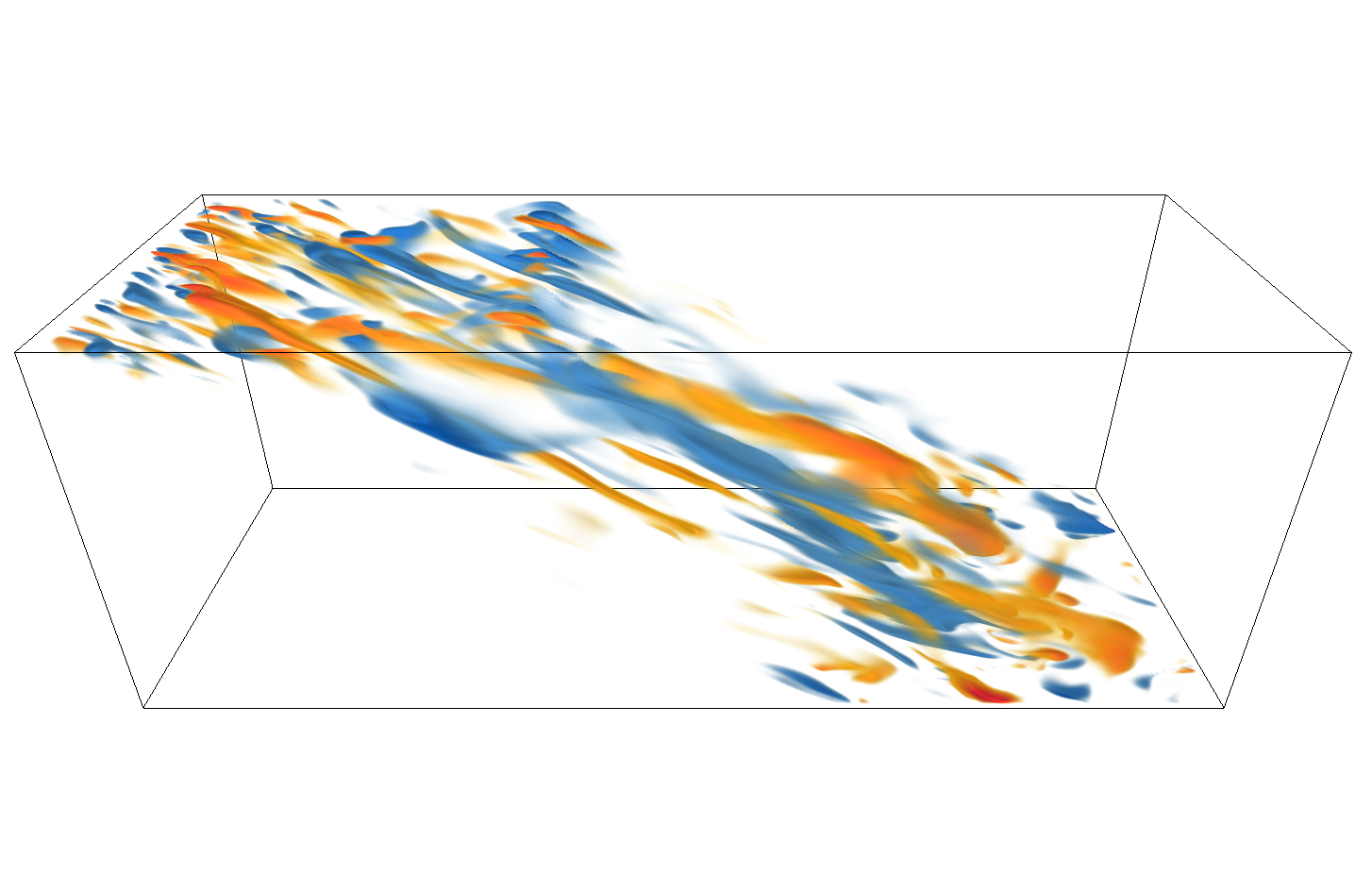}};
		\node at (2.0,-0.2) { $Z$};
	        \node at (6.5, -2.1) { $\leftarrow \, \mathrm{North} \qquad \qquad \qquad  Y \qquad \qquad \qquad \mathrm{ South} \, \rightarrow$};
		\node at (9.25, 1.5) {East};
		\node at (10.4, 0.45) {West};
		\node at (-2.2, 1.6) {(c)};
		\node at ( 2.8, 1.6) {(d)};
	\end{tikzpicture}
\caption{
	As in fig. \protect\ref{fig:theta_slice_30deg} but for 
	$\widetilde{Ra} = 120$, $\vartheta_f = 60^\circ$.
	}
\label{fig:theta_slice_60deg}
\end{figure}
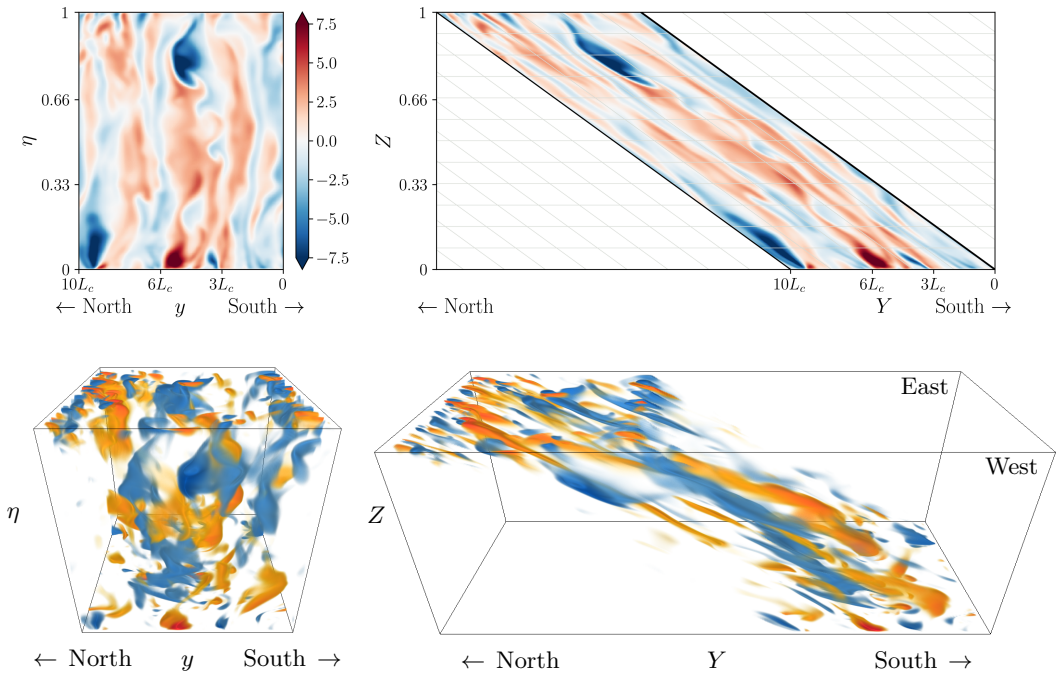

\section{Simulation results}
\label{sec:sim} 

Table~\ref{tab:simulation_grid} summarizes the simulation suites performed at $\sigma =1$
and colatitudes $\vartheta_f =\{0^\circ, 15^\circ, 30^\circ, 45^\circ, 60^\circ\}$ with 
increasing reduced Rayleigh number $\wRa\in(10,120)$. All simulations are performed in a
domain of size $10 L_c\times10 L_c \times 1$ considered sufficient to capture the entire 
range of unstable modes and also permit ample room for an inverse cascade 
\citep{mS06,kJ12,sM21}. Here $L_c=4.815$ denotes the critical convective wavelength for
N-S rolls at a specified  $\vartheta_f$; the physical N-S horizontal to axial height
domain aspect ratio is given by $10 L_c E_f^{1/3}$. Spatial discretizations are selected
such that all spatial scales of the motion down to the Kolmogorov dissipation scale
$\ell_k \sim \epsilon_u^{-1/4}$ are resolved (see eq. \eqref{eqn:dissrate}). All
simulations are integrated for time intervals sufficiently long to ensure the barotropic
flow saturates, with some cases demanding
marching the equations for as long as 900 convective time scales, corresponding to 60 vertical eddy turnover times.
Convergence to the statistically stationary state was assessed by comparing the deviation between averages 
on subsets of the data for both the Nusselt number and the three components of the kinetic energy.

\subsection{Flow morphology} 
\label{sec:Morphology}
% ... FIGURE 5
\begin{figure}
\centering
	\begin{tikzpicture}
	\node at (0,0)[anchor=south west]{
		\includegraphics[width=0.90\linewidth, trim={2.5cm 1.0cm 4.5cm 1.5cm}, clip]{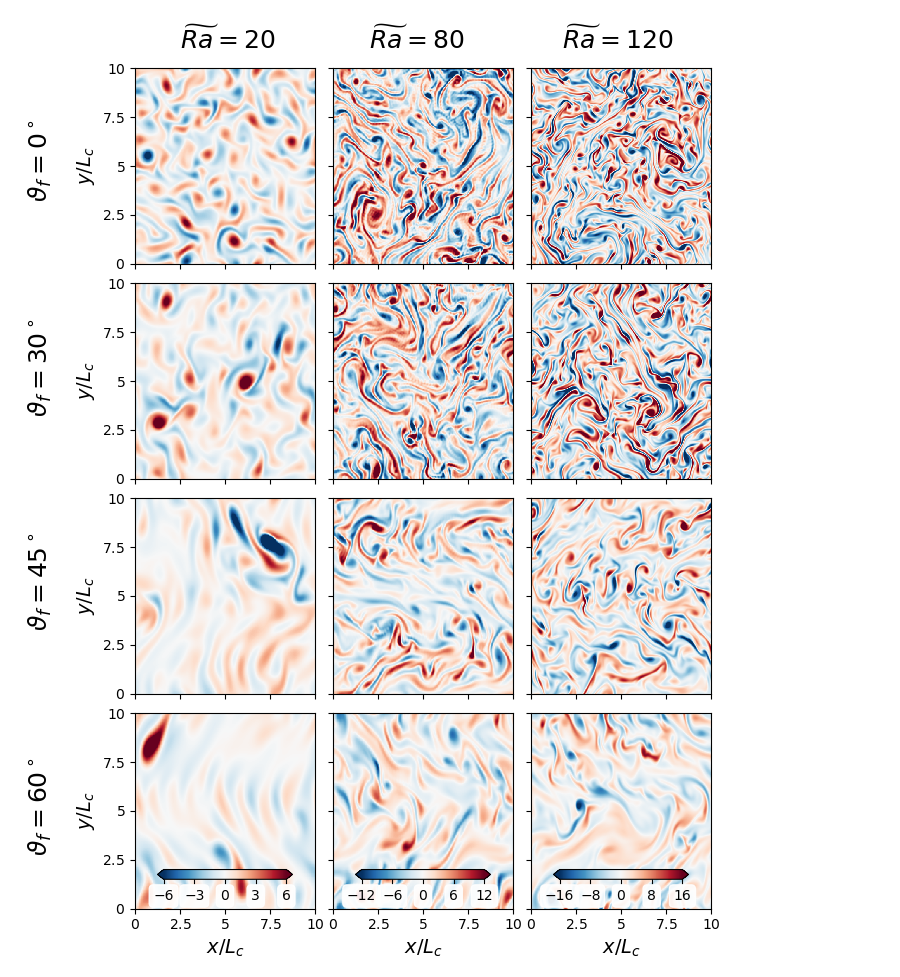}};
	\node at  (2.65,17.2)[anchor=south]{\Large$\widetilde{\rayleigh}=20$};
	\node at  (6.45,17.2)[anchor=south]{\Large$\widetilde{\rayleigh}=80$};
	\node at (10.25,17.2)[anchor=south]{\Large$\widetilde{\rayleigh}=120$};
	\node at  (2.65,0.2)[anchor=north]{\large$x/L_c$};
	\node at  (6.50,0.2)[anchor=north]{\large$x/L_c$};
	\node at (10.35,0.2)[anchor=north]{\large$x/L_c$};
	\node at (0.3, 2.80)[anchor=south, rotate=90]{\large$y/L_c$};
	\node at (0.3, 7.00)[anchor=south, rotate=90]{\large$y/L_c$};
	\node at (0.3,11.20)[anchor=south, rotate=90]{\large$y/L_c$};
	\node at (0.3,15.40)[anchor=south, rotate=90]{\large$y/L_c$};
	\node at (-0.4, 2.80)[anchor=south, rotate=90]{\Large$\vartheta_f=60^\circ$};
	\node at (-0.4, 7.00)[anchor=south, rotate=90]{\Large$\vartheta_f=45^\circ$};
	\node at (-0.4,11.20)[anchor=south, rotate=90]{\Large$\vartheta_f=30^\circ$};
	\node at (-0.4,15.40)[anchor=south, rotate=90]{\Large$\vartheta_f=0^\circ$};
        \end{tikzpicture}
\caption{
	Snapshot of the simulation grid in $\wRa$-$\vartheta_f$ parameter space.
	In all cases the temperature fluctuation $\theta$ is shown in the $(x,y)$ 
	plane at the edge of the upper thermal boundary layer. Columns (rows)
	represent fixed $\wRa$ ($\vartheta_f$). The color maps are adjusted for each
	value of $\wRa$, as indicated in the bottom row.
}
\label{fig:theta_vs_Ra_vartheta_xy_peak}
\end{figure}
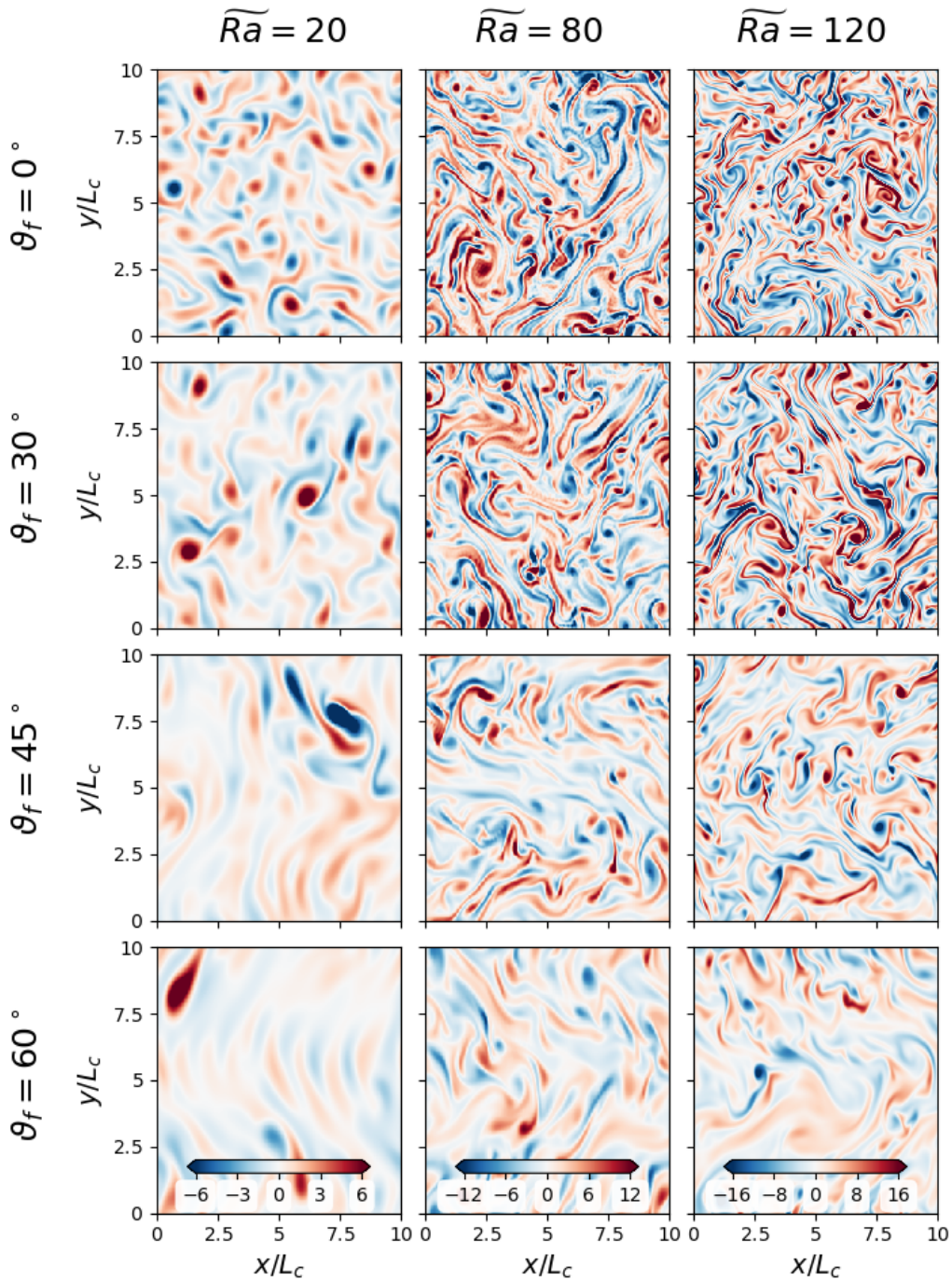

The formulation pursued here operates in a virtual cubic computational fluid domain
with coordinates $(x,y,\eta)$. This domain must be remapped to the physical domain,
a parallelepiped where the true vertical coordinate aligns with gravity and the
rotation axis is tilted according to the colatitude of the $f$-plane
(see fig.~\ref{fig:Schematic}). To set the stage for our visualizations,
figs.~\ref{fig:theta_slice_30deg} and \ref{fig:theta_slice_60deg} demonstrate the
remapping of the virtual $(x,y,\eta)$ space to the physical $(X,Y,Z)$ space for
two cases: $\widetilde{Ra}=20$, $\vartheta_f=30^\circ$ and $\widetilde{Ra}=120$,
$\vartheta_f=60^\circ$ (recall $\wRa_c \approx 8.69$ is the critical onset value
for all $\vartheta_f < 90^\circ$). The visualizations are not to scale given the
anisotropic rescaling of the $\textit{O}(1)$ vertical and $\textit{O}(E_f^{1/3})$
horizontal dimensions. Whether laminar or turbulent, coherent flow structures that
appear vertically aligned in the virtual domain (plots (a) and (c)) appear as axially
aligned in the physical domain (plots (b) and (d)).

% ... FIGURE 6
\begin{figure}
\centering
	\begin{tikzpicture}
	\node at (0,0)[anchor=south west]{
		\includegraphics[width=0.90\linewidth, trim={2.5cm 1.0cm 4.5cm 1.5cm}, clip]{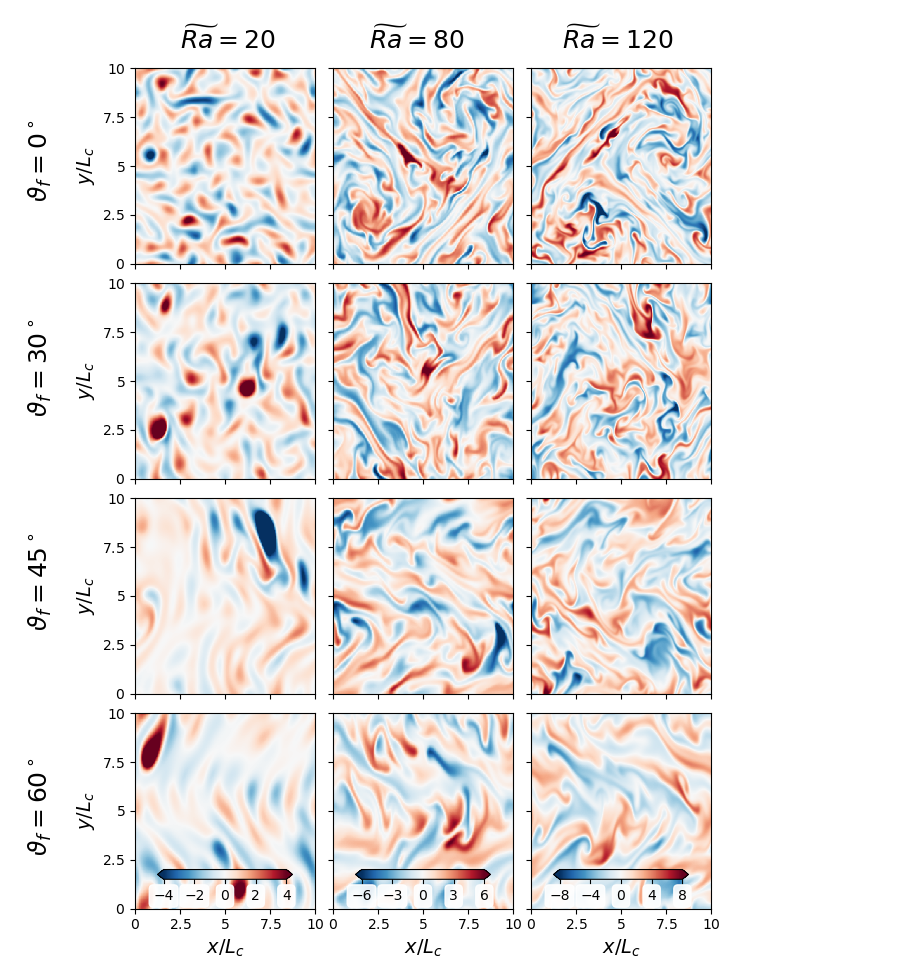}};
	\node at  (2.65,17.2)[anchor=south]{\Large$\widetilde{\rayleigh}=20$};
	\node at  (6.45,17.2)[anchor=south]{\Large$\widetilde{\rayleigh}=80$};
	\node at (10.25,17.2)[anchor=south]{\Large$\widetilde{\rayleigh}=120$};
	\node at  (2.65,0.2)[anchor=north]{\large$x/L_c$};
	\node at  (6.50,0.2)[anchor=north]{\large$x/L_c$};
	\node at (10.35,0.2)[anchor=north]{\large$x/L_c$};
	\node at (0.3, 2.80)[anchor=south, rotate=90]{\large$y/L_c$};
	\node at (0.3, 7.00)[anchor=south, rotate=90]{\large$y/L_c$};
	\node at (0.3,11.20)[anchor=south, rotate=90]{\large$y/L_c$};
	\node at (0.3,15.40)[anchor=south, rotate=90]{\large$y/L_c$};
	\node at (-0.4, 2.80)[anchor=south, rotate=90]{\Large$\vartheta_f=60^\circ$};
	\node at (-0.4, 7.00)[anchor=south, rotate=90]{\Large$\vartheta_f=45^\circ$};
	\node at (-0.4,11.20)[anchor=south, rotate=90]{\Large$\vartheta_f=30^\circ$};
	\node at (-0.4,15.40)[anchor=south, rotate=90]{\Large$\vartheta_f=0^\circ$};
        \end{tikzpicture}
\caption{As in fig.~\protect\ref{fig:theta_vs_Ra_vartheta_xy_peak} 
	but at the midplane location $\eta =0.5$.}
\label{fig:theta_vs_Ra_vartheta_xy_mid}
\end{figure}

\begin{figure}
\centering
\includegraphics[width=\textwidth, trim={0cm 3cm 0cm 0cm}, clip]{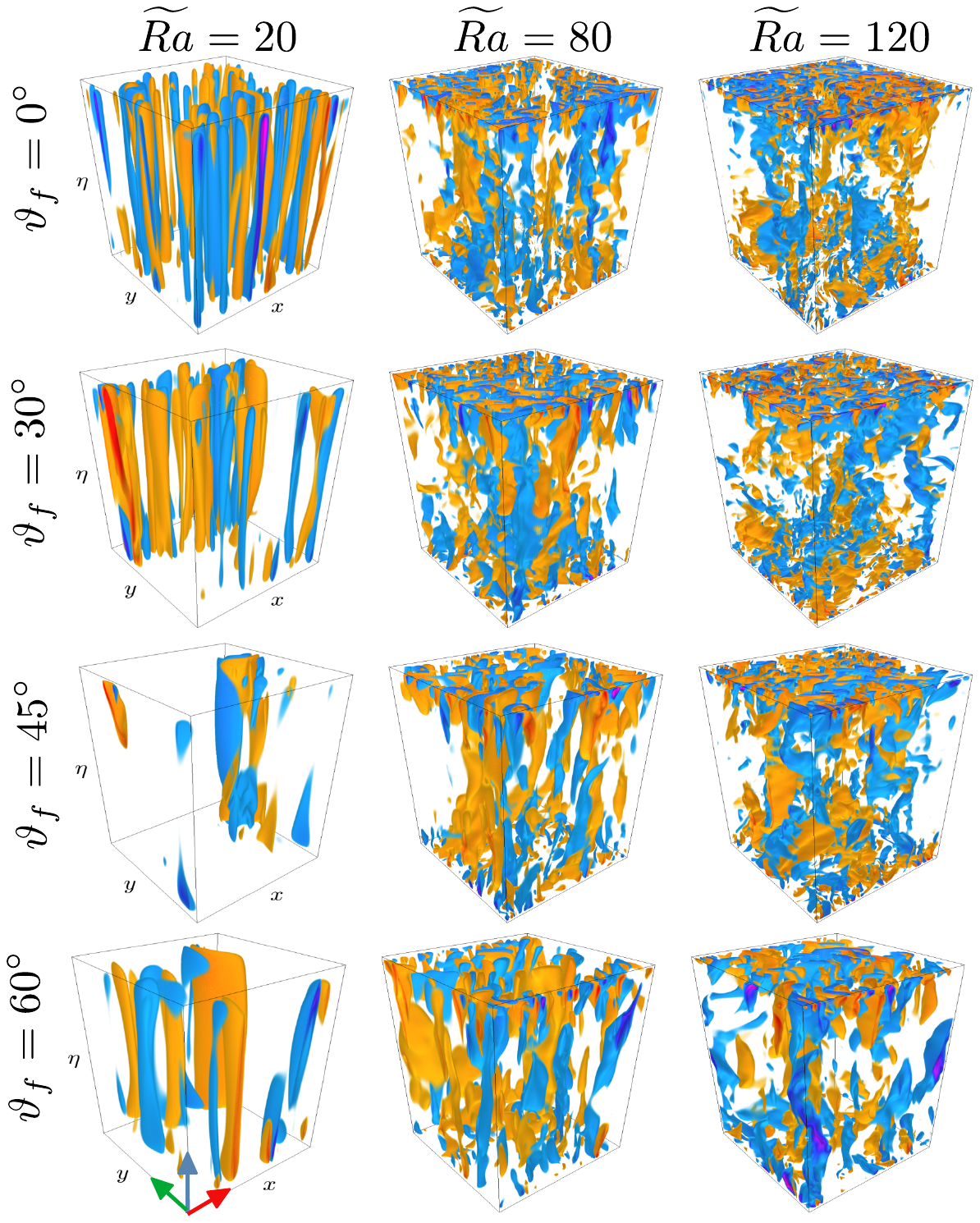}
\caption{
	Volume renderings of the temperature fluctuation $\theta$ for 
	various values of $\wRa$ (columns) and $\vartheta_f$ (rows) in the virtual
	computational domain. The axes in the lower left show the zonal 
	($x$, red), meridional ($y$, green), and axial ($\eta$, blue) directions. The physical domain is obtained via shearing in the
	$(y,\eta)$ plane.
        } 
\label{fig:volume_renderings_theta}
\end{figure}

An overview of the flow morphology as a function of colatitude $\vartheta_f$ and
$\widetilde{Ra}$ is given in figs.~\ref{fig:theta_vs_Ra_vartheta_xy_peak}, 
\ref{fig:theta_vs_Ra_vartheta_xy_mid} and \ref{fig:volume_renderings_theta}, where
snapshots of the fluctuating temperature field $\theta$ are shown in horizontal 
cross-sections at the thermal boundary layer, horizontal cross-sections at the 
midplane, and 3D volume renderings in the virtual domain, respectively.
Observations at fixed colatitude $\vartheta_f$ and increasing $\widetilde{Ra}$ 
(i.e., plots at fixed row from left to right) reveal flows of increasing spatial
complexity, i.e., diminishing spatial scales from left to right in all figures.
It is also found that the axial vorticity $\zeta$ (not shown) is strongly
correlated with $\theta$ within the thermal boundary layer. Here, compact
coherent structures and filamentary sheets that are anomalously warm (red
color scale) with negative cyclonicity or cold (blue) with positive 
cyclonicity are observed (fig.~\ref{fig:theta_vs_Ra_vartheta_xy_peak}). The
spatial complexity is maintained by continual vortical interactions in the
form of mergers between like-signed vortex pairs and propagation of oppositely-signed
vortex pairs. Given the axial extent of the coherent structures, this behaviour
is associated with strong lateral thermal mixing clearly evidenced by the broadening
of spatial scales in visualizations of the midplane
(fig.~\ref{fig:theta_vs_Ra_vartheta_xy_mid}). Volume renderings illustrated in
fig.~\ref{fig:volume_renderings_theta} reveal a transition, as $\wRa$ increases, from columnar structures that span the layer depth (leftmost column), to a plume regime where coherent columns span the layer only intermittently (second column), and into the geostrophic
turbulence regime where columnar structures are entirely absent (rightmost column).

% ... FIGURE 7

\begin{figure}
\centering
	\begin{tikzpicture}
		\node at (0,0) [anchor=south west]{
\includegraphics[width=0.92\linewidth,
		trim={2.0cm 1.5cm 0 1cm}, clip]{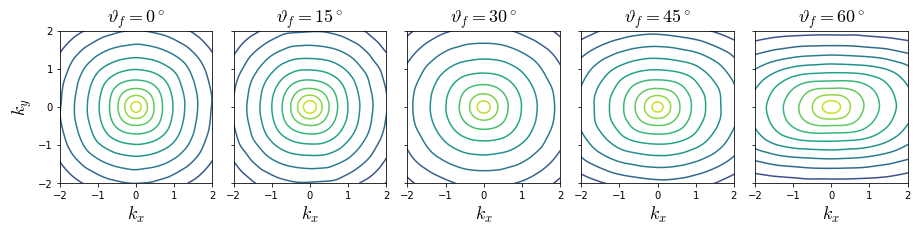}};
	\node at (0.1,0.2) [anchor=east] {\small $-2$};
	\node at (0.1,0.73) [anchor=east] {\small $-1$};
	\node at (0.1,1.26) [anchor=east] {\small $0$};
	\node at (-0.3,1.26) [anchor=south, rotate=90] {$k_y$};
	\node at (0.1,1.79) [anchor=east] {\small $1$};
	\node at (0.1,2.32) [anchor=east] {\small $2$};
	\node at (0.16,0.13) [anchor=north] {\small $-2$};
	\node at (0.70,0.13) [anchor=north] {\small $-1$};
	\node at (1.25,0.13) [anchor=north] {\small $0$};
	\node at (1.80,0.13) [anchor=north] {\small $1$};
	\node at (2.31,0.13) [anchor=north] {\small $2$};
	\node at (0.16,0.13) [anchor=north, shift={(2.5,0)}] {\small $-2$};
	\node at (0.70,0.13) [anchor=north, shift={(2.5,0)}] {\small $-1$};
	\node at (1.25,0.13) [anchor=north, shift={(2.5,0)}] {\small $0$};
	\node at (1.80,0.13) [anchor=north, shift={(2.5,0)}] {\small $1$};
	\node at (2.31,0.13) [anchor=north, shift={(2.5,0)}] {\small $2$};
	\node at (0.16,0.13) [anchor=north, shift={(  5,0)}] {\small $-2$};
	\node at (0.70,0.13) [anchor=north, shift={(  5,0)}] {\small $-1$};
	\node at (1.25,0.13) [anchor=north, shift={(  5,0)}] {\small $0$};
	\node at (1.80,0.13) [anchor=north, shift={(  5,0)}] {\small $1$};
	\node at (2.31,0.13) [anchor=north, shift={(  5,0)}] {\small $2$};
	\node at (0.16,0.13) [anchor=north, shift={(7.5,0)}] {\small $-2$};
	\node at (0.70,0.13) [anchor=north, shift={(7.5,0)}] {\small $-1$};
	\node at (1.25,0.13) [anchor=north, shift={(7.5,0)}] {\small $0$};
	\node at (1.80,0.13) [anchor=north, shift={(7.5,0)}] {\small $1$};
	\node at (2.31,0.13) [anchor=north, shift={(7.5,0)}] {\small $2$};
	\node at (0.16,0.13) [anchor=north, shift={(10 ,0)}] {\small $-2$};
	\node at (0.70,0.13) [anchor=north, shift={(10 ,0)}] {\small $-1$};
	\node at (1.25,0.13) [anchor=north, shift={(10 ,0)}] {\small $0$};
	\node at (1.80,0.13) [anchor=north, shift={(10 ,0)}] {\small $1$};
	\node at (2.31,0.13) [anchor=north, shift={(10 ,0)}] {\small $2$};
	\node at (1.25,-0.2) [anchor=north] {$k_x$};
	\node at (1.25,-0.2) [anchor=north, shift={(2.5,0)}] {$k_x$};
	\node at (1.25,-0.2) [anchor=north, shift={(  5,0)}] {$k_x$};
	\node at (1.25,-0.2) [anchor=north, shift={(7.5,0)}] {$k_x$};
	\node at (1.25,-0.2) [anchor=north, shift={(10.,0)}] {$k_x$};
	\node at (1.25, 2.4) [anchor=south] {$\vartheta_f=0^\circ$};
	\node at (1.25, 2.4) [anchor=south, shift={(2.5,0)}] {$\vartheta_f=15^\circ$};
	\node at (1.25, 2.4) [anchor=south, shift={(  5,0)}] {$\vartheta_f=30^\circ$};
	\node at (1.25, 2.4) [anchor=south, shift={(7.5,0)}] {$\vartheta_f=45^\circ$};
	\node at (1.25, 2.4) [anchor=south, shift={(10.,0)}] {$\vartheta_f=60^\circ$};
	\end{tikzpicture}
\caption{
	Spectral power map for the baroclinic component of the streamfunction
	$\Psi'=\Psi-\lbr \Psi \rbr$ at $\wRa = 120$ at the midplane $\eta = 0.5$.
	For the upright case $\vartheta_f=0^\circ$, the contour lines are circular
	in the $(k_{x}, k_{y})$ plane. As the colatitude increases contours 
	deform into ellipses with energy levels decaying more rapidly in the
	meridional $k_{y}$ direction than the zonal $k_{x}$ direction, i.e.,
	more power increasingly resides in meridionally aligned modes. To eliminate
	noise and generate smooth spectral contours, the time-averaged spectra are
	Gaussian-filtered before computing the contour lines.
	}
\label{fig:psi_full_spectrum_bc}
\end{figure}

The linear theory discussed in Section \ref{sec:linear_stability} and illustrated
in fig.~\ref{F:Marg} indicates an increasing stabilization of small-scale fluid
structures that depart from meridional alignment. Thus there is a tendency for
such structures to be most active at lower horizontal wavenumbers  (i.e., larger
spatial scales). This is evident in the spatial broadening of non-meridional fluid
structures in the horizontal cross-sections as $\vartheta_f$ increases at fixed
$\widetilde{Ra}$ (see columns top to bottom in 
figs.~\ref{fig:theta_vs_Ra_vartheta_xy_peak} and \ref{fig:theta_vs_Ra_vartheta_xy_mid}).
This broadening is also clearly demonstrated by the circular to elliptical deformation
of the level contours of two-dimensional spectral power maps of $\theta$ in the
$(k_x,k_y)$ plane computed at the layer mid-plane at $\wRa=120$ and increasing
$\vartheta_f$ (fig.~\ref{fig:psi_full_spectrum_bc}). It can be seen that the level
contours of the fully nonlinear solution strongly correlate with the level-set
contours for the growth rate obtained from linear stability theory (fig.~\ref{F:Marg}b),  thereby highlighting the role of linear instability in
driving the spectrum of convectively unstable modes at a fixed $\widetilde{Ra}$.
The distinction between the spectral power maps 
(fig.~\ref{fig:psi_full_spectrum_bc}) and linear growth rate map 
(fig.~\ref{F:Marg}b) resides in the presence of a large scale condensate that
produces power in wavenumbers at which the interior layer is linearly stable
to such convective modes. Aside from this distinction, the remaining similarities
suggest that linear theory, through the topology of the growth rates contours,
offers a way to reductively understand the observed nonlinear baroclinic dynamics.
Specifically, the growth rate contours are imprinted on the turbulent dynamics and,
to first order, the elliptical structure of the contours can be captured by the
elliptic-to-circle mapping $k_y\mapsto k_y/\eta_3$. 

% ... FIGURE 8

% ... FIGURE 9
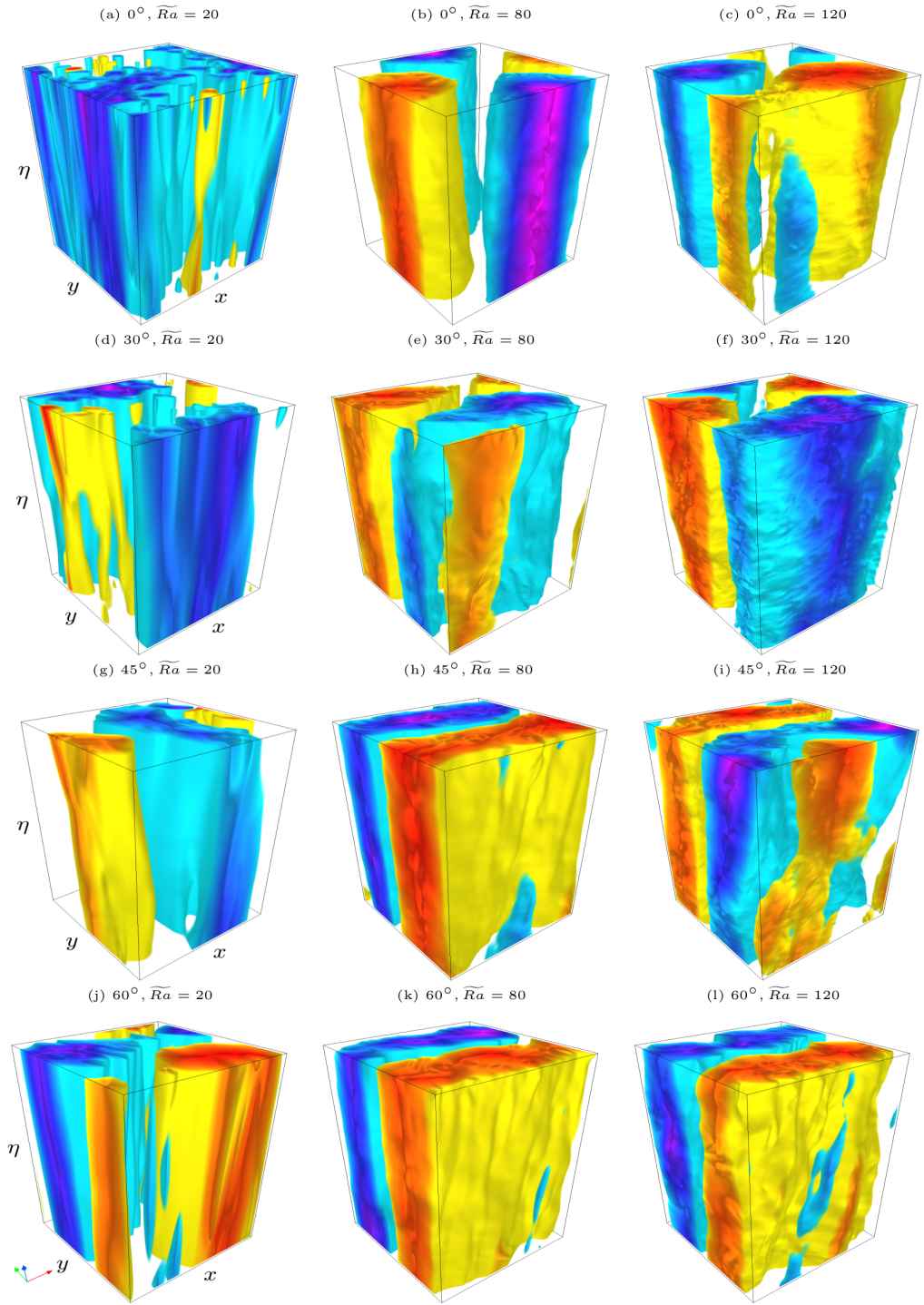
\begin{figure}
\centering
	\begin{tikzpicture}
	\node at (0,13.5) [anchor=west] 
		{\includegraphics[width=0.95\textwidth,
		trim={0 2.3cm 0 2.8cm}, clip]{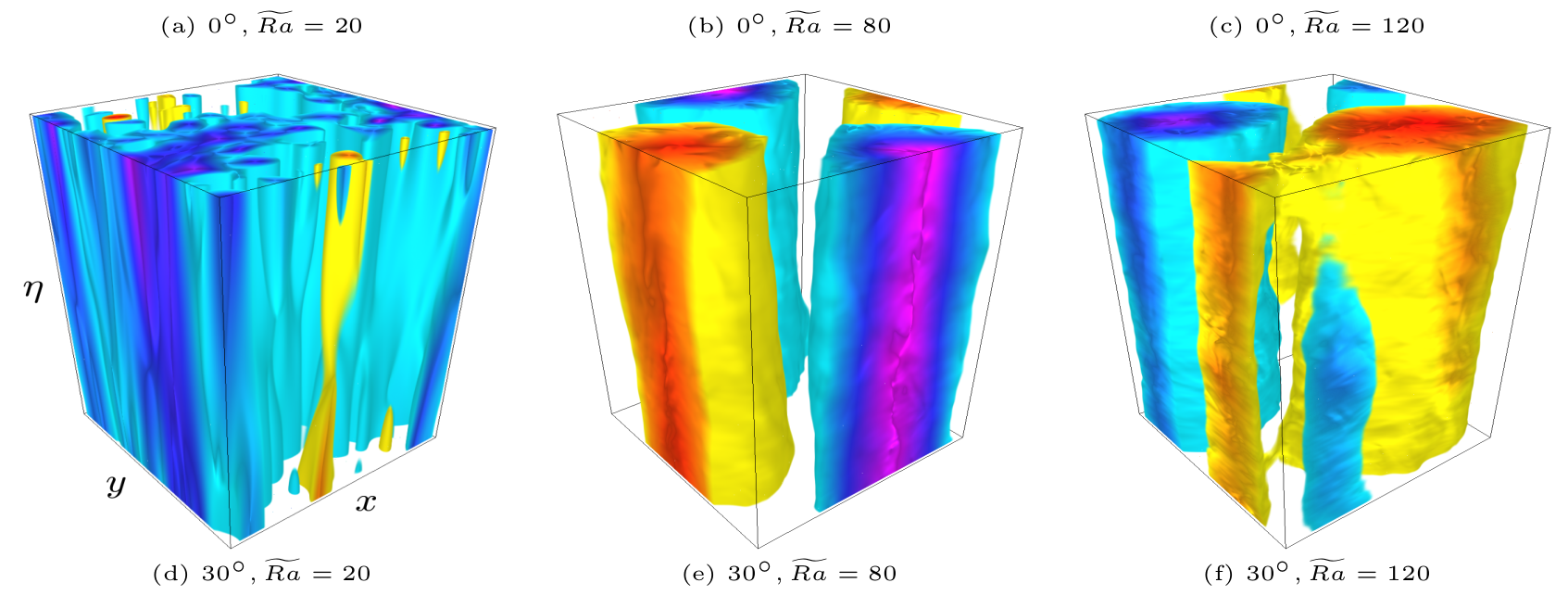}};
	\node at (0, 9) [anchor=west] 
		{\includegraphics[width=\textwidth,
		trim={0 2.3cm 0 0}, clip]{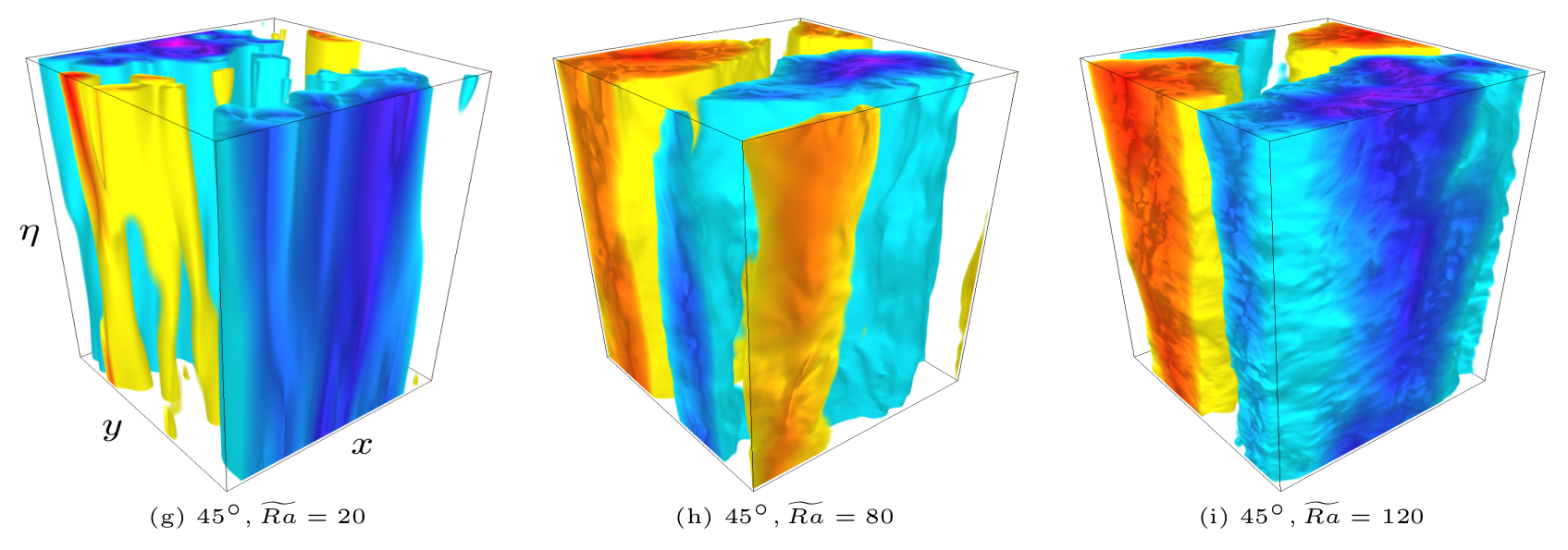}};
	\node at (0, 4.5) [anchor= west] 
		{\includegraphics[width=\textwidth,
		trim={0 2.0cm 0 0}, clip]{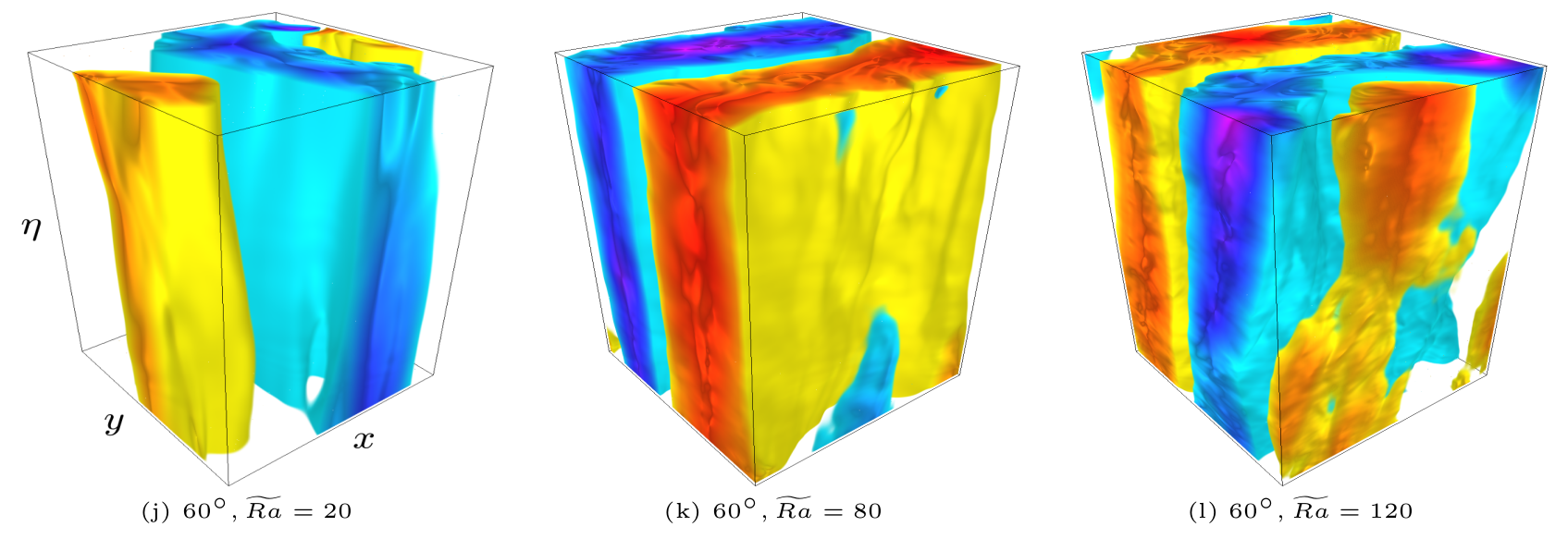}};
	\node at (0, 0) [anchor= west] 
		{\includegraphics[width=\textwidth,
		trim={0 0.6cm 0 0 0}, clip]{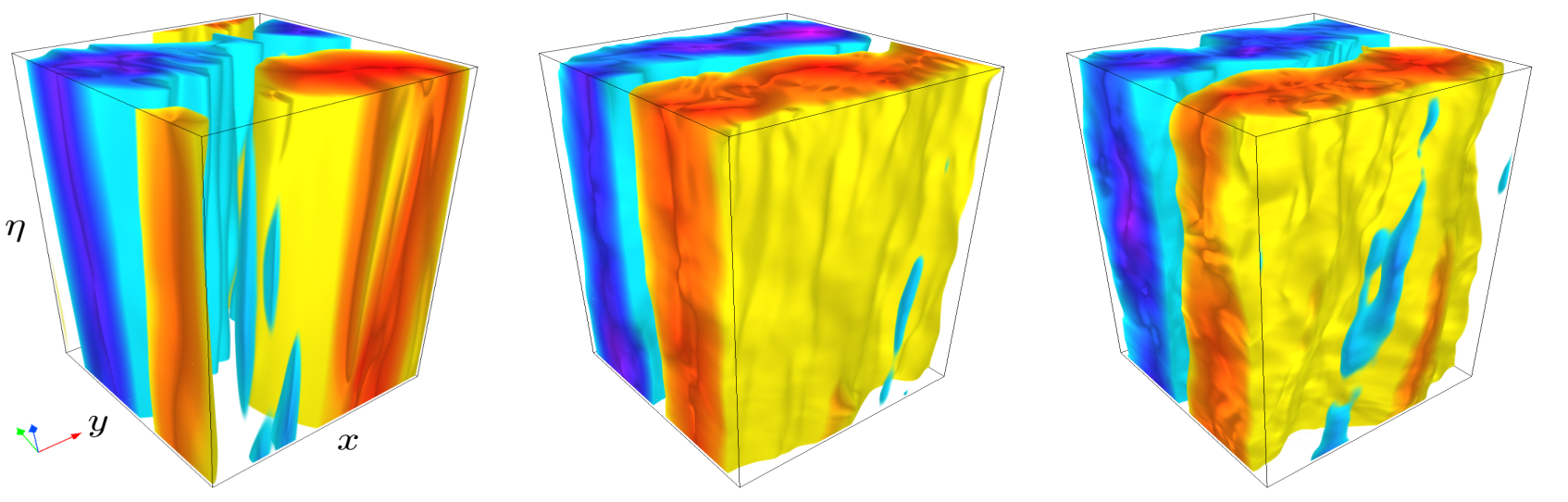}};
	\node at (0, 0.0) [anchor=south, rotate=90] {\Large$\vartheta_f=60^\circ$};
	\node at (0, 4.5) [anchor=south, rotate=90] {\Large$\vartheta_f=45^\circ$};
	\node at (0, 9.0) [anchor=south, rotate=90] {\Large$\vartheta_f=30^\circ$};
	\node at (0,13.5) [anchor=south, rotate=90] {\Large$\vartheta_f= 0^\circ$};
	\node at (2.3,15.5) [anchor=south] {\Large$\widetilde{\rayleigh}= 20$};
	\node at (6.7,15.5) [anchor=south] {\Large$\widetilde{\rayleigh}= 80$};
	\node at (11.,15.5) [anchor=south] {\Large$\widetilde{\rayleigh}=120$};
	\end{tikzpicture}
\caption{
	Volume renderings of the streamfunction $\Psi$ in the virtual
	$(x,y,\eta)$-domain for various values of $\wRa$ (columns) and $\vartheta_f$
	(rows). Illustrated is the evolution of the condensate from a large
	scale vortex (LSV) to a zonal jet (ZJ) as $\vartheta_f$ increases. 
}
\label{fig:volume_renderings_psi}
\end{figure}
 
 \subsection{Inverse cascade and the barotropic manifold} 
 \label{sec:InvCasBM}
 \begin{table}
\centering
\begin{tabular}{p{4em} p{4em} p{4em} p{4em} p{4em} p{4em} p{4em} p{4em}}
\hline
\hline
$\wRa$ / $\vartheta_{f}$ & $0^{\circ}$ & $15^{\circ}$ & $30^{\circ}$ & $40^{\circ}$ & $45^{\circ}$ & $50^{\circ}$ & $60^{\circ}$ \\
\hline
$\hphantom{1}10$  & NoLSF  & NoLSF  & NoLSF  & NoLSF     & NoLSF    & NoLSF    & NoLSF \\
$\hphantom{1}20$  & NoLSF  & NoLSF  & NoLSF  & NoLSF     & NoLSF    & NoLSF    & NoLSF \\
$\hphantom{1}40$  & LSV    & LSV    & LSV    & B(LSV)    & B        & B        & ZJ    \\
$\hphantom{1}60$  & LSV    & LSV    & LSV    & B(LSV)    & B        & B(ZJ)    & ZJ    \\
$\hphantom{1}80$  & LSV    & LSV    & LSV    & B         & B(ZJ)    & ZJ       & ZJ    \\
$100$             & LSV    & LSV    & LSV    & B(ZJ)       & ZJ       & ZJ       & ZJ    \\
$120$             & LSV    & LSV    & LSV    & ZJ       & ZJ       & ZJ       & ZJ    \\
\hline
\hline
\end{tabular}
\caption{Large--scale flow structures in the statistically steady state from the
	$\wRa$-$\vartheta_{f}$ simulation grid. Zonal jets are denoted by ``ZJ'',
	a large scale dipolar vortex is denoted by ``LSV'', and no large scale flow
	by ``NoLSF''.   Intermittent bistable cases, where the flow state oscillates
	between East-West jets and large scale vortices are denoted by the letter
	``B''. When a bistable case is dominated in time by a particular regime,
	that regime is indicated in parentheses.}
\label{tab:large_scale_flow}
\end{table}

It is now well-established that turbulent rotating convection generates nonlocal
inverse energy transfer. This is known to occur for upright convection at
$\vartheta_f =0^\circ$ \citep{kJ12,cG14,aR14,sS14,sM21} and more recently demonstrated
for tilted convection $\vartheta_f > 0^\circ$  \citep{novi2019, barker2020}. Here,
details are provided on the nature of the inverse cascade within the asymptotic
quasi-geostrophic limit $\varepsilon\equiv E_f^{1/3}\rightarrow 0$ on the $f$-plane. 
A particular focus is placed on the barotropic manifold within which this energy
transfer takes place (Section~\ref{sec:baroman} and \citet{kJ12,aR14,bF14}). 

% ... FIGURE 10
\begin{figure}
\centering
	\includegraphics[trim={1.8cm, 0cm, 1cm, 0cm}, clip,width=1.\textwidth]{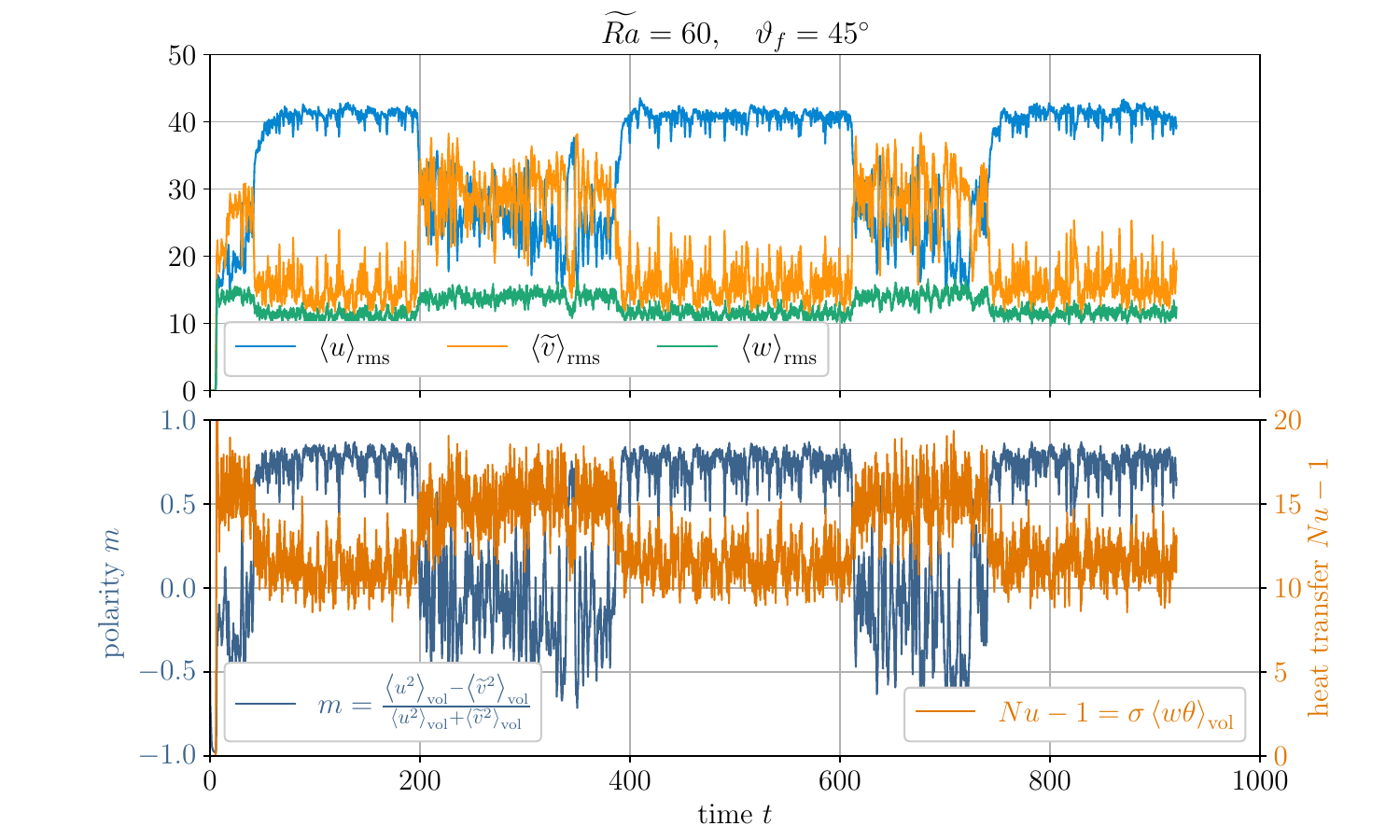}
\caption{
	Upper panel: Root-mean-square (rms) velocity components for an
	intermittent (bistable) large-scale flow with $\wRa = 60$ at
	colatitude $\vartheta_f = 45^\circ$, where
	\begin{math}
	\left \langle u \right \rangle_{\mathrm{rms}}
	= \sqrt{\left \langle u^2 \right \rangle_{\mathrm{vol}}}
	\end{math} 
	and similarly for the other two components. Bottom panel:
	Time series of the flow polarity 
	\begin{math}
	m\equiv\left(
	\left\langle u^2 \right\rangle_\mathrm{vol} - 
	\left\langle \widetilde{v}^2 \right\rangle_\mathrm{vol}
	\right) / \left(
	\left\langle u^2 \right\rangle_\mathrm{vol} +
	\left\langle \widetilde{v}^2 \right\rangle_\mathrm{vol} 
	\right)
	\end{math}
	and of the heat flux as measured by 
	$Nu-1 \equiv \sigma \left\langle w \theta \right\rangle$. 
	When the zonal velocity $u=-\partial_y \Psi$ is large compared to 
	meridional velocity $\widetilde{v}=\partial_x \Psi$ the flow is
	in a zonal jet state, characterised by the polarity $m$ taking
	values close to unity.  When the quantities are similar in 
	magnitude the flow exhibits a large-scale dipolar vortex, for
	which $m$ fluctuates around 0.  
}
\label{fig:rms_vs_t}
\end{figure}

The emergence of large-scale structures is evident in all flow fields but is most
explicitly observable within the geostrophic streamfunction field $\Psi$ as
illustrated in the volume rendering in the $\widetilde{Ra}$-$\vartheta_f$ parameter
space (fig.~\ref{fig:volume_renderings_psi}). Table~\ref{tab:large_scale_flow}
classifies the observed large-scale condensate in our simulation suite into four
categories: no observed large scale flow (`NoLSF'), a large-scale vortex dipole
(`LSV'), a zonal flow (`ZJ') consisting of an eastward and westward propagating
jet, and perhaps remarkably, an intermediate large-scale flow exhibiting
bistability between a LSV and a zonal jet (`B'). No meridional jets are observed.
For our simulation suite the transition to a state dominated by a large
scale condensate first occurs within the range $20< \wRa < 40$. At polar, ie.,
low colatitudes 
$0^\circ \lesssim \vartheta_f \lesssim 30^\circ$, 
it is observed that 
as $\widetilde{Ra}$ increases the inverse cascade results in a LSV. At large
colatitudes, $\vartheta_f \gtrsim 60^\circ$, zonal jets are the preferred state.
Owing to increasing computational expense, the simulation suite is restricted
to $\vartheta \le 60^\circ$. At mid-to-low latitudes 
$30^\circ \lesssim \vartheta_f < 60^\circ$, an intermediate bistable regime is
observed displaying an on/off switching intermittency between the LSV and the
zonal jet state. More precisely, at fixed $\wRa$ and increasing $\vartheta_f$
(i.e., from left to right in Table~\ref{tab:large_scale_flow}), a bistable large
scale flow is observed that evolves from spending a longer period of time in the 
LSV state (denoted by B(LSV)) to one spending a longer time in the zonal jet
state (denoted by B(ZJ)). From Table~\ref{tab:large_scale_flow}, it is found that
the interval at fixed colatitude (columns) for the existence of the bistable
regime narrows substantially as $\wRa$ increases, although the finer details of
the conjectured closure of the bistability wedge have not been determined.  An example
of bistability is illustrated in fig.~\ref{fig:rms_vs_t} in terms of the rms
$u$-velocity as a function of time for the case $\wRa=60$, $\vartheta_f=45^\circ$.
For this case, the time intervals with a LSV or ZJ state are approximately equal
and $\textit{O}(100)$ in length while the transition time between the two states
is rapid, of $\textit{O}(10)$.  It can also be seen that when $u_{rms}\sim \widetilde{v}_{rms}$
in the time series the flow is in the LSV state while a zonal jet is observed when
$u_{rms}\gtrsim \widetilde{v}_{rms}$. A signature of mode switching is also evident in $w_{rms}$
with slightly higher values in the LSV state. Similar switching between a jet and
a vortex state has also been seen in two-dimensional barotropic turbulence in
rectangular periodic domains with aspect ratio $1<L_x/L_y\lesssim 1.1$ \citep{fB2009, zu23}.

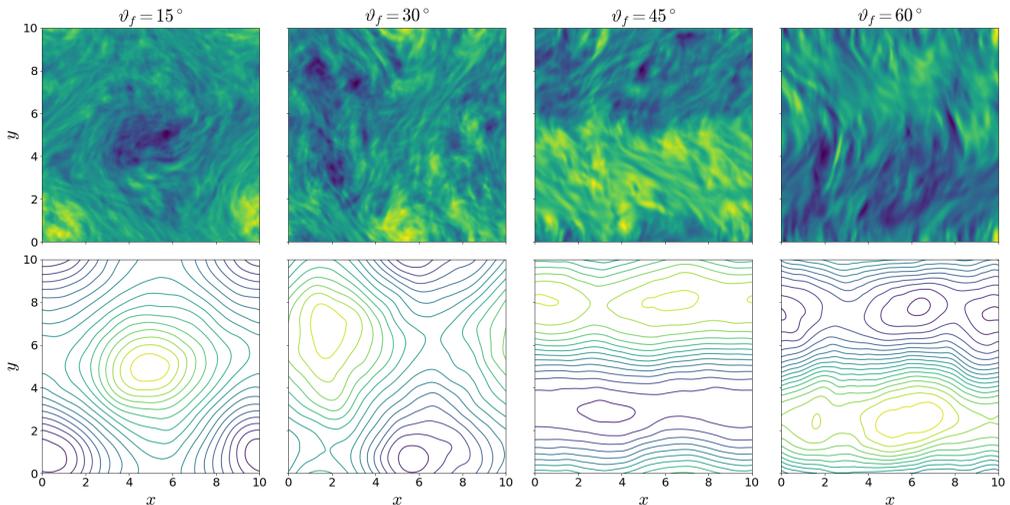
\begin{figure}
\centering
	\begin{tikzpicture}
	\node at (0,0) [anchor=south west]{
\includegraphics[width=0.92\linewidth,
	trim={1.5cm 1.5cm 0 1.45cm}, clip]
	{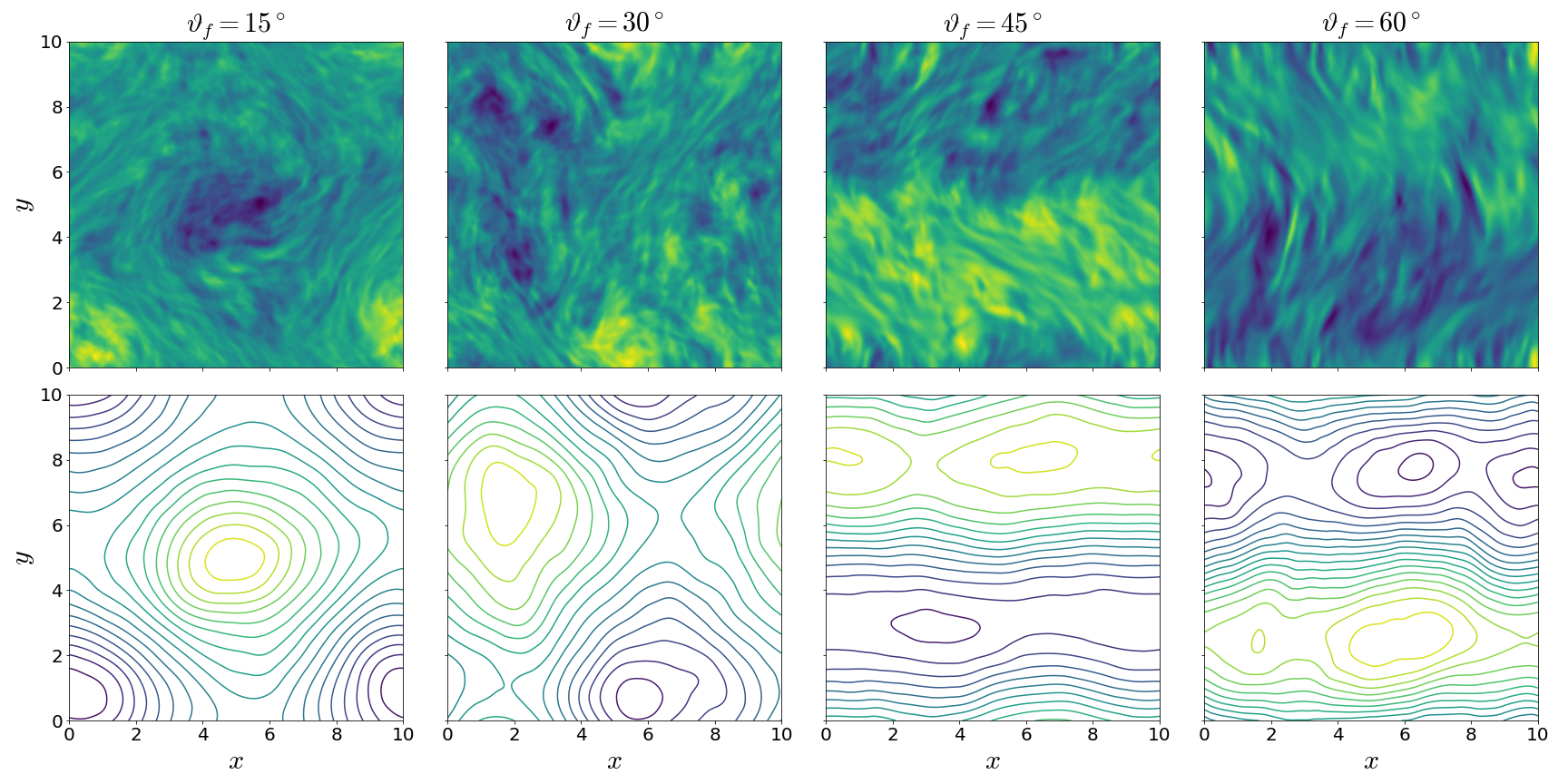}};
        \node at (0.1,1.8) [anchor=south, rotate=90]{$y$};
        \node at (0.1,4.6) [anchor=south, rotate=90]{$y$};
        \node at (1.75, 0.1) [anchor=north]{$x$};
        \node at ( 4.8, 0.1) [anchor=north]{$x$};
        \node at (7.85, 0.1) [anchor=north]{$x$};
        \node at (10.9, 0.1) [anchor=north]{$x$};
        \node at (1.75, 5.85) [anchor=south]{$\vartheta_f=15^\circ$};
        \node at ( 4.8, 5.85) [anchor=south]{$\vartheta_f=30^\circ$};
        \node at (7.85, 5.85) [anchor=south]{$\vartheta_f=45^\circ$};
        \node at (10.9, 5.85) [anchor=south]{$\vartheta_f=60^\circ$};
	\end{tikzpicture}
\caption{
	Axially-averaged vertical vorticity $\lbr\zeta\rbr$ (top row) and
	contours of the streamfunction $\lbr \Psi\rbr $ (bottom row) at
	$\wRa = 120$ for
	$\vartheta_f \in \{ 15^{\circ}, 30^{\circ}, 45^{\circ}, 60^{\circ} \}$.
	As the colatitude $\vartheta_f$ increases the dominant large scale
	vortex structure gives way to an East-West jet.
	}
\label{fig:vorticity_Ra120}
\end{figure}

Section~\ref{sec:baroman} details an analysis of the barotropic manifold and highlights
its central role as the location of the inverse energy transfer mechanism. The asymptotic
approach taken here reveals what is obscured in DNS studies of the iNSE on the $f$-plane,
i.e., the large-scale dynamics is principally captured by the axially-averaged vertical
vorticity $\left\langle \zeta \right\rangle = \nabla^{\prime 2}_\perp\left\langle \Psi \right\rangle$.
Moreover, the increased smoothness of the barotropic streamfunction
$\left\langle \Psi\right\rangle 
$ provides a clear measure for inferring the structure of the large-scale flow 
(fig.~\ref{fig:vorticity_Ra120}). At fixed $\widetilde{Ra}=120$ and increasing
$\vartheta_f$, one observes the structured evolution from LSV to ZJ
with embedded vortices (viewed as closed streamlines) within the transitional
region between the eastward and westward propagating jets (\citet{aF2017}; see also Supplementary Movie 1).
We mention that whole sphere simulations may exhibit both large scale structures simultaneously, with a deep large scale vortex at the pole and deep large scale zonal flow at larger colatitudes, with both structures statistically invariant under rotations around the polar axis \citep{linJFM21}.
% ... FIGURE 11

As noted in Section~\ref{sec:baroman} the inviscid, unforced, barotropic vorticity
equation at the pole $\vartheta_f=0^\circ$ is invariant under the rotations
$\mathcal{R}_\phi$, indicating no directional preference for the inverse energy
transfer. At the domain scale this would favor a LSV, i.e., the condensate with
the greatest degree of isotropy. However, as evident from the observation of zonal
jets, this symmetry is evidently broken once $\vartheta_f\ne0$, both at the equation
level and at the solution level. The source of this symmetry breaking is the baroclinic
forcing and anisotropic viscous dissipation terms, RHS of \eqref{eq:KEB}.
The viscous diffusion operator, given by $\nabla^{\prime2}_\perp=\pd{xx}+\eta^{-2}_3 \pd{yy}$,
implies enhanced dissipation in the  meridional direction by the factor $\eta^{-2}_3$
and hence a greater tendency to quench (i.e., smooth) meridional variations. One may
suppose that this propensity may favor meridional jets but no N-S jets are observed.
Thus the dominant source of symmetry breaking must lie with the baroclinic production terms. 

% ... FIGURE 12 bis 
\begin{figure}
\centering
\includegraphics[width=0.49\linewidth]{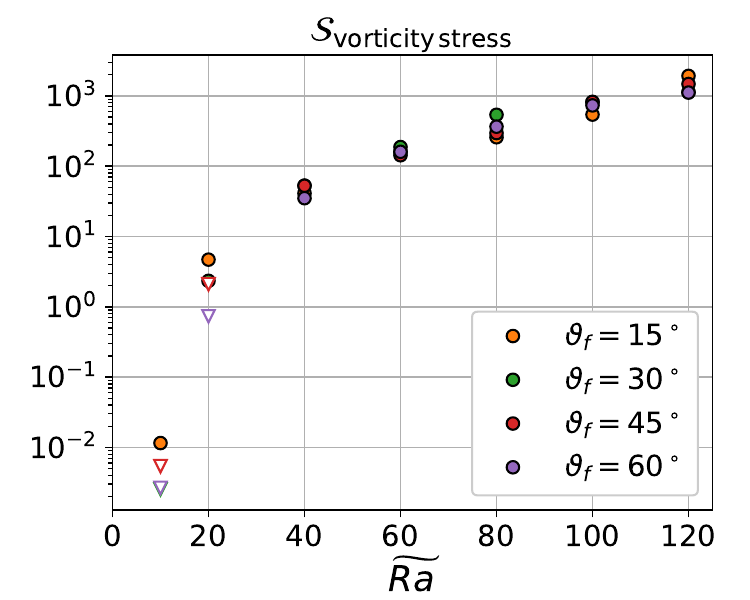}
\includegraphics[width=0.49\linewidth]{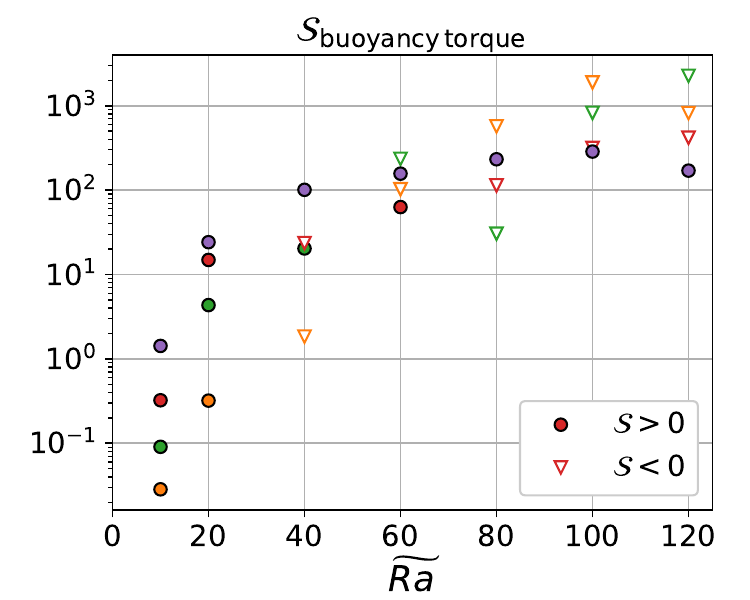}
\caption{
Contributions to the barotropic kinetic energy budget $\Sstress$ (left panel) and $\Sbuoy$ (right panel), as defined in equation~(\ref{eq:KEB}), color-coded by colatitude. In this semi-log representation, filled circles correspond to positive values whereas hollow triangles represent negative values. } 
\label{fig:baroclinic_fluxes_new}
\end{figure}
% ... FIGURE 12
\begin{figure}
\centering
\includegraphics[width=\linewidth]{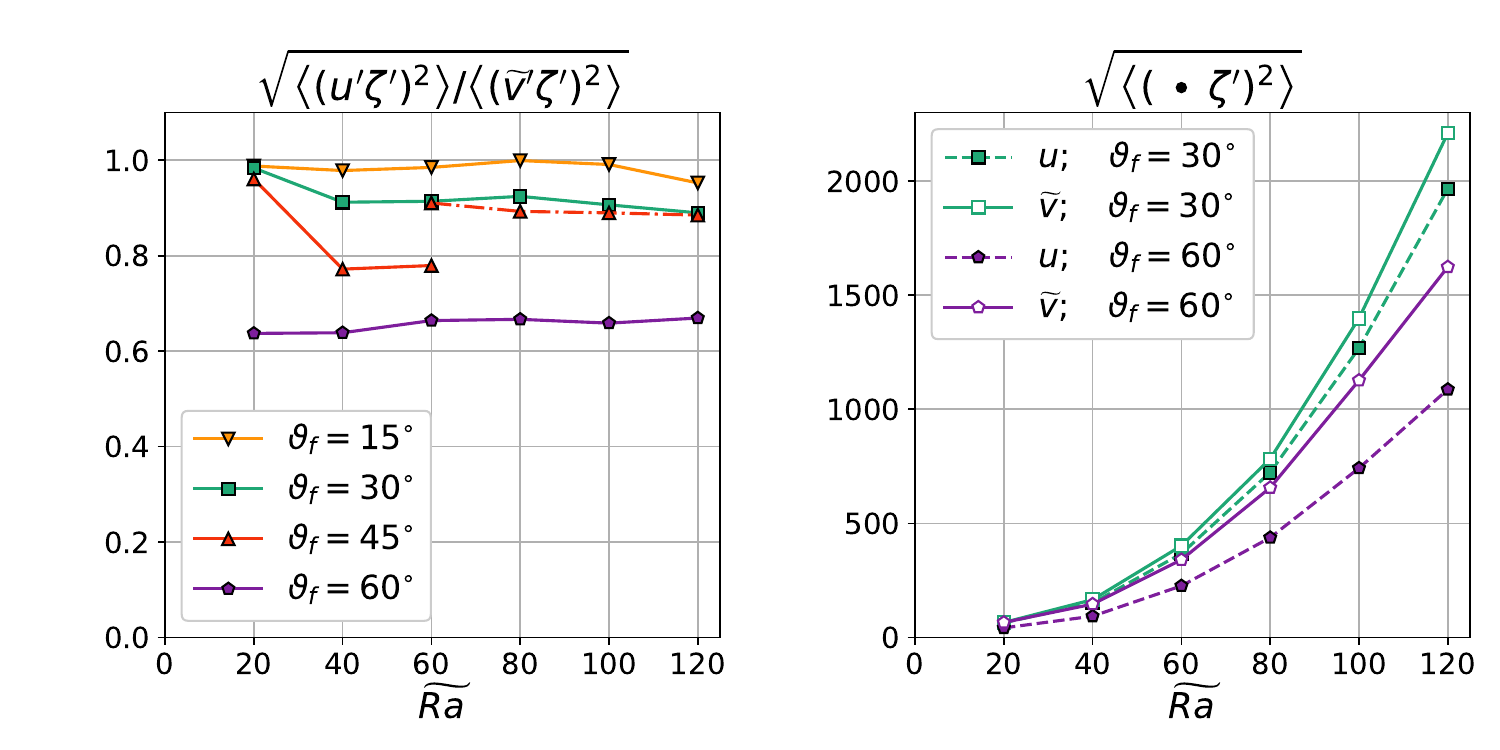}
\caption{
        Zonal ($\langle u' \zeta'\rangle_{\rm rms}$) and meridional
        ($\langle \tilde{v}' \zeta'\rangle_{\rm rms}$) vorticity fluxes as a
	function of $\widetilde{Ra}$ for various $\vartheta_f$, where the primes denote the baroclinic component obtained by subtracting the axial average along the rotation axis $\eta$ (eqs.~(\ref{subeqs:baroclinic-barotropic})); all quantities have been expressed using the streamfunction:
	$u=-\partial_y \Psi$, $\widetilde{v} = \partial_x \Psi$, 
	$\zeta=\nabla'^2_\perp\Psi$.
	Left panel: flux ratio.
	Right panel: individual fluxes for $\vartheta_f=30^\circ$, $60^\circ$.
	Data obtained from averaging over several volume samples reveal
	systematic ordering by colatitude.
} 
\label{fig:baroclinic_fluxes}
\end{figure}

Figure~\ref{fig:baroclinic_fluxes_new} presents the results of a barotropic kinetic energy budget analysis indicating the magnitude of the baroclinic source terms $\mathcal{S}_j$ in \eqref{eq:KEB} that drive the inverse cascade and generate the condensate. Positive (negative) signatures denote an energy source (sink). It is observed that $\Sstress$ (panel (a)), derived from the vortical stresses, is an energy source, and is positive for all $\vartheta_f$ and a monotonically increasing function of $\wRa$. 
Further parsing this result, we see that the relative magnitude of the baroclinic vorticity stresses in (\ref{eqn:BVE_b}b, \ref{eqn:BVE_c}) indicates a propensity in favor of driving zonal jets if $\langle \tilde{v}' \zeta'\rangle_{\rm rms}  > \langle u' \zeta'\rangle_{\rm rms}$.
This finding is corroborated in fig.~\ref{fig:baroclinic_fluxes}(a) where the ratio
of the fluxes $\langle u' \zeta'\rangle_{\rm rms}/\langle \tilde{v}' \zeta'\rangle_{\rm rms}$
is computed and shown to be less than unity for sufficiently large $\vartheta_f$
and $\wRa$; fig.~\ref{fig:baroclinic_fluxes}(b) shows the individual fluxes as a function of $\wRa$ for different values of $\vartheta_f$. 

The production of meridional kinetic energy by the buoyancy torque is illustrated in fig.~\ref{fig:baroclinic_fluxes_new}(b). 
In spite of significant temporal fluctuations, this term acts on average as a sink with an absolute magnitude that increases with $\wRa$ when $\wRa\gtrsim 80$ but decreases with increasing $\vartheta_f$, indicating increased suppression of the buoyancy torque at small colatitude.

% ... FIGURE 13
\begin{figure}
\centering
\includegraphics[width=0.49\linewidth, trim={2.5cm 6.8cm 2.7cm 7cm}, clip]{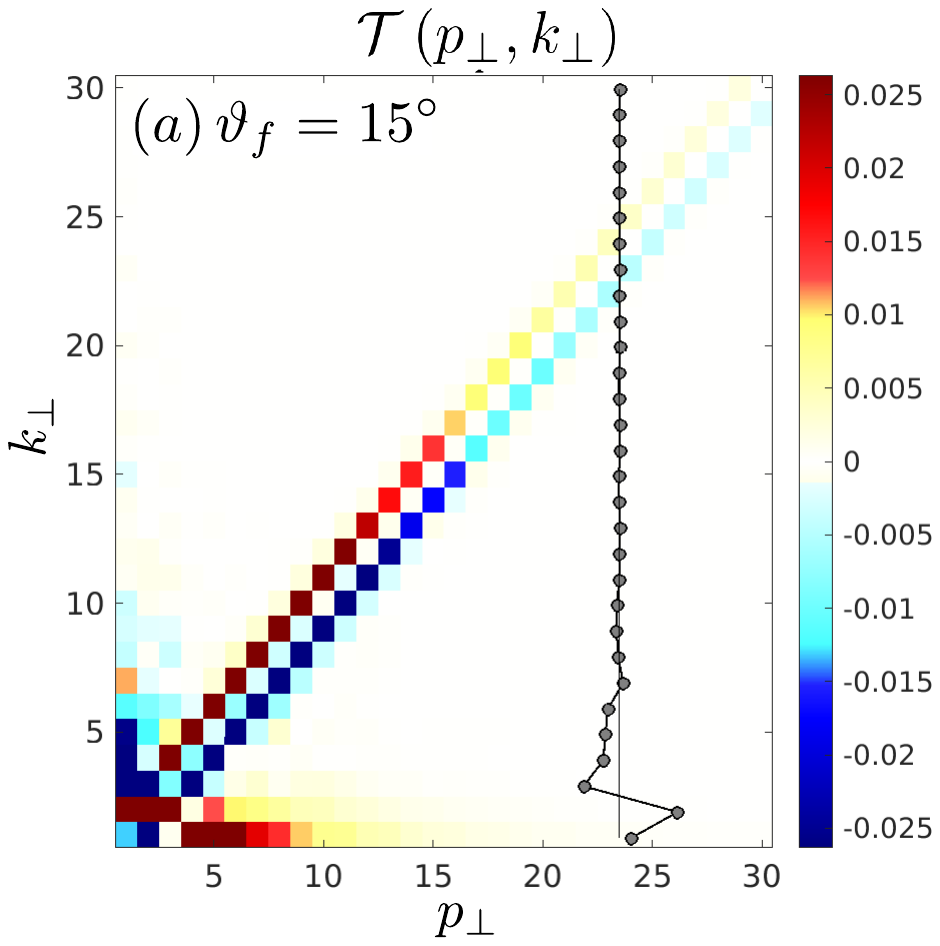}
\includegraphics[width=0.49\linewidth, trim={2.5cm 6.8cm 2.7cm 7cm}, clip]{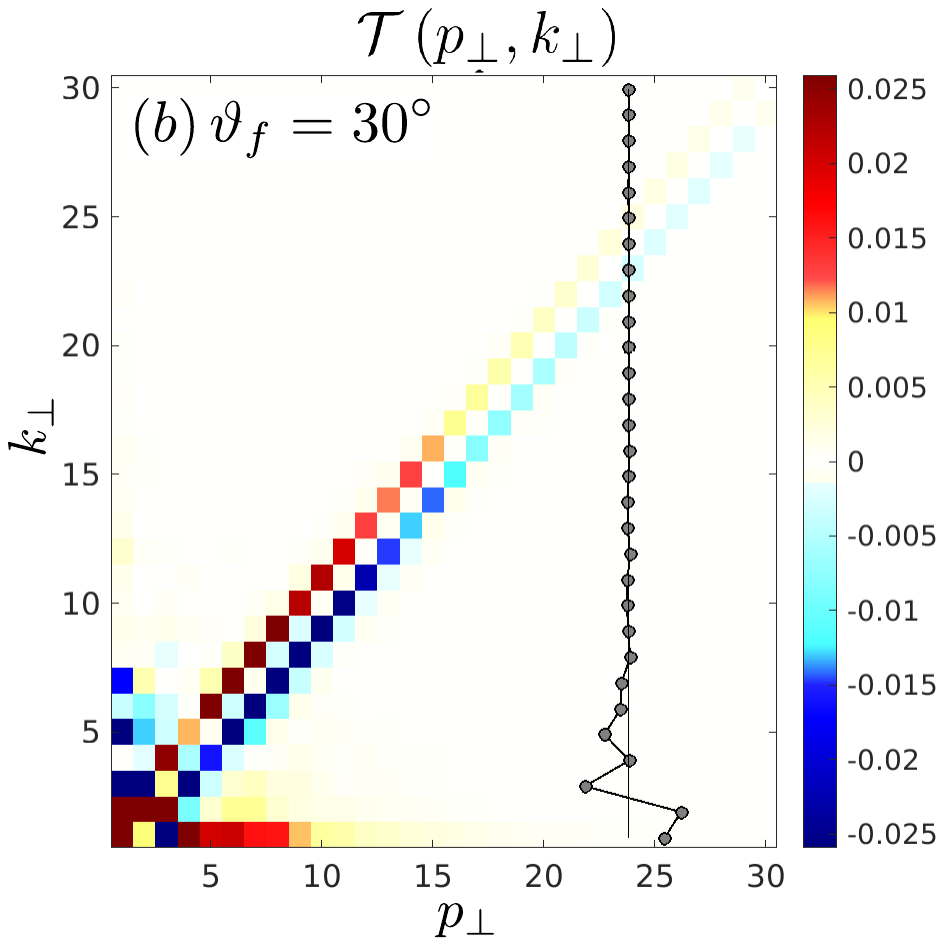}
\includegraphics[width=0.49\linewidth, trim={2.5cm 6.8cm 2.7cm 7cm}, clip]{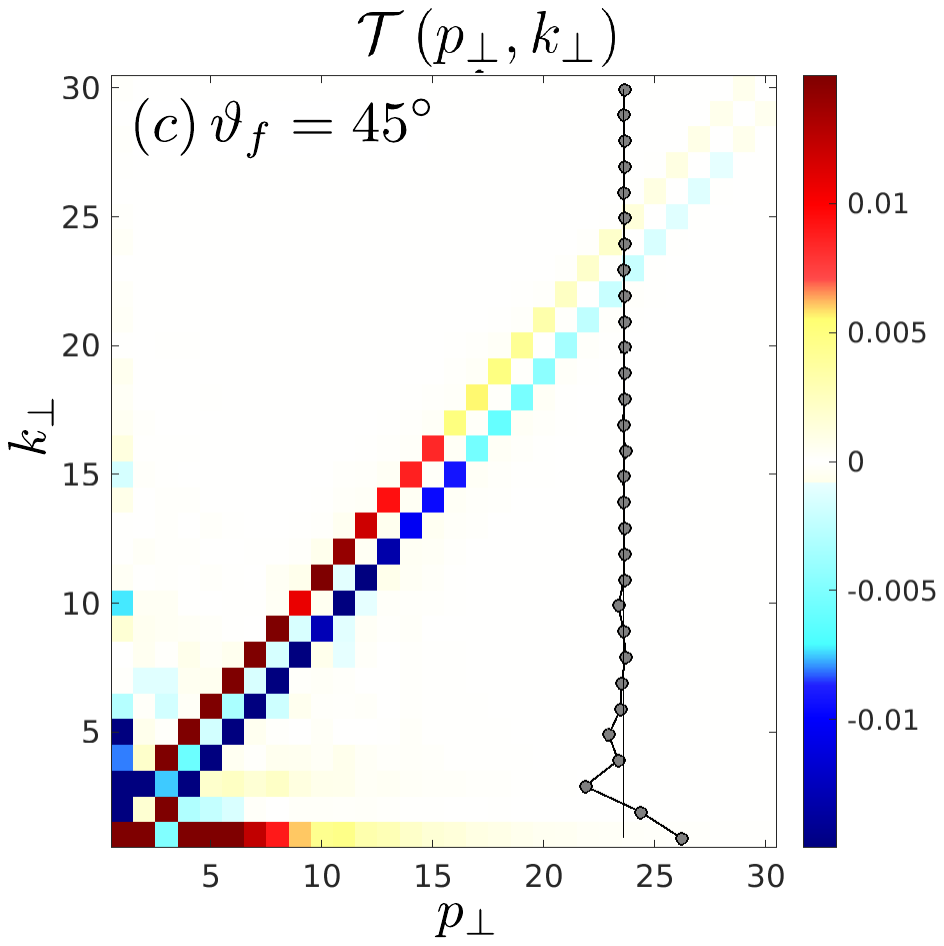}
\includegraphics[width=0.49\linewidth, trim={2.5cm 6.8cm 2.7cm 7cm}, clip]{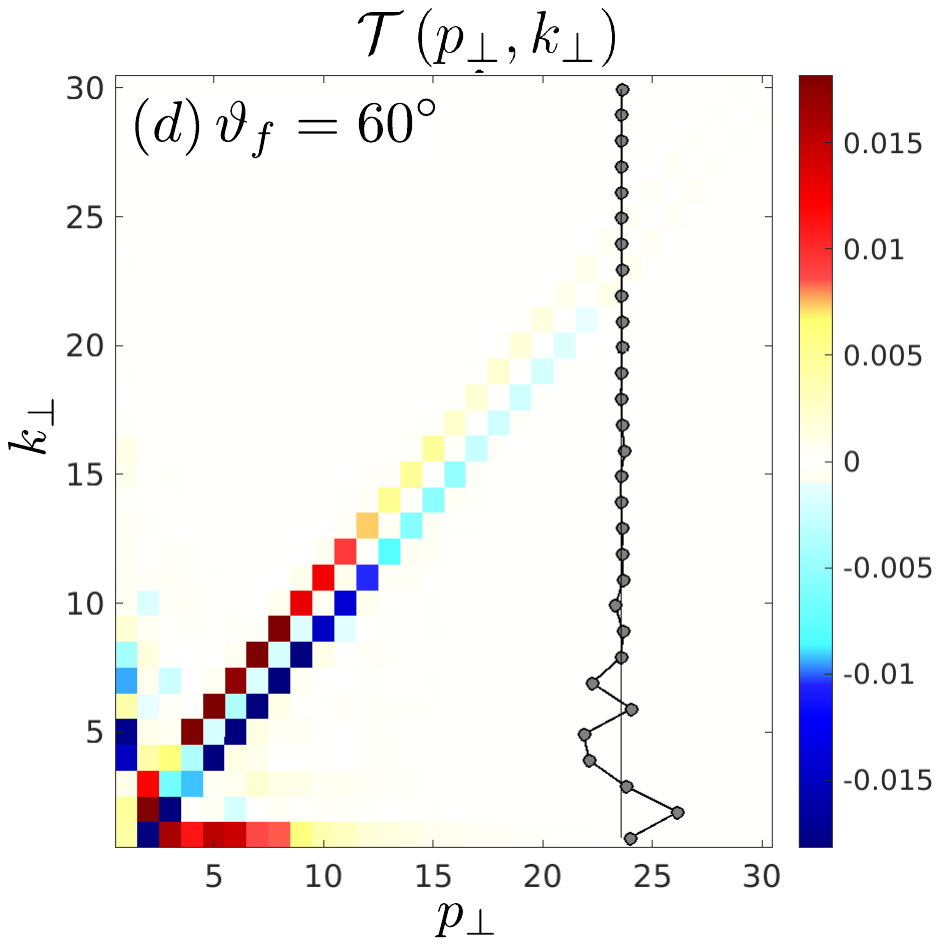}
\caption{
	Barotropic transfer diagram for $\wRa = 120$ and
	$\vartheta_f = \{ 15^{\circ}, 30^{\circ},45^{\circ}, 60^{\circ} \}$.
	The diagonals indicate a local forward cascade of energy from lower
	to higher wavenumbers. The nonlocal inverse energy transfer is
	characterised by the lower left corner of the plot, where a range
	of low wavenumbers, $p_\perp \lesssim 10$, all contribute to energy
	in the box scale $k_{\perp box} = 1$. The line at the right of the
	figure sums each row to produce $\mathcal{T}_{k_\perp}$, the total contribution to
	wavenumber $k_\perp$. Negative values indicate energy taken from
	wavenumber $p_\perp$ and added to wavenumber $k_\perp$. }
\label{fig:bt_transfer_diagram_Ra120}
\end{figure}

% ... FIGURE 14
\begin{figure}
\centering
\includegraphics[width=\linewidth]{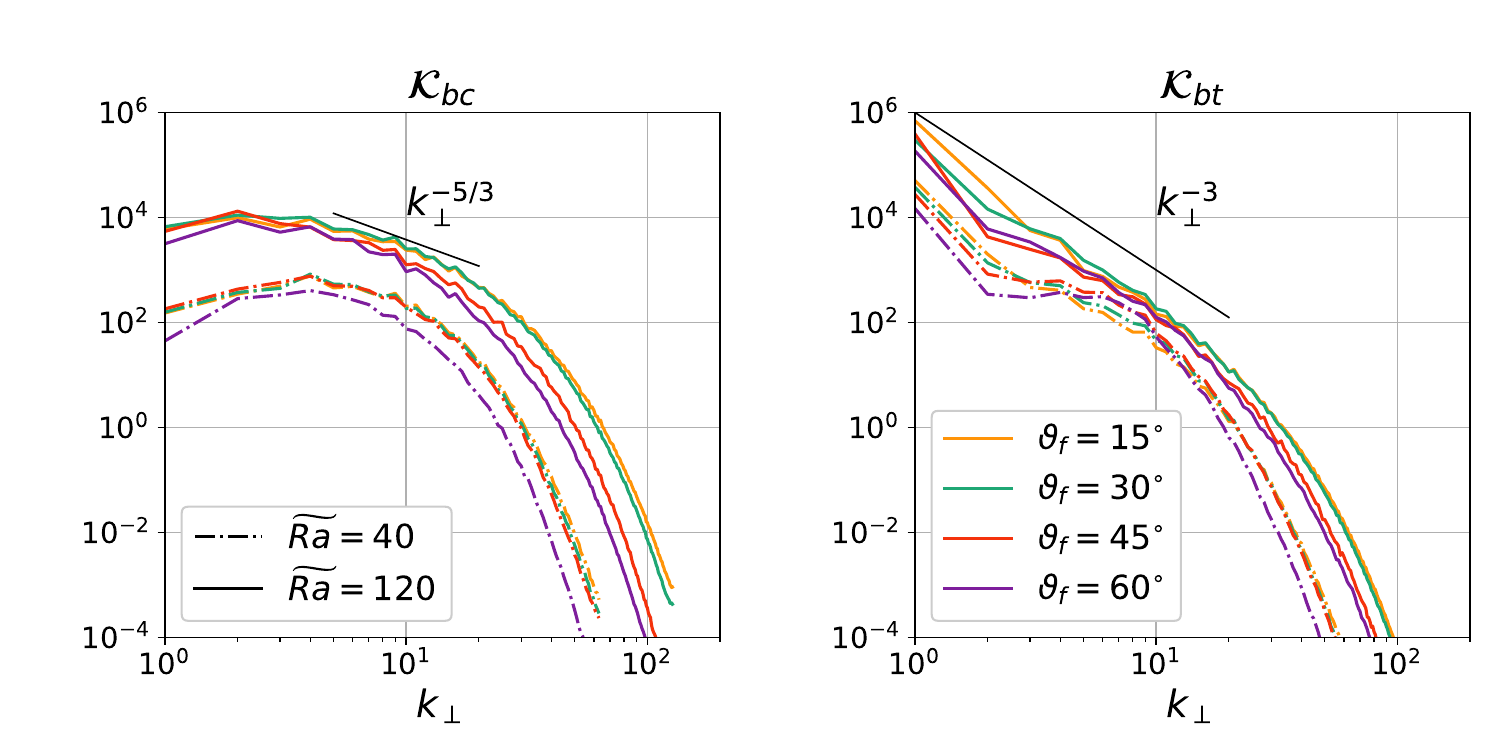}
\caption{
	Baroclinic and barotropic 
	kinetic energy spectra $\mathcal{K}_{bc}(k_\perp)$ (left panel)
	and $\mathcal{K}_{bt}(k_\perp)$ (right panel) for $\widetilde{Ra}=40$ (dash-dotted lines)
	and $\widetilde{Ra}=120$ (solid lines) at different color-coded colatitudes $\vartheta_f$. 
    Primes denote baroclinic components, as defined in eqs.~(\ref{subeqs:baroclinic-barotropic}).
	}
\label{fig:energy_spectra}
\end{figure}

\subsection{Barotropic energy transfer}

Figure~\ref{fig:bt_transfer_diagram_Ra120} shows sample transfer maps at $\wRa=120$ of the barotropic transfer function $\mathcal{T}(p_\perp,k_\perp)$ detailing the transfer of energy from all wavenumbers in a cylindrical shell of magnitude $p_\perp$ to wavenumbers in a shell of magnitude $k_\perp$. The transfer function satisfies the symmetry property $\mathcal{T}(p_\perp, k_\perp) =- \mathcal{T}(k_\perp, p_\perp)$ \citep[for details see][]{aR14}. Thus $\mathcal{T}(p_\perp,k_\perp) >0$ indicates energy deposition from shell $k_\perp$ to $p_\perp$ and vice versa when $\mathcal{T}(p_\perp,k_\perp)<0$. The overall takeaway is that irrespective of the morphology of the condensate, it is found that $\mathcal{T}(p_\perp,k_\perp)$ shares the same characteristics for all colatitudes $\vartheta_f$.  Here, all wavenumber shells are box-normalized with the box scale $L_{box}=10$ such that ${k}_{box}=2\pi/L_{box}\mapsto 1$ implying that the characteristic normalized convective wavenumber is ${k}_{\perp c}\approx10$. As in upright rotating convection the barotropic energy transfer has a strong signal close to the diagonal $p_\perp = k_\perp$ indicating a local transfer of energy and showing that energy is extracted from shells $k_\perp<p_\perp$ ($\mathcal{T} < 0$) and deposited in shells $k_\perp>p_\perp$ ($\mathcal{T} > 0$). This process is indicative of a direct local enstrophy cascade to small scales.
Also observed is a strong off-diagonal signal adjacent to the wavenumber axes indicating a nonlocal transfer of energy from shells $k_\perp < p_\perp$ (specifically $p_\perp\in(4,9)$) and deposited in shell $k_\perp\approx 1$ that corresponds to the box-scale condensate.

The total energy flux $\mathcal{T}_{k_\perp}=\sum_{p_\perp} \mathcal{T}(k_\perp, p_\perp)$ in a given shell $k_\perp$ is presented in the adjacent line plots within each subplot of fig.~\ref{fig:bt_transfer_diagram_Ra120}. This situation is typical and shows that above the characteristic convective instability wavenumber $k_{\perp c}\approx10$ the net energy flux into a wavenumber within the direct cascade is zero. Here, the enstrophy flux is carried downscale until it is dissipated. Below the convective instability wavenumber ($k_{\perp c} \lesssim 10$), the nonlocal inverse energy transfer dominates the direct cascade and a negative flux of energy $\mathcal{T}_{k_\perp}<0$ is observed for $k_\perp\in(3,8)$ and a positive flux for $k_\perp\le 2$. This behaviour reflects the extraction of energy from intermediate scales and its deposition in the large scales.

\subsection{Energy spectra}

Figure~\ref{fig:energy_spectra} shows a typical sample of the barotropic and the midplane baroclinic kinetic energy spectra $\mathcal{K}_{bt}(k_\perp)$ and $\mathcal{K}_{bc}(k_\perp)$ at $\wRa=40$ (dash-dotted lines) and $\wRa=120$ (solid lines).
The barotropic spectrum $\mathcal{K}_{bt}(k_\perp)$ peaks at the domain scale wavenumber (i.e., $k_{\perp box}=1$) and is indicative of an inverse cascade operating on length scales above the convective injection scale. The spectrum exhibits a steep power law scaling closer to $k^{-4}$ than the $k^{-3}$ scaling suggested in earlier work \citep{lSM94,SW99,chertkov2007dynamics,aR14} representing the impact of nonlocal energy transfer into the box-scale condensate and masking the predicted $-5/3$ power law for an inverse energy cascade. Steeper power laws cannot be excluded \citep{vK25}.
Moreover, irrespective of the type of large scale condensate,  $\mathcal{K}_{bt}(k_\perp)$ also reveals an insensitivity to the colatitude $\vartheta_f$. 

In contrast, the baroclinic spectrum $\mathcal{K}_{bc}(k_\perp)$ peaks around the wavenumber of the optimal linear growth rate of N-S convective rolls that scales according to $k_{\perp box}\propto\wRa^{-1/8}$ \citep{sC61,tO23}. In line with the observation from  linear theory that, by comparison, a departure of the alignment of a convective mode from the meridional direction results in the contraction in the range of unstable wavenumbers with increasing $\vartheta_f$ (fig.~\ref{F:Marg}), the total baroclinic power at each $k_\perp$ decreases with $\vartheta_f$. No evidence of a power law indicative of a dissipation-free inertial range is observed in the direct cascade of baroclinic energy \citep[cf.][]{tO23}. Specifically, if the dynamics on small scales remain geostrophically balanced with $Ro_\ell\ll1$, then overcoming the effect of the Taylor-Proudman constraint requires that viscosity remains influential, a conclusion consistent with the recent finding of \cite{tO23}. In summary, the combined barotropic and midplane baroclinic spectra provide evidence for a bi-directional energy cascade, i.e., an inverse cascade of barotropic energy and a direct cascade of baroclinic energy. 

% ... FIGURE 15
\begin{figure}
\centering
\includegraphics[width=\linewidth]{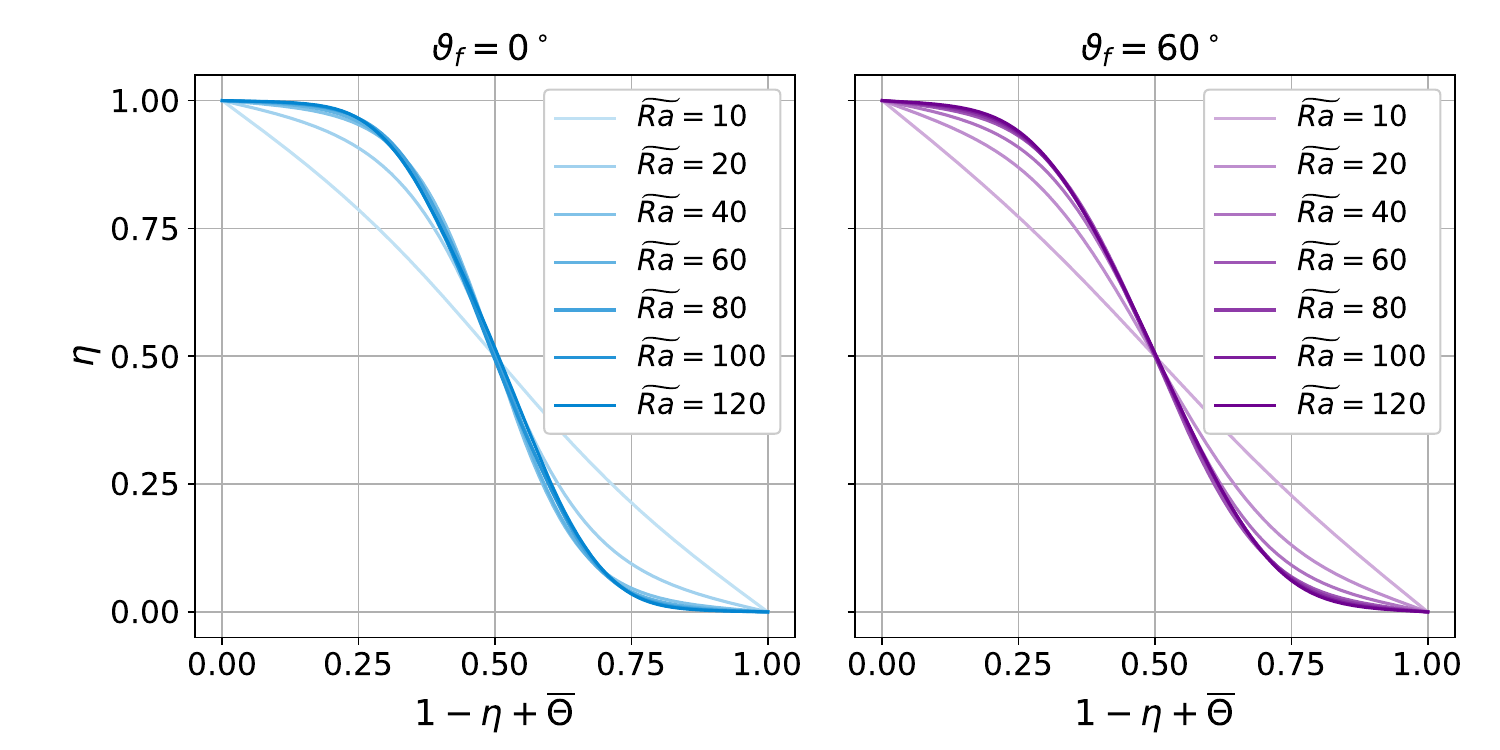}
\caption{
	Time-averaged mean temperature profiles $\overline{\Theta}$ for
	$\vartheta_f = 0^\circ$ (left) and $\vartheta_f = 60^\circ$ (right)
	for different values of  $\wRa$ indicated by color. The midplane
	mean temperature gradient saturates as $\wRa$ increases. 
	}
\label{fig:mean_temperature_vs_ra}
\end{figure}

% ... FIGURE 16
\begin{figure}
\centering
\includegraphics[width=.6\linewidth]{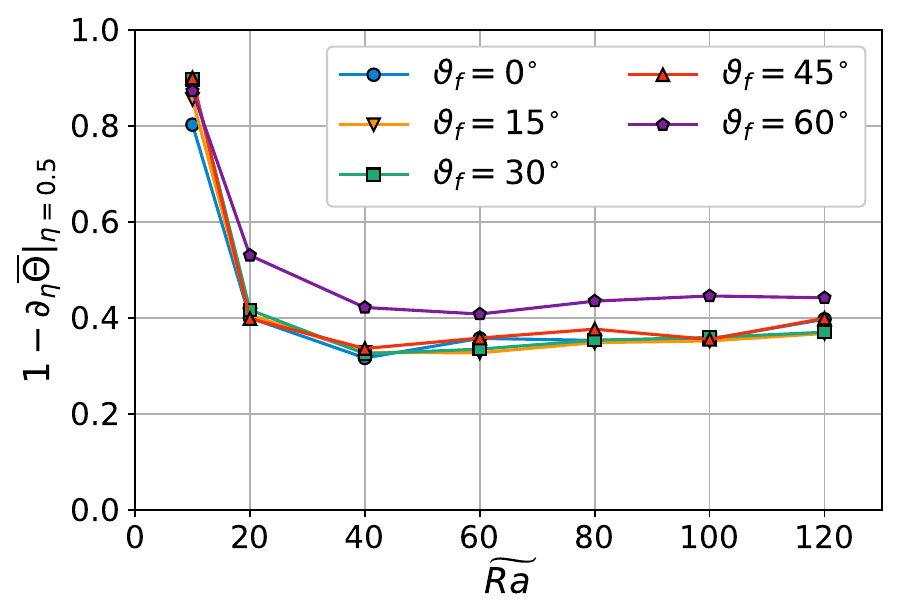}
\caption{
	Time-averaged gradient of the mean temperature $\partial_\eta \left( 1
	- \eta + \overline{\Theta} \right)$ at the midplane $\eta = 0.5$ as a
	function of $\wRa$ for different values of $\vartheta_f$ showing that
	the mean gradient saturates with increasing $\wRa$ at all colatitudes.
}
\label{fig:dz_mean_temperature}
\end{figure}

\subsection{Mean temperature profiles}

In contrast to non-rotating RBC, lateral mixing due to vortical interactions of thermal columns and plumes plays a key role in controlling the behaviour of the
mean temperature gradient in rotating RBC \citep{kJ96}. Coherent columnar structures spanning the
layer depth act as highly efficient conduits for heat transport and models have
demonstrated that the midplane mean temperature gradient follows $\partial_\eta
\overline{\Theta}\vert_{1/2}\propto \wRa^{-1}$ when such structures dominate the flow \citep{mS06,iG10}. This initial evolution towards an isothermal interior as $\wRa$ increases is illustrated in
figs.~\ref{fig:mean_temperature_vs_ra} and \ref{fig:dz_mean_temperature}. Figure~\ref{fig:mean_temperature_vs_ra} shows that the mean temperature profile $\overline{\Theta}$ is symmetric with respect to the midplane $\eta=1/2$ despite the loss of the symmetry $\mathcal{R}_\eta$. Moreover, irrespective of $\vartheta_f$, in the geostrophic turbulent state, $\wRa\gtrsim 40$, buoyant plumes generated within the thermal boundary layers with both cyclonic and anticyclonic signatures stir fluid elements laterally as they attempt to transport heat across the layer. Additionally, the plumes themselves 
are continually eroded and even annihilated by vortical interactions during their
journey across the layer. As a result, in the statistically stationary state,
the thermal deposition by lateral mixing has a tendency to anomalously warm (cool) the ambient temperature in the lower (upper) layers and thereby sustain an unstable mean temperature profile \citep{kJ96}. 
Figure~\ref{fig:dz_mean_temperature} tracks 
the mean temperature gradient $-1 + \partial_\eta \overline{\Theta}$ evaluated at
the midplane as a function of $\wRa$ for various $\vartheta_f$. In all
cases the mean temperature gradient saturates at approximately $-0.4$, a value
evidently insensitive to the colatitude of the $f$-plane. This finding is consistent with recent investigations of upright convection ($\vartheta_f=0$) using DNS~\citep{sS14} or quasi-geostrophic simulations using the reduced equations \citep{mS06,kJ12,sM21,tO23}. It differs, however, from recent $f$-plane simulations at $E=10^{-6}$ \citep{kannan26} in which $Nu$ does not reach the diffusivity-free scaling $Nu\sim{\wRa}^{3/2}$ found here and in spherical shell simulations with a similar Ekman number \citep{wang21}.

\subsection{Global transport} 

\begin{figure}
    \centering
    \includegraphics[width=1.\textwidth]{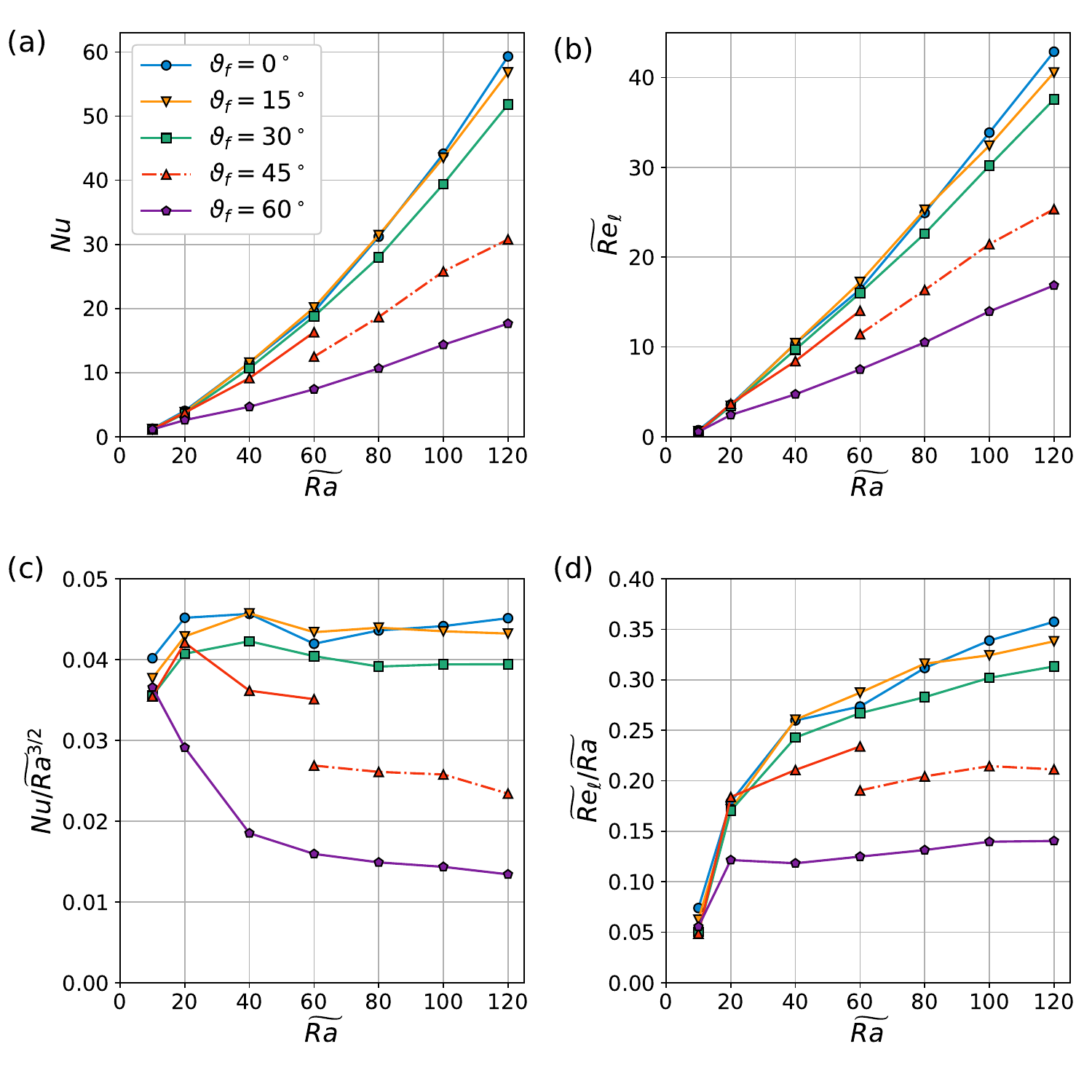}
    \caption{
	    Dependence of (a) the Nusselt number $Nu$ and (b) the Reynolds
	    number $\convectiveReynolds$ defined in equation~(\ref{def:convectiveReynolds})
	    on $\wRa$ for different values
	    of $\vartheta_f$. (c,d) Compensation with the respective dissipation-free
	    scaling laws $\wRa^{3/2}$ and $\wRa$. For $\vartheta_f=45^\circ$,
	    the solid line corresponds to flows dominated by large scale
	    vortex structures while the dash-dotted line corresponds
	    to flows dominated by a zonal jet.
	    }
    \label{fig:nure_vs_rayleigh}
\end{figure}

\begin{figure}
\centering
\includegraphics[width=\linewidth]{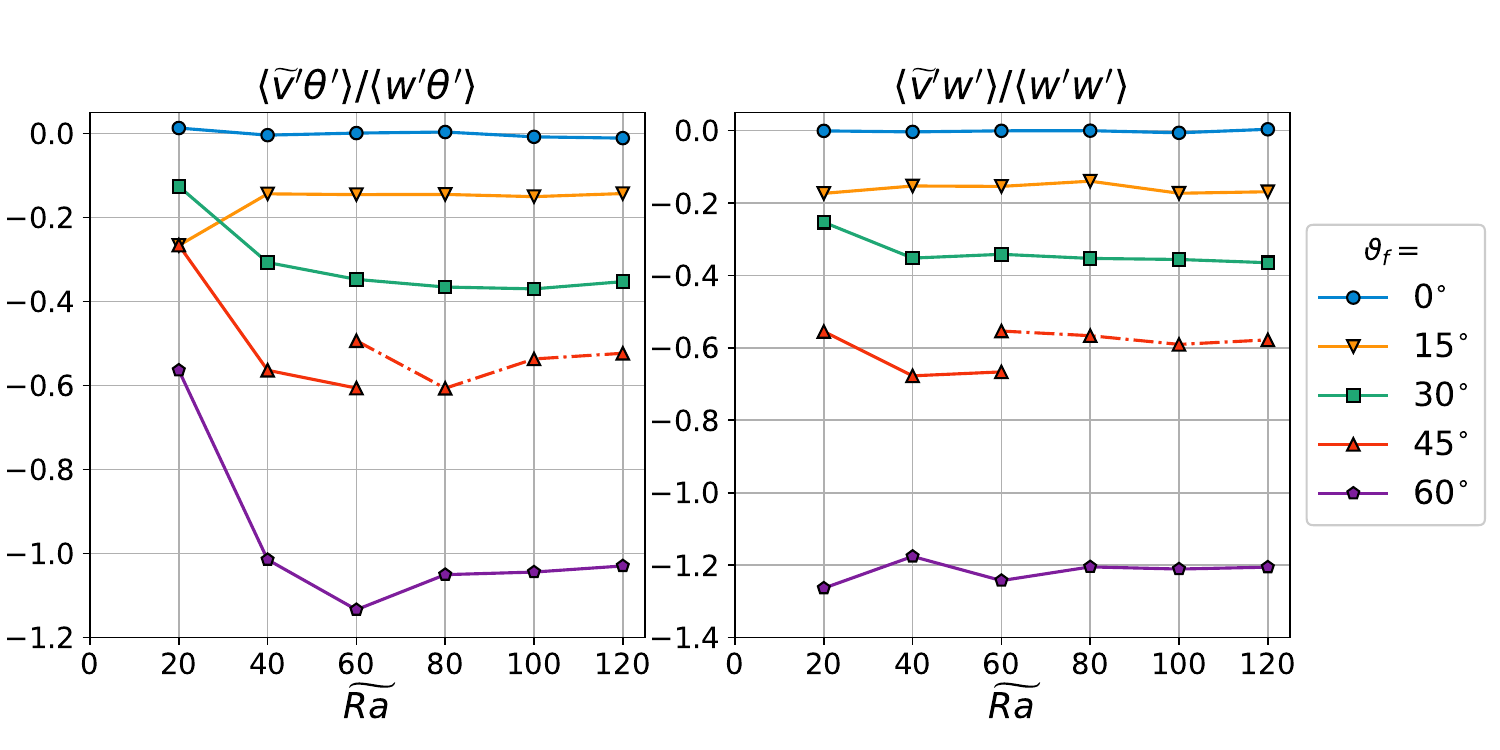}
\includegraphics[width=\linewidth]{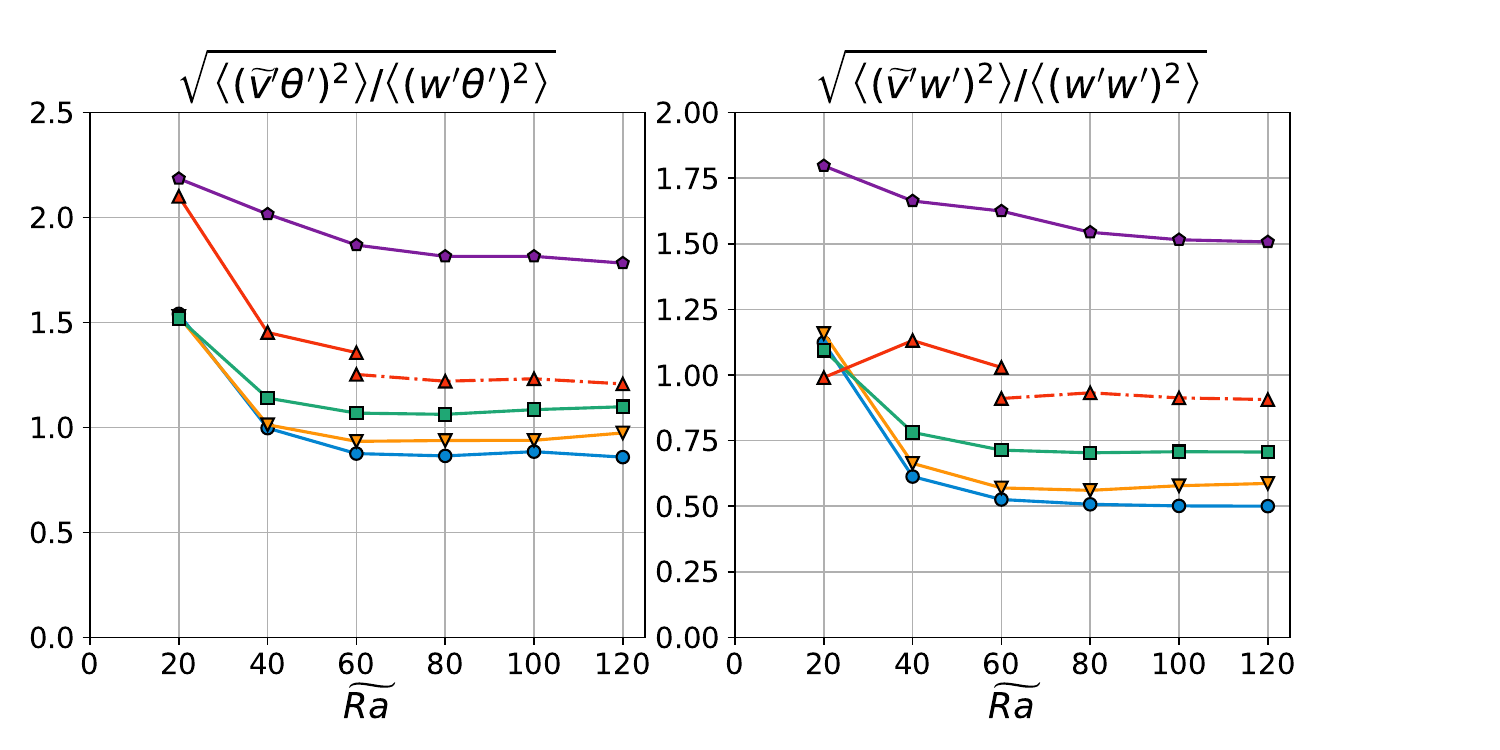}
\caption{
	Ratio of meridional to vertical transport of baroclinic temperature
	(left) and vertical velocity (right) showing monotonic dependence
	on colatitude $\vartheta_f$ as in figs.~\protect\ref{fig:baroclinic_fluxes} and \ref{fig:nure_vs_rayleigh};
	primes denote baroclinic components, as defined in eqs.~(\ref{subeqs:baroclinic-barotropic}).
	Top row: mean values; bottom row: rms values.
	} 
\label{fig:baroclinic_fluxes_v_vs_w}
\end{figure}

Figure~\ref{fig:nure_vs_rayleigh} shows the global heat and momentum transfer via the quantities $Nu$ and $\convectiveReynolds$ shown as functions of $\wRa$ for different values of $\vartheta_f$. 
The Nusselt numbers obtained here for the upright case $\vartheta_f=0$ are similar to those reported elsewhere \citep{vanKan24}, see also \citet{leng26} for their stress-free case with $E=10^{-5}$.
Both $Nu$ and $Re_\ell$ are found to be monotonically increasing functions of $\wRa$ while, at fixed $\wRa$, both are monotonically decreasing functions of $\vartheta_f$. This decrease can be directly associated with the breaking of the
horizontal rotational symmetry $\mathcal{R}_\phi$ that results in (i) increased suppression of active convective modes that are not meridionally aligned and therefore do not participate in heat transport, and (ii) the production of meridional heat and momentum fluxes $\langle \overline{\widetilde{v}\theta} \rangle $ and $\langle \overline{\widetilde{v} w}\rangle$ 
that diminish the available flux for vertical flux transport. This is shown in
fig.~\ref{fig:baroclinic_fluxes_v_vs_w} illustrating the increasing importance of
the meridional flux and its ratio with respect to the vertical flux. 

% ... FIGURE 17

Asymptotic dissipation-free scaling laws for the global heat and momentum transfer
at $\vartheta_f=0^\circ$, as measured by the Nusselt number $Nu$ and the vertical
Reynolds number $\convectiveReynolds=w_{\rm rms}$, have been deduced from the CIA balance
\citep{AHJPRR20} and are given by $Nu\sim \sigma (\wRa/\sigma)^{3/2}$ and
$\convectiveReynolds\sim\wRa/\sigma$. The realization of these scaling exponents is known to
be impacted by the presence of a large scale condensate. Specifically, for
$\vartheta_f=0^\circ$ large exponents are observed due to the presence of a LSV
\citep{sM21,tO23}. To this end we have computed nonlinear regression fits to $Nu$
and $\convectiveReynolds$, assuming power law scaling behaviour; the results of these fits are
listed in Table~\ref{tab:NuReScaling}. The resulting power law exponent for $Nu$
spans a range both greater and smaller than the dissipation-free exponent of $3/2$.
This behaviour is consistent with the observed weak departure from the dissipation-free
scaling law given by $Nu \sim \wRa^{1.56}$ at $\vartheta=0^\circ$ (Table~\ref{tab:NuReScaling}).
At the same time the power law exponent for $\convectiveReynolds$ remains greater than unity,
indicating enhancement of momentum transport by the condensate that is large enough
to offset the attenuation linked to increasing colatitude. Notably, at $\vartheta=0^\circ$,
a strong departure from the dissipation-free scaling law is observed with
$\convectiveReynolds \sim \wRa^{1.33}$ (Table~\ref{tab:NuReScaling}).

% ... FIGURE 18

\begin{table}
\centering
\begin{tabular}{|c|c|c||c|c|}
\hline
\multicolumn{1}{|c}{} & \multicolumn{2}{|c||}{$Nu\sim\alpha \wRa^\beta$} & 
\multicolumn{2}{c |}{$\convectiveReynolds\sim\gamma \wRa^\delta$} \\
\hline
\hline
$\vartheta_f$ & \hphantom{2}$\alpha$ & \hphantom{2}$\beta$ & \hphantom{2}$\gamma$ &  \hphantom{2}$\delta$\\
\hline
\hline
$\hphantom{1}0^\circ$  & 0.033 & 1.56  & 0.071  & 1.33 \\
$15^{\circ}$ & 0.044 & 1.49  & 0.114 & 1.23 \\
$30^{\circ}$ & 0.045 & 1.47  & 0.101 & 1.23 \\
$45^{\circ}$  & 0.056 & 1.32  & 0.098  & 1.16 \\
$60^{\circ}$ & 0.042 & 1.26 & 0.059 & 1.18 \\
\hline
\hline
\end{tabular}
\caption{
Nonlinear regression fits to power laws for heat transport, $Nu\sim \alpha \wRa^\beta$, 
	and momentum transport, $\convectiveReynolds \sim \gamma \wRa^\delta$, 
	computed for $\wRa\geq 60$.
For reference the dissipation-free scaling laws are given by $Nu\sim \wRa^{3/2}$
and $\convectiveReynolds \sim \wRa^1$, respectively. 
	}
\label{tab:NuReScaling}
\end{table}

\section{Conclusion}
\label{sec:conc} 

Reduced equations describing convection on the tilted $f$-plane in the geo- and astrophysically relevant limit of $E_f\ll Ro\ll1$ have been derived and studied via numerical simulations. The misalignment between radial gravitational acceleration and the axis of rotation results in the  propensity for columnar fluid structures of width $\textit{O}(E_f^{1/3} H)$ to align axially with the local rotation vector due to the Taylor-Proudman constraint. This property demands the problem be formulated in a non-orthogonal coordinate system where the upright coordinate aligns with the rotation axis. This approach  circumvents the numerical impact of asymptotically small $\textit{O}(E_f^{1/3} H)$ length scales appearing in the local vertical direction due to the rotational alignment. For geophysical and astrophysical bodies with Ekman number $E_f\in (10^{-12}, 10^{-18})$ this implies a horizontal to vertical scale ratio of $\ell/H\in (10^{-4},10^{-6})$.

A systematic application of asymptotic perturbation theory utilizing $\varepsilon\equiv E_f^{1/3}$ as an expansion parameter resulted in the non-hydrostatic quasi-geostrophic equations on the $f$-plane, the reduced equations~(\ref{eqn:qgf}). These equations hold the additional advantage of filtering out dynamics of secondary importance that, if retained, result in strong spatio-temporal numerical constraints. This includes the filtering of fast inertial waves and Ekman boundary layers that exist on $\textit{O}(E_f^{1/3} H)$ isotropic and $\textit{O}(E_f^{1/2} H)$ vertical scales, respectively. Visualizations of numerical simulations of the quasi-geostrophic equations show a progression in the complexity of the flow morphology, from axially aligned columnar flows to geostrophic turbulence with axially aligned small scale coherent structures, all appearing upright in the virtual non-orthogonal coordinate system. 
Small scale convective motions within the rotating layer also drive a nonlocal upscale energy transfer resulting in the appearance of a domain-scale condensate, either in the form of a large scale vortex dipole or zonal jet. We demonstrated that this occurs owing to the existence of a barotropic manifold devoid of linear thermal (baroclinic) forcing and characterised by \textit{axially} independent vortical dynamics.
 The barotropic dynamics are governed by the axially averaged vertical vorticity equation forced by zonal and meridional vortical fluxes, $\overline{u'\zeta'}$ and $\overline{\tilde{v}'\zeta'}$, meridional buoyancy torques, and damped by anisotropic diffusion, eq.~(\ref{eqn:BVE}). In the absence of forcing and dissipation the manifold conserves area-averaged energy and enstrophy that respectively cascade inversely and directly among spatial scales. The degree of anisotropy in the magnitude of the zonal and meridional vortical fluxes dictates the form of the large scale condensate. This process occurs as a consequence of the breaking of the rotational symmetry $\mathcal{R}_\phi$ in horizontal planes and the reflection symmetry $\mathcal{R}_\eta$ across the midplane that holds in the upright case $\vartheta_f=0$. A large scale vortex is favored for isotropy among the fluxes that occurs for $f$-planes close to the pole ($0^\circ \le \vartheta_f \lesssim 30^\circ$), whereas a zonal jet is favored when the meridional vorticity flux dominates as occurs at large colatitudes ($\vartheta_f \gtrsim 60^\circ$). At intermediate colatitudes ($30^\circ \le \vartheta_f \lesssim 60^\circ$), a newly discovered bistable state with abrupt switches between these two states is observed. This regime shrinks with increasing $\wRa\equiv Ra E_f^{4/3}$.  Similar behaviour is observed in convection with an imposed magnetic field \citep[][]{jN22,mC23} and resembles that in 2D turbulence in anisotropic domains \citep{fB2009, zu23}.

Simulations show that the heat and momentum transport, as measured by the Nusselt number $Nu$ 
and the Reynolds number $\convectiveReynolds = w_{\rm rms}$, both exhibit a power-law scaling $\sim c \wRa^\beta$, with an exponent $\beta$ 
	in a good agreement with the diffusivity-free prediction for geostrophic turbulence $\beta_{GT}=3/2$ in polar regions characterized by small to moderate colatitudes
	$0^\circ \lesssim \vartheta_f \lesssim 30^\circ$.
	This exponent decreases monotonically with increasing $\vartheta_f$, and departs from the $3/2$ value for $\vartheta_f \geq 45 ^\circ$ (figs.~\ref{fig:nure_vs_rayleigh}(a,b)).
The behaviour of the prefactor may be understood as a consequence of rotational alignment of convective
structures that generate a poleward meridional flux transport, i.e., $\overline{\widetilde{v}' \theta}>0$ and $\overline{\widetilde{v}' w}>0$, thereby diminishing the available flux for vertical heat and momentum transport.
This effect is substantially enhanced by the presence of a large scale condensate, be it LSV or jet, as shown in figs.~\ref{fig:nure_vs_rayleigh}(c,d). 
Moreover compared to the dissipation-free Reynolds number scaling exponent $\beta_{Re}=1$, we find
that $\beta_{Re}$ is substantially larger than unity at all small to moderate colatitudes (fig.~\ref{fig:nure_vs_rayleigh}(d)),
an effect that is absent in $\beta_{Nu}$ (fig.~\ref{fig:nure_vs_rayleigh}(c)).
Thus viscosity continues to play an essential role in momentum transport even at substantial $\wRa$.
This effect is not unexpected given that in the absence of bottom friction viscosity provides the
primary mechanism for the saturation of the condensate \citep{jN24,vK25}.

 As with upright rotating RBC, we observe that the mean temperature gradient in the bulk saturates due to lateral mixing thereby sustaining an unstably stratified interior with  $- 1 + \partial_\eta \overline{\Theta}\vert_{1/2} \approx -0.4$. This saturation is a characteristic attribute of rotating RBC \citep{kJ96,mS06,kJ12,sS14}. We remark, however, that these findings are at odds with the studies of \citet{aB14} and \citet{barker2020} who predict a dimensional scaling $(\partial_Z \overline{\Theta})^*\sim \lb F^2 \Omega^4/H^4\rb^{1/5}$ for an applied heat flux $F$. This scaling may be recast non-dimensionally in terms of the flux Rayleigh number $Ra_F \equiv Ra Nu$ as 
$\partial_\eta \overline{\Theta}\vert_{1/2} \propto (\sigma/Ra^3_F E_f^4)^{1/5}$ illustrating 
	a continual approach to an isothermal interior as the thermal forcing increases. This result fundamentally derives from a single-mode linear theory for the optimal growth rate $\sigma_{l}(k_{\perp opt}; Ra)$ whose convective amplitude $v_k$ is determined by a balance with quadratic nonlinearities via the nonlinear growth rate estimate $\sigma_{l} (k_\perp) \equiv v_k k_{\perp opt}$ \citep{dS79}. As in the single-mode theory of \cite{iG10}, applicable to coherent columnar structures, this approach necessarily omits the impact of nonlinear lateral mixing and is therefore unable to capture the ability of $\partial_\eta \overline{\Theta}$ to saturate. We argue, therefore, that the scaling theory suggested in \citet{aB14} and \citet{barker2020} does not apply
    within the strongly forced, geostrophic turbulence regime that is investigated in the present study.

	The local area $f$-plane used here omits several important features captured in extended plane-layer or spherical domains. For instance, the narrow gap approximation may also be relaxed to entertain deeper layers where the effects of compressibility gain importance.  The impact of differential rotation captured through spatial variation of the system rotation or the topographic $\beta$-effect at mid-latitudes (or $\gamma$-effect at the poles) may also be explored through quasi-geostrophic $\beta$- or $\gamma$-convection \citep{kJ06,mC13,Chan13,miquelPRF18}. These effects ultimately constrain the spatial extent of the large scale condensate identified here. It is of interest that a recent DNS investigation of a deep spherical shell has highlighted many similarities with the present plane-layer environment
\citep{gastine_aurnou_2023}. In addition to generalization to more realistic
geometries, future research avenues with the present $f$-plane model would advantageously 
include additional physical ingredients, such as Ekman pumping~\citep{jCKMSV16,troJFM24}, internal heating~\citep{bouillautPNAS2021, 
hadjerciJFM2024} and magnetism~\citep{calkinsJFM15, tobiasJFM2021}, of great relevance to geo- and 
astrophysical flows. Extension of the approach to rapidly rotating spheres and ellipsoids would be welcome but will have to overcome the nonuniformity in the characteristic convective scale as a function of the colatitude $\vartheta_f$, possibly by leveraging promising global mapping methods recently introduced by~\citet{ellisonJCP2022} and \citet{ellisonJCP2023}.

\section*{Acknowledgements}
This work began under the direction of Keith Julien, who passed away on April 14, 2024. The coauthors have attempted to complete the manuscript in line with Keith's high scientific standards. 
KJ was supported by the Division of Mathematical Sciences at the National Science Foundation (NSF) through grant numbers DMS-2009319 and DMS-2308338. 
BM was supported by the Agence Nationale de la Recherche (ANR) through grant number ANR-23-CE30-0016-01.
MAC was supported by the NSF Geophysics Program through grant numbers EAR-1945270 and EAR-2201595. 
EK was supported by NSF through grant numbers DMS-2009563 and DMS-2308337. 
This project was also granted access to computational resources of TGCC under the allocation 2024-A0162A10803 made by GENCI, and to resources of PMCS2I
(P\^ole de Mod\'elisation et de Calcul en Sciences de l’Ing\'enieur de l’Information) of Ecole Centrale de Lyon.
Volume renderings were produced using the software Vapor~\citep{vapor1, vapor2}.
\section*{Declaration of Interests}
The authors report no conflict of interest.

\section*{Author ORCIDs}
{\noindent
B. Miquel, https://orcid.org/0000-0001-6283-0382;\\ 
M.A. Calkins, https://orcid.org/0000-0002-2830-5661;\\
K. Julien, https://orcid.org/0000-0002-4409-7022;\\
E. Knobloch, https://orcid.org/0000-0002-1567-9314.}
\section*{CC-BY license}
This research was funded, in whole or in part, by Agence Nationale de la Recherche (Grant ANR-23-CE30-0016-01). A CC-BY public copyright license has been applied by the authors to the present document and will be applied to all subsequent versions up to the Author Accepted Manuscript arising from this submission, in accordance with the grant’s open access conditions.

\appendix

% APPENDIX A
\section{Derivation of the Reduced $f$NHQG Equations}
\label{Appdx:A}
The study of \citet{kJ06} suggests a multiple scales asymptotic approach 
in both space and time for reducing the iNSE to the quasi-geostrophic PDE
system (\ref{eqn:qgf}) valid in the limit $\varepsilon\rightarrow 0$. 
Here, we show that the same multiple scales strategy goes through provided all fluid variables are expressed as functions of non-orthogonal coordinates with iso-surfaces advantageously 
aligned with both rotation and the fluid layer's bounding surfaces.
This appendix is dedicated to a more detailed derivation of the system (\ref{eqn:qgf})
than the broad sketch provided in Section~\ref{sec:prelim}.

\begin{figure}
\centering
	\includegraphics[width=0.47\linewidth]{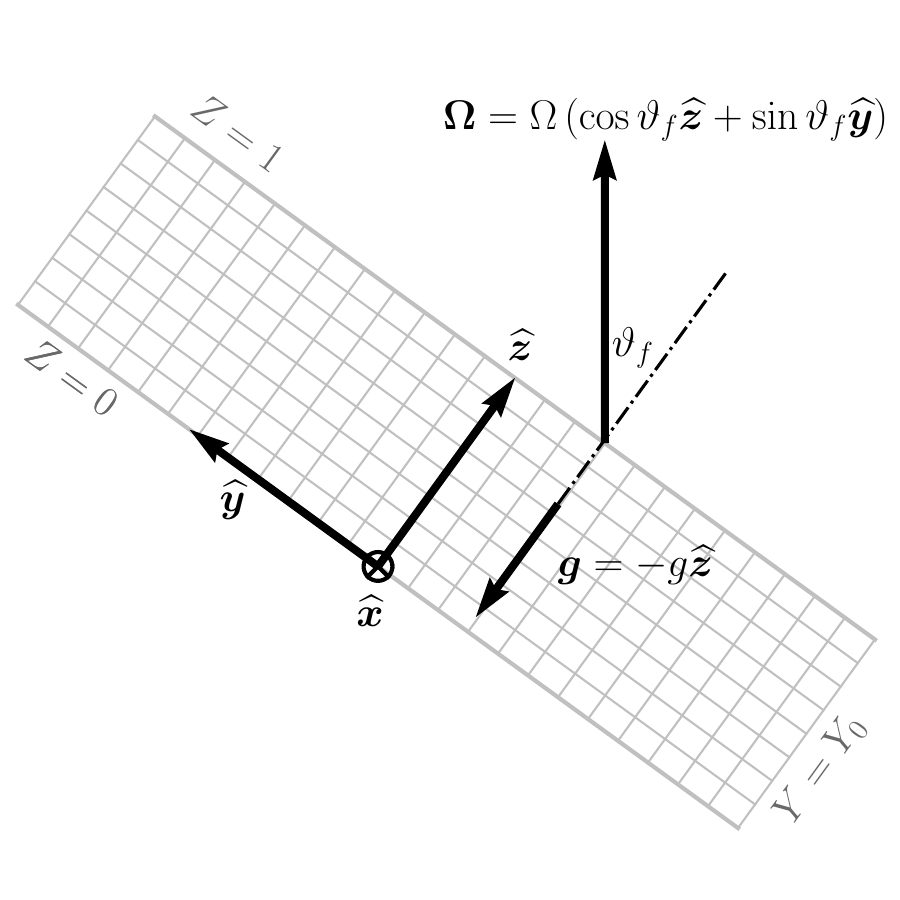}
	\includegraphics[width=0.47\linewidth]{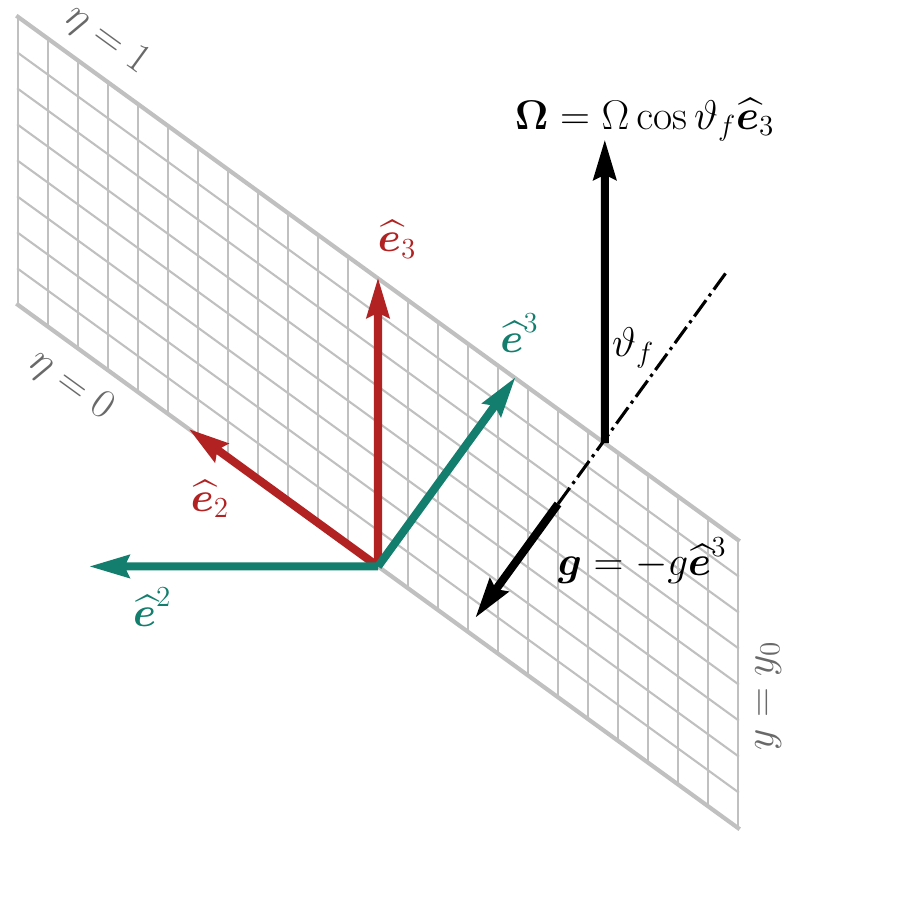}
	\caption{
		Left panel: Cartesian coordinates and basis vectors. Right panel: non-orthogonal coordinates and covariant ($\baseCov{i}$) and contravariant ($\baseContra{i}$) basis vectors. }
	\label{appFig:geometry}
\end{figure}

Our starting point is a flat layer of fluid of constant depth $H$,
subjected to locally vertical gravity $\boldsymbol{g}=-g\haz$ (i.e. normal to the bounding surfaces)
and rotating at a rate $\Omega$ about an axis at an angle $\vartheta_f$ relative to gravity (see fig.~\ref{appFig:geometry}).
The preference for rotationally aligned dynamics through the Taylor-Proudman
constraint suggests the introduction of non-orthogonal coordinates, obtained by shearing
the Cartesian coordinates $(X,Y,Z)$ such that the position vector $\boldsymbol{r}$
is written
\begin{equation}
\boldsymbol{r} = X\, \hx + Y\, \hy + Z\, \haz\,. 
\end{equation}
The sheared coordinates, denoted $(x,y,\eta)$ in the main text, are defined by:
\begin{subequations}
\label{app_eq:sheared_coordinates}
\begin{align}
	X&=x\\
	Y&=y+\gamma \eta\\
	Z&= \eta
\end{align}
\end{subequations}
with  $\gamma \equiv \tan\vartheta_f$. In the context of this appendix, we rely on the conciseness of tensorial
notation for our derivation. Thus, coordinates are denoted
\begin{equation}
\left( \coords^1, \coords^2, \coords^3\right) = \left(x,y,\eta\right),
\end{equation}
where the superscript consistently refers to contravariant components when employed with a vector, and should not be confused with exponentiation. 
In addition, we use 
Einstein's implied summation convention over indices that appear exactly one time as a superscript and one time as a subscript in a given expression. With this notation,
spatial derivatives are written $\partial_i=\partial/\partial \coords^i$ ($1\leq i \leq3$).
The covariant base vectors $(\baseCov{i})_{1\leq i \leq 3}$ are
defined by $\baseCov{i} = \partial_i \boldsymbol{r}$, such that:
\begin{subequations}
\label{app_eq:covariant_basis}
\begin{align}
\boldsymbol{\widehat{e}}_1& = \frac{\partial \boldsymbol{r}} {\partial x} = \hx\,,\\
\boldsymbol{\widehat{e}}_2& = \frac{\partial \boldsymbol{r}} {\partial y} = \hy\,,\\
\boldsymbol{\widehat{e}}_3& = \frac{\partial \boldsymbol{r}} {\partial \eta} = \haz + \gamma \hy \, .
\end{align}
\end{subequations}
The dual, contravariant base vectors $(\boldsymbol{\widehat{e}}^i)_{1\leq i \leq 3}$ 
are defined by the orthogonality property:
\begin{equation}
\baseCov{i} \boldDot \baseContra{j} = \delta_{ij} ,
\end{equation}
where $\delta$ is the Kronecker delta symbol. It follows that
\begin{equation}
\label{app_eq:contravariant_basis}
\begin{pmatrix}
\boldsymbol{\widehat{e}}^1 \\ 
\boldsymbol{\widehat{e}}^2 \\ 
\boldsymbol{\widehat{e}}^3
\end{pmatrix}
	= \begin{pmatrix}
		\hx \\ \hy - \gamma \haz \\ \haz
	\end{pmatrix} .
\end{equation}
Projections between the covariant and the contravariant representations are
readily computed with the completely covariant $\mathsfbi{G}_{\bullet\bullet}$
and completely contravariant $\mathsfbi{G}^{\bullet\bullet}$ metric tensors:
\begin{subequations}
\begin{align}
\mathsfbi{G}_{\bullet\bullet}& = \left( G_{ij} \right)_{1\leq i,j \leq 3} 
			       = \left( \baseCov{i} \boldDot \baseCov{j}  \right)_{1\leq i,j \leq 3} = 
\begin{pmatrix}
	1 & 0 & 0 \\ 
	0 & 1 & \tan \vartheta_f \\ 
	0 & \tan \vartheta_f & \frac{1}{\cos^2 \vartheta_f}
\end{pmatrix}\, , \\
	\mathsfbi{G}^{\bullet\bullet}& = \left( G^{ij} \right)_{1\leq i,j \leq 3} 
			       = \left( \baseContra{i} \boldDot \baseContra{j}  \right)_{1\leq i,j \leq 3} = 
\begin{pmatrix}
	1 & 0 & 0 \\ 
	0 & \frac{1}{\cos^2 \vartheta_f} & -\tan \vartheta_f \\ 
	0 & - \tan \vartheta_f & 1
\end{pmatrix}\, .
\end{align}
\end{subequations}
The conversion between covariant and contravariant representations of a vector, 
$\boldsymbol{u} = u_i \baseContra{i} = u^i \baseCov{i}$,
is handily captured by contraction with the appropriate metric tensor:
\begin{equation}
	u_i = G_{ij} u^j, \quad \text{and} \quad u^i = G^{ij} u_j \,.
\end{equation}
This gives rise to the following particularly useful representations of the gradient operator
\begin{subequations}
\begin{align}
	\nabla& = \nabla^{\bullet} =\baseContra{i}\partial_i =  \hx \pd{x} +
	\baseContra{2} \partial_{y} +  \haz \partial_{\eta}\,, \\
	&= \nabla_{\bullet} = G^{ij} \baseCov{j}\partial_i = \hx \pd{x} + 
	\hy \left( \frac{1}{\eta^2_3} \partial_{y} -\gamma \pd{\eta} \right) +  
	\baseCov{3} \left( \pd{\eta}-\gamma \pd{y}\right)\, ,
\end{align}
\end{subequations}
where superscripted $\nabla^{\bullet}$ and subscripted $\nabla_{\bullet}$ gradient operators denote the contravariant and covariant representations, respectively.
The scalar Laplace operator is given by:
\begin{equation}
	\label{app_eq:laplace_operator}
	\Delta = G^{ij} \partial_i \partial_j 
	 = \partial_{1}\partial_{1} + \frac{1}{\cos^2 \vartheta_f} \partial_{2}\partial_{2}  - 2 \tan\vartheta_f \partial_{2}\partial_3 
	+ \partial_{3}\partial_{3}\, .
\end{equation}
% FIGURE A1

In the Boussinesq limit, the dimensional incompressible Navier-Stokes equations 
governing the flow in a plane layer delimited by $0\leq \eta \leq H$ are:
\begin{subequations}
\begin{gather}
	\partial_t \boldsymbol{u}  + \boldsymbol{u \cdot \nabla u} 
	+ 2\Omega \cos \tangle \baseCov{3} \times \boldsymbol{u}
	= - \frac{\nabla p}{\rho_0} + \alpha g \temperature \baseContra{3}
	+ \nu \Delta \boldsymbol{u}\,, \\
	\partial_t \temperature  + \boldsymbol{u \cdot \nabla} \temperature 
	= \kappa \Delta \temperature\,, \\
	\divergence \boldsymbol{u} = 0 \, .
\end{gather}
\end{subequations}
Central to the nondimensionalisation of these equations is the intrinsic rotational scale, readily expressed by introducing the small parameter
\begin{equation}
	\varepsilon = \ekmanf^{1/3} \equiv \left( \frac{\nu}{2H^2 \Omega \cos \vartheta_f} \right)^{1/3}\,.
\end{equation}
Accordingly, lengths are measured in units of the characteristic horizontal scale $\ell=\varepsilon H$,
so that the fluid domain becomes $0 \leq \eta \leq \varepsilon^{-1}$.
Time and velocity are measured in units of the associated diffusive time  $\ell^2/\nu$ 
and characteristic velocity $\nu/\ell$. Finally, temperature is measured in units of the temperature difference $\Delta_T$ across the layer, and decomposed into a vertical profile $\Theta(\eta)$
and asymptotically small fluctuations in the horizontal direction $\varepsilon \theta(\boldsymbol{r})$.
The momentum equation becomes:
\begin{subequations}
\begin{gather}
	\partial_t \boldsymbol{u}  + \boldsymbol{u \cdot \nabla u} 
	+ \varepsilon^{-1} \baseCov{3} \times \boldsymbol{u}
	= -\varepsilon^{-1} \nabla p + \frac{\widetilde{Ra}}{\sigma}  \theta\, \baseContra{3} + \Delta \boldsymbol{u}\,, \\
	\divergence \boldsymbol{u} = 0\,.
\end{gather}
\end{subequations}
After projection onto the covariant basis $\left(\baseCov{1}, \baseCov{2}, \baseCov{3}\right)$ these equations become:
\begin{subequations}
	\begin{align}
	\partial_t u_1 + \left( \boldsymbol{u \cdot \nabla u} \right)_1 
		- \varepsilon^{-1} u^2& = -\varepsilon^{-1} \partial_1 p + \Delta u_1\,, \\
	\partial_t u_2 + \left( \boldsymbol{u \cdot \nabla u} \right)_2 
		+ \varepsilon^{-1} u^1& = -\varepsilon^{-1}\partial_2 p + \Delta u_2\,,\\
	\partial_t u_3 + \left( \boldsymbol{u \cdot \nabla u} \right)_3 
		& = -\varepsilon^{-1} \partial_3 p + \frac{\widetilde{Ra}}{\sigma} \theta + \Delta u_3\,,\\
		0 & = \partial_i u^i \,.
	\end{align}
\end{subequations}
\subsection{Asymptotic expansion}
Each variable is expressed as a power series in $\varepsilon$, with
the expansion order written in parentheses to avoid 
confusion with tensorial indices:
\begin{subequations}
\begin{align}
	\boldsymbol{u}& =  \boldsymbol{u}_{(0)} + 
						 \boldsymbol{u}_{(1)} \varepsilon + 
						 \boldsymbol{u}_{(2)} \varepsilon^2 + \dots
					 \\
	p & = p_{(0)} + 
	                           p_{(1)}\varepsilon + 
				   p_{(2)}\varepsilon^2 + \dots \\
	\theta & =  \theta_{(0)}	  + \theta_{(1)} \varepsilon
								  + \theta_{(2)} \varepsilon^2
								  + \dots .
\end{align}
\end{subequations}
In the axial direction, in addition to rapid variations in $\xi^3$ that correspond 
to the fast scale $O(\varepsilon H)$, we introduce a slow scale $\chi^3 = \varepsilon \xi^3$ 
such that 
\begin{equation}
	\label{app_eq:multiple_scales}
	\partial_3 \mapsto \partial_3 + \varepsilon D_3, \quad \text{with} \quad  D_3=\frac{\partial}{\partial \chi^3} .
\end{equation}

\subsection{Leading order geostrophy}
At leading order, the momentum equation is dominated by the geostrophic balance between
the small scale pressure gradient and the Coriolis force that can be written formally as $\mathcal{L}_{G} \left(\boldsymbol{u}_{(0)}, p_{(0)} \right)=0$, 
or more explicitly:
\begin{subequations}
	\label{app_eq:leading_order_geostrophy}
\begin{align}
	-  u^2_{(0)} +\partial_1  p_{(0)}&=0\,, \\
	   u^1_{(0)} +\partial_2  p_{(0)}&=0\,, \\
		\partial_3  p_{(0)} &=0 \,, \\
	\partial_1  u^1_{(0)}
					    + \partial_2 u^2_{(0)}
					    + \partial_3 u^3_{(0)} &=0 \,. 
\end{align}
\end{subequations}
As for upright rotating convection, pressure plays the role of the streamfunction 
and we write $p_{(0)}=\Psi$:
\begin{equation}
	\boldsymbol{u}_{(0)} 
	= - \curl \left( \Psi \baseContra{3} \right) + u^3_{(0)} \baseCov{3}
= -\partial_2\Psi \baseCov{1} + \partial_1\Psi \baseCov{2} + u^3_{(0)} \baseCov{3}
\end{equation}
or, in the contravariant representation: 
\begin{equation}
 \boldsymbol{u}_{(0)} = -\partial_2 \Psi \baseContra{1} 
	+ \left( \partial_1 \Psi + \tan \tangle u^3_{(0)}\right) \baseContra{2}
	+ \left( \tan\tangle \partial_1 \Psi + \frac{1}{\cos^2\tangle} u^3_{(0)}\right) \baseContra{3} .
\end{equation}
Second, along the rotation axis $\baseCov{3}$, variations 
with the fast scale $\coords^3$ are prohibited by the Taylor-Proudman theorem and, at 
leading order, variables can only evolve on the ``slow'' scale $\chi^3$:
\begin{equation}
	\partial_3  \Psi = 0,\quad 
	\partial_3  u^3_{(0)} = 0 .
\end{equation}
It follows that the Laplace operator, given in equation~(\ref{app_eq:laplace_operator}), becomes:
\begin{equation}
	\label{app_eq:laplace_operator_rescaled}
	\Delta =  \partial_{1}\partial_{1} + \frac{1}{\cos^2 \vartheta_f} \partial_{2}\partial_{2}  - 2 \varepsilon\tan\vartheta_f \partial_{2}D_3 
	+\varepsilon^2 D_{3}D_{3}
\end{equation}
and is thus dominated by its horizontal component:
\begin{equation}
	\label{app_eq:laplace_operator_horizontal}
	\Delta =  \nabla_\perp^{\prime 2} + O(\varepsilon),\quad \text{with} \quad \nabla_\perp^{\prime 2} = \partial_{1}\partial_{1} + \frac{1}{\cos^2 \vartheta_f} \partial_{2}\partial_{2} \,,
\end{equation}
where the prime emphasizes derivatives on fast scales, as in the main text. 

We close this paragraph by observing that the axial vorticity---defined as the third covariant component of the curl of velocity, as in the main text---is readily obtained from the streamfunction as
\begin{equation}
	\zeta \equiv \baseCov{3} \bdot \curl\, \ub = \nabla_\perp^{\prime 2} \Psi\,.
	\label{AppEq:covariant_axial_vorticity}
\end{equation}
By contrast, the (hatted) isotropic operator $\widehat{\nabla}_\perp^{\prime 2} \equiv \partial_{11} + \partial_{22}$ enters the definition of the third contravariant component of vorticity
\begin{equation}
	\omega^3 \equiv \baseContra{3} \bdot \curl\, \ub = \widehat{\nabla}_\perp^{\prime 2} \Psi\,.
\label{AppEq:contravariant_axial_vorticity}
\end{equation}

\subsection{Solvability condition}
Governing equations for the dynamics of $(\Psi,u^3_{(0)})$ are obtained at the next order, and require 
the computation of advection by the flow, which we present now.
Advection by the flow is dominated by its horizontal component
\begin{equation}
	\label{app_eq:leading_horizontal_velocity}
	\boldsymbol{u}_\perp = \boldsymbol{u} - \left( \contra{3} \boldsymbol{\cdot u} \right) \cov{3} = \boldsymbol{u} - u^3 \cov{3} = u^1 \cov{1} + u^2 \cov{2}\,,
\end{equation}
and is written
\begin{subequations}
\begin{align}
	\boldsymbol{Q} =& \boldsymbol{u\cdot \nabla u}  \\
		       =& \boldsymbol{u}_{\perp,(0)} \boldsymbol{\cdot \nabla u}_{(0)} + O(\varepsilon)  \\
		       =& \left( u^1_{(0)} \partial_1  + u^2_{(0)} \partial_2 \right) u^k_{(0)} \cov{k} + O(\varepsilon)\\
		       =& J\left[\Psi, u^k_{(0)}\right]\baseCov{k} + O(\varepsilon) \\
		       =& J\left[\Psi, u_{k,(0)}\right]\baseContra{k} + O(\varepsilon) .
\end{align}
\end{subequations}
At the next order, we compute the projection onto the covariant basis $\cov{i}$:
\begin{equation}
	\label{app_eq:next_order}
	\mathcal{L}_{G}\left(\boldsymbol{u}_{(1)}, p_{(1)}\right) =\begin{pmatrix}
	-\partial_t u_{1,(0)} - Q_{1,(0)}+ \horizResLap u_{1,(0)} \\
	-\partial_t u_{2,(0)} - Q_{2,(0)} + \horizResLap u_{2,(0)} \\
	-\partial_t u_{3,(0)} - Q_{3,(0)} + \frac{\wRa}{\sigma} \theta + \horizResLap u_{3,(0)} \\
					    - D_3  u^3_{(0)}
	\end{pmatrix}.
\end{equation}
The associated solvability condition is obtained 
by demanding that the right hand side $\left( r_1, r_2, r_3, r_4\right)$ of
equation~(\ref{app_eq:next_order}) is orthogonal to the kernel of the adjoint operator $\mathcal{L}_G^\dagger = - \mathcal{L}_G$:
\begin{equation}
	\forall P^*, W^{*}:\quad \int d\mathcal{V} \left( -\partial_2'P^* r_1 + \partial_1' P^* r_2 + W^{*} r_3 + P^*r_4 \right)=0,
\end{equation}
yielding
\begin{multline}
	\forall P^*, W^*:\quad \int d\mathcal{V} \Bigg(
-\partial_2' P^* \left[
		-\partial_t u_{1,(0)} -Q_{1,(0)}+ \horizResLap u_{1,(0)}
	\right]  \\
	+ \partial_1' P^* \left[
	-\partial_t u_{2,(0)} - Q_{2,(0)} + \horizResLap u_{2,(0)} 
	\right] \\
        + W^* \left[
		 - \partial_3 \Psi
	-\partial_t u_{3,(0)} - Q_{3,(0)} + \frac{\rayleigh}{\sigma} \theta + \horizResLap u_{3,(0)} 
	\right] \\
 - P^* \partial_3 u^3_{(0)} 
\Bigg) = 0.\label{solve}
\end{multline}

\subsection{Governing equation for the axial velocity}
The expression above must hold for $P^*=0$, in which case the third line naturally 
yields a governing equation for the third contravariant velocity component:
 $u_{3,(0)} = \baseCov{3} \boldsymbol{\cdot u}_{(0)}$
\begin{equation}
	\label{eq:Usub3}
	\partial_t u_{3,(0)} + J \left[ \Psi, u_{3,(0)} \right]+ D_3 \Psi =
	\frac{\wRa}{\sigma} \theta + \horizResLap u_{3,(0)}\,.
\end{equation}
In the main text, we write $u_{3,(0)}$ as $\axialVelocity$ for conciseness. Equation (\ref{eq:Usub3}) 
is thus the governing equation for axial velocity (\ref{eqn:qgfW}) in the $f$NHQGE set.

\subsection{Governing equation for the axial vorticity}
Integration of \eqref{solve} by parts yields:
\begin{equation}\label{eq:Psi}
	\partial_t\left( \partial_2 u_1 - \partial_1 u_2 \right) + \partial_2 Q_1 - \partial_1 Q_2 + D_3 u^3 = \horizResLap \left( \partial_2 u_1 - \partial_1 u_2\right).
\end{equation}
Expressing the velocity components using the geostrophic pressure, one obtains:
\begin{multline}
	\label{AppEq:axial_velocity_with_u3}
	\partial_t\left( - \partial_{22} \Psi - \partial_{11} \Psi - \partial_1 \tan \tangle u^3 \right) - \partial_2 J[\Psi, \partial_2\Psi] - \partial_1 J [\Psi, \partial_1 \Psi ] + D_3 u^3 \\ = \horizResLap \left( - \partial_{22} \Psi - \partial_{11} \Psi - \partial_1 \tan \tangle u^3 \right) .
\end{multline}
To eliminate $u^3$, we compute $\cos^2 \tangle \tan\tangle$ times (\ref{eq:Usub3}):
\begin{multline}\label{eq:cos2tanUsub3}
	\partial_t \left( u^3 \tan\tangle + \sin^2 \tangle \partial_1\Psi\right)+ J[\Psi, \tan\tangle u^3 + \sin^2 \tangle \partial_1\Psi] + \cos^2 \tangle \tan \tangle D_3\Psi \\ = \cos^2\tangle \tan \tangle \frac{\wRa}{\sigma} \theta
	+\horizResLap \left( u^3 \tan\tangle + \sin^2 \tangle \partial_1\Psi\right)
\end{multline}
and observe that:
\begin{equation}
	\partial_2Q_1 - \partial_1Q_2 + \cos^2 \tangle \tan \tangle \partial_1 Q_3 = - \cos^2 \tangle J[\Psi, \horizResLap \Psi].
\end{equation}
Computing $\left[ (\ref{eq:Psi})+\partial_1 (\ref{eq:cos2tanUsub3})\right]/\cos^2\tangle $ now yields equation~(\ref{eqn:qgfPSI}), the governing 
equation for axial vorticity:
\begin{equation}
	-\partial_t \horizResLap \Psi 
	-J[\Psi,\horizResLap \Psi] +D_3 \axialVelocity  
	= \tan\tangle \frac{\wRa}{\sigma} \partial_1 \theta - \nabla_\perp^{\prime 4} \Psi.
\end{equation}
The system is closed using the governing equations for the mean and fluctuating temperature. At leading order, one shows using incompressibility that temperature advection is 
\begin{multline}
	\boldsymbol{u \cdot \nabla} \left( -\chi^3 + \overline{\Theta}(\chi^3) + \varepsilon \theta(\boldsymbol{r})\right)
	=  \varepsilon u^3_{(0)}  \left( D_3 \overline{\Theta} -1 \right) \\
	+ \varepsilon \partial_1 \left( u^1_{(0)} \theta \right) + \varepsilon \partial_2 \left( u^2_{(0)}\theta\right) 
	+ \varepsilon^2 D_3\left(u^3_{(0)} \theta\right)\,,
\end{multline}
leading to the governing equations for temperature
\begin{gather}
	\varepsilon^{-2}\partial_t \overline{\Theta} + D_3 \overline{u^3_{(0)} \theta} = \frac{1}{\sigma}D_{33}\overline{\Theta}\,,\\
	\partial_t \theta + \boldsymbol{\nabla}_\perp \boldDot \left( \boldsymbol{u}_\perp \theta \right) + u^3_{(0)} \left( D_3 \overline{\Theta} -1 \right) = \frac{1}{\sigma}\horizResLap \theta.
\end{gather}

\section{Boundary Conditions}
\label{Appdx:boundaryConditions}

The impenetrability condition on top and bottom boundaries $\xi^3=0,1$ is naturally expressed using the contravariant vertical velocity
	\begin{equation} 
		u^3=\ub \bdot \baseContra{3} = 0\,,
		\label{AppEq:contravariant_impenetrability}
\end{equation} while 
the covariant velocity
\begin{equation}
\label{AppEq:covariant_impenetrability}
	u_3 = \ub \bdot \baseCov{3} = \gamma \partial_1 \Psi\,.
\end{equation}
Consequently, the temperature equation becomes
\begin{equation}
	 \partial_t \theta + J[\Psi, \theta] = \frac{1}{\sigma}   \nabla^{\prime 2}_\perp  \theta\,,
\end{equation}
yielding the variance equation 
\begin{equation}
		 \frac{1}{2} \partial_t\overline{ \theta^2} = - \frac{1}{\sigma}\overline{\left\| \nabla^\prime_\perp \theta \right\|^2} \,,
\end{equation}
proving that horizontal temperature fluctuations are smoothed out along the bounding surfaces at long times:
\begin{equation}
	\lim_{t\to\infty} \left.\overline{  \theta^{2}}\right|_{\xi^3=0,1} = 0\,.
\end{equation}
Hence, upon setting $\theta=0$, enforcing impenetrability (\ref{AppEq:covariant_impenetrability}), and reordering terms,
the $f$NHQG equations (\ref{eq:Usub3},\ref{AppEq:axial_velocity_with_u3}) become:
\begin{subequations}
\begin{align} 
	\label{AppEq:D3Psi}
	D_3\Psi&=-\gamma\left(\partial_t \partial_1\Psi 
	+ J[ \Psi,  \partial_1\Psi ]
	- \nabla^{\prime 2}_\perp \partial_1 \Psi\right)\,,\\
	\label{AppEq:D3U3}
	D_3U^3&= \partial_t \omega^3 + J[ \Psi, \omega^3] - \nabla^{\prime 2}_\perp \omega^3\,,
\end{align}    
\end{subequations}
where we recall that $\omega^3= \left(\partial_{11}+\partial_{22}\right) \Psi$, as in (\ref{AppEq:contravariant_axial_vorticity}).
For upright rotating convection~\citep{kJ98a,kJ12} characterized by $\gamma=0$, the first relation (\ref{AppEq:D3Psi}) enforces $D_3 \Psi=0$, 
which in turns implies that bounding 
surfaces are implicitly stress-free ($\baseContra{3}\bdot \bnabla (\baseCov{1,2} \bdot \ub) =0$).
This simplification is lost in the tilted case $\gamma\ne 0$ where
\begin{equation}
	\baseContra{3} \bdot \bnabla u_{1,2}
	= \left(-\varepsilon^{-1}\gamma\partial_2 + D_3\right) u_{1,2}\ne 0\,,
\end{equation} 
or, equivalently:
\begin{equation}
	\label{AppEq:stressFree_BC}
	 \left(-\varepsilon^{-1}\gamma\partial_2 + D_3\right) \left( \partial_{11} + \partial_{22} \right) \Psi + \partial_1 D_3 u^3\ne 0\,.
\end{equation}
Hence Ekman boundary layers are inevitably present along all mechanical boundaries as discussed further by \citet{troJFM24}.
The impact of the resulting Ekman pumping on the turbulent regime is within the scope of future work.

\section{Energetics}\label{appC}

Inspection of the $f$NHQG system reveals that the horizontally averaged kinetic 
energy, computed from either the covariant or the contravariant components
\begin{equation}
	\label{AppEq:kineticEnergy_covariant}
	\overline{\mathcal{E}}_{K} \equiv \frac{1}{2}\overline{ \ub\bdot\ub} 
	= \frac{1}{2}  \cos^2\vartheta_f \left(
	                        \overline{\left\|  \nabla^\prime _\perp   \Psi\right\|^2} +\overline{U_3U_3}
	                        \right)
\end{equation}
satisfies
\begin{equation}
 \partial_t\overline{\mathcal{E}}_{K} 
	+  \cos^2\vartheta_f D_3 \left( \overline{u^3 \Psi }\right)  =\frac{\wRa}{\sigma}\overline{u^3 \theta }
	- \cos^2\vartheta_f\left(  \overline{ \left( \nabla^{\prime 2}_\perp   \Psi \right)^2} + \overline{\left\| \nabla_\perp^{\prime} U_3\right\|^2}\right)\,.
\end{equation}
Integrating with respect to depth (denoted with angled brackets, as in the main text, 
e.g. Section~\ref{sec:baroman}, yields the corresponding volume-averaged expression:
\begin{equation}
 \partial_{t}\left\langle \overline{\mathcal{E}}_{K}\right\rangle 
	=\frac{\wRa}{\sigma^2}  \left( Nu -1 \right)
	- \cos^2\vartheta_f \left \langle \overline{ \left( \nabla^{\prime 2}_\perp   \Psi \right)^2} + \overline{\left\| \nabla_\perp^{\prime} U_3\right\|^2} \right \rangle
\end{equation}
from which the steady-state power integral of the first kind, $ \epsilon_{u} = \frac{\wRa}{\sigma^2}\lb Nu -1 \rb $,
is deduced, indicating a balance between kinetic energy dissipation and convective energy production.
	An alternative, yet rigorously equivalent, expression for the kinetic energy~\eqref{AppEq:kineticEnergy_covariant} can be formulated based on the contravariant velocity component $\ub\bdot \baseContra{3} = u^3 = w$: 
\begin{equation}
	\label{AppEq:kineticEnergy_contravariant}
	\overline{\mathcal{E}}_{K} = \frac{1}{2}\overline{ \ub\bdot\ub} 
	= \frac{1}{2}  \left(
			\overline{\left( \partial_1 \Psi\right)^2} +
			\overline{\left( \partial_2 \Psi \right)^2 }
			+ 2 \gamma \overline{u^3 \partial_1 \Psi} 
			+ \frac{\overline{u^3 u^3}}{\cos^2 \vartheta_f} 
	                \right)\,.
\end{equation}

The buoyant potential energy
\begin{equation}
	\overline{\mathcal{E}}_{P} = \frac{1}{2}\lb  \overline{ \theta^{2}} +\varepsilon^{-2}\overline{ \Theta}^2\rb
\end{equation}
satisfies 
\begin{equation}
  \partial_t \overline{\mathcal{E}}_{P}  + 
	D_3 \lb \overline{ \Theta }\, \overline{u^3 \theta}  \rb   
 =  \overline{u^3 \theta}+ 
	\frac{1}{\sigma} \lb\overline{ \Theta} D_{33}\overline{ \Theta} - \overline{\left\| \nabla^{\prime}_\perp  \theta \right\|^2} \rb
\end{equation}
 which upon integration with respect to depth yields
\begin{equation}
 \partial_t \left \langle \overline{\mathcal{E}}_{P} \right \rangle  
	= \frac{1}{\sigma}
	\left[ 
	     \left( Nu -1 \right)
	   - \left \langle 
		     \lb D_3\overline{ \Theta}\rb^2
                  +  \overline{\left\| \nabla^{\prime}_\perp  \theta \right\|^2} 
	      \right \rangle
	\right] \,.
\end{equation}
In steady state, the  power integral of the second kind, $\epsilon_{\vartheta} = Nu -1$, is obtained. Notably,
$ \epsilon_{u} = \frac{\wRa}{\sigma^2} \epsilon_{\vartheta}$.

\bibliographystyle{jfm}

\bibliography{JFM_FPLANE.bib}

\end{document}